НАЦІОНАЛЬНА АКАДЕМІЯ НАУК УКРАЇНИ

ІНСТИТУТ ФІЗИКИ

На правах рукопису

## ЛУЦИК ПЕТРО МИКОЛАЙОВИЧ

# ЕЛЕКТРОННІ ПРОЦЕСИ НА ГРАНИЦІ РОЗДІЛУ ОРГАНІЧНИХ НАПІВПРОВІДНИКІВ

спеціальність 01.04.15 – фізика молекулярних та рідких кристалів

дисертація на здобуття наукового ступеня

кандидата фізико-математичних наук

Науковий керівник:

Верцімаха Ярослав Іванович,

кандидат фіз-мат наук,

старший науковий співробітник

Київ – 2007

# ЗМІСТ

## ПЕРЕЛІК УМОВНИХ СКОРОЧЕНЬ

| | |
|---|---|
| ГР | границя розділу |
| ГС | гетероструктура |
| ІЧ | інфрачервоний |
| ККД | коефіцієнт корисної дії |
| НП | напівпровідник |
| СЕ | сонячний елемент |
| СП | спектр поглинання |
| СТ-стан | стан з переносом заряду (charge transfer state) |
| УФ | ультрафіолетовий |
| ФЕ | фото-ерс |
| ФЛ | фотолюмінесценція |

--------------------

| | |
|---|---|
| AFM | атомно-силова мікроскопія (atomic force microscopy) |
| CuPc | фталоціанін міді (cuprum phthalocyanine) |
| $E_g$ | ширина забороненої зони |
| $E_f$ | рівень Фермі (робота виходу електрона) |
| FF | фактор заповнення |
| hν | енергія фотону |
| HTP | гексатіопентацен (hexathiopentacene) |
| $I_c$ | потенціал іонізації |
| $I_{ph}$ | сила фотоструму |
| $I_{кз}$ | сила струму короткого замикання |
| ITO | суміш оксидів індію та олова (indium tin oxide) |
| $L$ | довжина дифузії |
| MDMO–PPV | похідна поліфенілвінілену – (poly[2-methoxy-5-(3,7-dimethyloctyloxy)-1,4-phenylenevinylene) |
| MPP | метил-заміщений периленовий барвник – (N, N`-dimethyl perylene-tetracarboxylic acid diimide) |
| P3HT | похідна політіофену – (poly{3-hexylthiophene}) |
| PCBM | похідна фулерену $C_{60}$ – (methanofullerene [6,6]-phenyl C61-butyric acid methyl ester) |
| PEDOT:PSS | шар електропровідного полімеру – (poly(3,4-ethylenedioxythiophene) poly{styrenesulfonate}) |
| PbPc | фталоціанін свинцю (lead phthalocyanine) |
| Pc | фталоціанін (phthalocyanine) |
| Pn | пентацен (pentacene) |
| $S$ | швидкість поверхневої рекомбінації |
| $SnCl_2Pc$ | $SnCl_2$ фталоціанін (dichlorotin-phthalocyanine) |
| SnPc | фталоціанін олова (tin phthalocyanine) |
| TiOPc | фталоціанін титанілу (titanyl phthalocyanine) |
| TTT | тетратіотетрацен (tetrathiotetracene) |
| $T_S$ | температура підкладки при термічному напиленні |
| $T_A$ | температура відпалу |
| VOPc | фталоціанін ванаділу (vanadyl phthalocyanine) |
| $V_D$ | висота потенціального бар'єру |
| $U_{xx}$ | напруга холостого ходу (напруга розімкненого кола) |
| ZnPc | фталоціанін цинку (zinc phthalocyanine) |

--------------

| | |
|---|---|
| $\alpha$ | коефіцієнт поглинання |
| $\beta$ | квантова ефективність фотогенерації |
| $\Delta E_C$ | розрив зон провідності |
| $\Delta E_V$ | розрив валентних зон |
| $\chi$ | спорідненість до електрона |
| $\varphi$ | фото-ерс |

<div align="center">ВСТУП</div>

**Анотація**

Дисертація присвячена дослідженню властивостей границі розділу фоточутливих органічних напівпровідників. В рамках роботи вивчено вплив відпалу та температури підкладки під час напилення на енергетичну структуру екситонів, оптичні та фотовольтаїчні властивості плівок $SnCl_2$ фталоціаніну, метил-заміщеного периленового барвника та гексатіопентацену.

Визначено оптимальні умови отримання анізотипних гетероструктур на основі метил-заміщеного периленового барвника та ізотипних гетероструктур на основі пентацену, які є перспективними для розробки органічних сонячних елементів.

Встановлено, що отримані результати для анізотипних гетероструктур узгоджуються з енергетичною діаграмою на основі моделі Андерсона, а для ізотипних гетероструктур описуються моделлю Ван Опдорп.

<u>Ключові слова</u>: органічні напівпровідники, гетероструктури, границя розділу, органічні сонячні елементи, екситони з переносом заряду та екситони Френкеля.

**Аннотация**

Диссертация посвящена исследованию свойств границы раздела фоточувствительных органических полупроводников. В рамках работы изучено влияние отжига и температуры подложки во время напыления на энергетическую структуру экситонов, оптические и фотовольтаические свойства пленок $SnCl_2$ фталоцианина, метил-замещенного периленового красителя и гексатиопентацена.

Определены оптимальные условия получения анизотипных гетероструктур на основе метил-замещенного периленового красителя и изотипных гетероструктур на основе пентацена, которые являются перспективными для разработки органических солнечных элементов.

Установлено, что полученные результаты для анизотипных гетероструктур согласуются с энергетической диаграммой на основе модели Андерсона, а для изотипных гетероструктур описываются моделью Ван Опдорп.

Ключевые слова: органические полупроводники, гетероструктуры, граница раздела, органические солнечные элементы, экситоны с переносом заряда и экситоны Френкеля.

**Summary**

The thesis is mainly devoted to investigation of optical and photovoltaic properties of iso-type and aniso-type heterostructures based on photosensitive layers of organic semiconductors. The basic attention in this work was focused on determination of optimal conditions for preparation of organic thin-film structures of methyl substituted perylene pigment, hexathiopentacene and $SnCl_2$ phthalocyanine. These structures are perspective for preparation of photosensitive heterostructures and can be used for development of effective photoconverters of solar energy.

To solve this task the influence of annealing and substrate temperature during thermal deposition of the films on morphology and surface structure, optical and photovoltaic properties of thin films of methyl substituted perylene pigment; hexathiopentacene and $SnCl_2$ phthalocyanine was studied. The surface (crystallites size, roughness, surface recombination rate) and bulk (diffusion length of excitons) parameters of the films under consideration were determined based on obtained results. These parameters can allow to determinate the efficiency of the films as the components of the heterostructures. Also energy structure of excitons in thin films of methyl substituted perylene pigment and $SnCl_2$ phthalocyanine was analyzed based on previous reference data and obtained experimental results. The contribution of Frenkel excitons and charge transfer excitons was considered based on the energy structures of investigated films.

The photovoltaic and optical properties of organic iso-type and aniso-type heterostructures prepared on the substrates with different temperatures were investigated. The analysis of obtained data for aniso-type heterostructures based on Anderson model and for iso-type heterostructures based on Van Opdorp model was carried out. Based on this analysis it was shown that organic aniso-type heterostructures based on methyl substituted perylene pigment and iso-type heterostructures based on pentacene can be described with the models proposed and developed for inorganic semiconductor heterojunctions.

The optimal conditions were determined for preparation of organic photosensitive heterostructures, which absorb most part of solar illumination in visible and near infrared range, effectively generate and separate charge carriers on the interface of heterojunction. Prepared under optimal conditions aniso-type heterostructures based on methyl substituted perylene pigment and iso-type heterostructures based on pentacene are perspective elements for development of organic photoconverters, including organic solar cells.

-----------------------------------------------------------------------------------

**Актуальність теми дисертаційного дослідження**

Дослідження фізичних явищ в тонкоплівкових гетероструктурах (ГС) на основі органічних напівпровідників (НП) пов'язані з можливістю створення ефективних органічних світлодіодів, транзисторів, сенсорів та сонячних елементів (СЕ) [1,2]. Для успішної розробки та підвищення ефективності органічних елементів НП техніки на основі тонкоплівкових ГС, незалежно від методів виготовлення, необхідне глибоке розуміння електронних процесів, що відбуваються як в об'ємі, так і на границі розділу (ГР) цих НП.

На даний час досліджено властивості ГР основних неорганічних НП, побудовано моделі та теорії гетеропереходів для випадку об'ємних та тонкоплівкових структур, з врахуванням енергетичної діаграми речовин, поверхневих станів, тощо [3]. Тим не менше проблеми створення та вдосконалення ГС залишаються актуальними. Одним з способів розв'язання цих проблем є пошук нових матеріалів та конструкцій ГС. При цьому особлива увага приділяється дослідженням ГС на основі органічних НП, фізичний базис знань для яких розвинутий недостатньо на теперішній час. Слід відмітити, що при застосуванні органічних ГС не ставиться задача замінити високоефективні елементи приладів на основі неорганічних НП. Органічні структури мають ряд переваг в порівняні з неорганічними елементами спеціального призначення. Так, наприклад, для органічних матеріалів простим та дешевим способом можна отримувати гнучкі тонкоплівкові структури великої площі, що робить органічні пристрої перспективними для практичного застосування [2]. Крім цього, враховуючи притаманну органічним матеріалам меншу ефективність поверхневої рекомбінації носіїв заряду (оскільки на ГР органічних НП відсутні розриви ковалентних зв'язків, які в неорганічних НП є центрами захоплення та рекомбінації носіїв заряду [4]), на основі органічних ГС можуть бути створені ефективні СЕ.

Отже, пошук нових органічних компонент ГС та детальне вивчення фотоелектричних процесів на ГР цих ГС необхідні для створення та впровадження ефективних органічних елементів оптоелектроніки. При цьому отримання органічних ГС з широкою областю фоточутливості, малою швидкістю поверхневої рекомбінації та високим запірним бар'єром на ГР компонент є актуальними задачами розробки ефективних органічних фотоперетворювачів сонячного випромінювання.

**Зв'язок роботи з науковими програмами, планами, темами**

Дисертаційна робота виконувалась згідно з планами відділу молекулярної фотоелектроніки Інституту фізики Національної Академії наук України в рамках науково-дослідних тем: 1.4.1 В/109 "Дослідження фотоелектронних властивостей нанокластерних структур на основі органічних композитних матеріалів" – номер держреєстрації 0104U000683 (2004-2006), „Фотоелектроніка багатофункціональних молекулярних композитів" (2007-2009).

**Мета і задачі дослідження**

**Метою** роботи було дослідження електронних процесів на ГР фоточутливих органічних шарів та ГС, отриманих на їх основі; пошук способів підвищення фоточутливості, а також встановлення зв'язку між енергетичною структурою та параметрами для досліджуваних органічних тонкоплівкових структур.

*Об'єктами досліджень* були термічно напилені тонкоплівкові структури НП: n-типу провідності – $SnCl_2$ фталоціанін ($SnCl_2Pc$), диметил перилен тетракарбоксил диімід або метил-заміщений периленовий барвник – methyl-substituted perylene pigment (MPP); p-типу провідності – гексатіопентацен (HTP), пентацен (Pn) та фталоціанін свинцю (PbPc), а також ізотипні (p-p – одного типу провідності) та анізотипні (p-n – різного типу провідності) двошарові ГС на основі досліджуваних органічних НП.

Для досягнення поставленої мети ставились та вирішувались наступні **задачі**:

– визначити основні закономірності електронних процесів та встановити способи покращення параметрів для шарів $SnCl_2Pc$, MPP та HTP, які є перспективними компонентами для розробки фоточутливих ГС;

– дослідити особливості фотовольтаїчних властивостей та визначити методи оптимізації фоточутливості анізотипних ГС на основі органічних НП і встановити зв'язок енергетичної структури з експериментальними значеннями фото-ерс (ФЕ);

– вивчити механізми формування ФЕ в ізотипних органічних ГС, оцінити можливість використання цих ГС для розширення спектральної області

фоточутливості та підвищення ефективності багатошарових фотоперетворювачів.

**Наукова новизна одержаних результатів** полягає в тому, що вперше:

– встановлено, що слабкі смуги в спектрах поглинання (СП) та фото-відбивання плівок $SnCl_2Pc$ (з максимумами при 1.34, 1.52 та 2.05 еВ) пов'язані з утворенням станів з переносом заряду (СТ-станів);

– запропоновано схему енергетичної структури екситонів для плівок МРР, яка враховує взаємодію між молекулами МРР в межах шару та між шарами;

– визначено довжину дифузії локалізованих екситонів ($L$) в термічно напилених плівках НТР, $SnCl_2Pc$ та МРР. При цьому в плівках МРР $L$ зростає вдвічі зі збільшенням температури підкладки ($T_S$) від 300 до 370 K при виготовленні плівок, тоді як в плівках НТР та $SnCl_2Pc$ $L$ практично не залежить від $T_S$;

– встановлено, що швидкість поверхневої рекомбінації носіїв заряду ($S$) на вільній поверхні плівок МРР значно більша, ніж біля ГР з ІТО електродом, що пов'язано з більш ефективною адсорбцією активних газів (напр., кисню) на вільній поверхні плівок, ніж зі сторони ІТО контакту. При цьому $S$ зменшується при відпалі або напиленні плівок МРР при температурі 370 K. Також показано, що ефективність рекомбінації на ГР досліджуваних ГС менша, ніж на вільній поверхні плівок $SnCl_2Pc$, МРР та НТР, що може бути обумовлено десорбцією активних газів з вільної поверхні плівок при виготовленні ГС;

– встановлено, що двошарові анізотипні органічні ГС якісно описуються за енергетичною моделлю Андерсона, яка не враховує поверхневі стани [1];

– показано, що в ізотипних ГС Pn/PbPc отриманих при $T_S = 300$ K на ГР формується велика кількість поверхневих станів і така ГС описується еквівалентною схемою двох діодів Шотткі послідовно з'єднаних назустріч один одному;

– встановлено, що підвищення $T_S$ від 300 до 370 К при напиленні шарів для ізотипних ГС Pn/PbPc дозволяє зменшити концентрацію поверхневих станів на ГР і розширити спектральну область фоточутливості в порівнянні з областю фоточутливості Pn.

**Практичне значення одержаних результатів** полягає в наступному:

– розроблено методи підвищення фоточутливості компонент ГС – тонких плівок органічних НП: МРР та НТР;

– виготовлено та досліджено нову ГС (PbPc/МРР), спектральна область фоточутливості якої більша, ніж раніше досліджених органічних ГС, і близька до спектральної області фоточутливості кремнієвих елементів;

– встановлено, що використання ізотипних ГС Pn/PbPc, отриманих при $T_S =$ 370 К, для створення p-p$^+$ переходів в багатошарових органічних СЕ, дозволяє збільшити ФЕ та розширити спектральну область фоточутливості цих елементів;

– визначені технологічні умови виготовлення органічних ГС, при використанні яких можна зменшити $S$ та розширити спектральну область фоточутливості багатошарових органічних СЕ на основі ГС.

**Особистий внесок здобувача**

Автор самостійно одержав нові експериментальні дані з дослідження оптичних та фотовольтаїчних властивостей тонкоплівкових структур, проводив обробку та узагальнення результатів. Постановка задачі, вибір об'єктів та методів дослідження; обговорення та інтерпретація результатів, а також формулювання висновків проведено спільно з науковим керівником Я.І. Верцімахою. Автор приймав безпосередню участь в підготовці наукових робіт до публікації. Вимірювання спектрів фото-модульованого відбивання та їх аналіз проводились спільно з Я. Місієвичем та А. Подгородецьким. Вимірювання спектрів фотолюмінесценції та поглинання, а також обговорення та інтерпретація отриманих результатів відбувались спільно з

К. Палевською та Ю. Свораковським. Зображення поверхні плівок на атомно-силовому мікроскопі отриманні О. Литвин. Автор щиро вдячний всім співавторам за плідну співпрацю.

**Апробація результатів дисертації** проходила на наступних наукових форумах: 5th International Conference on "Electronic Processes in Organic Materials" (ICEPOM-5) Kyiv, 2004; International Conference "Modern Problems of Condensed Matter Optics" (MPCMO), Kyiv, 2006; 6th International Conference on "Electronic Processes in Organic Materials" (ICEPOM-6), Gurzuf, Crimea, Ukraine, 2006; III Ukrainian Scientific Conference on Physics of Semiconductor (USCPS-3), Odesa, Ukraine, 2007 та на семінарах відділу молекулярної фотоелектроніки Інституту фізики НАН України.

**Публікації**

За темою дисертації опубліковано 12 наукових праць: з них 6 статей та 6 тез конференцій.

**Структура та обсяг дисертації**

Дисертація складається з вступу, п'яти розділів, висновків та списку використаної літератури з 171 найменувань. Текст дисертаційної роботи викладено на 142 сторінках, містить 56 рисунків та 8 таблиць.

У **першому розділі** представлено огляд літератури за темою дисертації. Охарактеризовано основні електронні процеси в органічних НП, розглянуто фотоелектричні властивості двошарових ГС, проаналізовано переваги, недоліки та проблеми створення органічних СЕ. Особливу увагу зосереджено на даних, що стосуються основних класів органічних НП досліджуваних в рамках цієї роботи.

В **другому розділі** описано технологію виготовлення зразків (плівок, структур ITO/органічний НП та ГС), наведено опис методів дослідження, особлива увага приділялась методиці вимірювання ФЕ досліджуваних структур. Також наведено критерії та обґрунтування вибору компонент для розробки фоточутливих органічних ГС.

В **третьому розділі** представлено результати дослідження структури поверхні, оптичних та фотовольтаїчних властивостей плівок $SnCl_2Pc$, MPP та HTP. Вивчено вплив відпалу та $T_S$ під час термічного напилення на енергетичну структуру екситонів для плівок $SnCl_2Pc$ та MPP. Визначено довжину дифузії екситонів ($L$) в досліджуваних плівках $SnCl_2Pc$, MPP та HTP.

В **четвертому** розділі представлені результати дослідження фотовольтаїчних властивостей анізотипних (p-n) ГС на основі шарів Pn, PbPc та описаних в попередньому розділі шарів $SnCl_2Pc$, MPP та HTP. Проведено аналіз отриманих результатів та побудовано енергетичні діаграми для досліджуваних анізотипних ГС на основі моделі Андерсона. Показано, що двошарові анізотипні ГС MPP/Pn, PbPc/MPP отримані при $T_S$ до 370K є перспективними елементами для розробки органічних фотоперетворювачів.

В **п'ятому** розділі наведено результати досліджень фотовольтаїчних властивостей ізотипних ГС на основі Pn та інших компонент p-типу провідності HTP та PbPc. Проведено аналіз отриманих результатів за моделлю Ван Опдорп для ізотипних ГС. Визначено оптимальні умови отримання фоточутливих ізотипних ГС на основі Pn, які є перспективними для розробки органічних CE.

# РОЗДІЛ 1

# ФОТОЕЛЕКТРИЧНІ ВЛАСТИВОСТІ ГЕТЕРОСТРУКТУР НА ОСНОВІ ОРГАНІЧНИХ НАПІВПРОВІДНИКІВ

1.1. Фотоелектричні явища в органічних напівпровідниках: поглинання світла та фотогенерація носіїв заряду

В останні десятиліття досягнуто значних успіхів у синтезі та дослідженні органічних НП – перспективного матеріалу для оптоелектронного та фотовольтаїчного застосування (польові транзистори, світлодіоди, СЕ, оптичні перемикачі, сенсори, тощо) [1,2]. Поглинання світла та фотогенерація носіїв заряду в органічних НП характеризується рядом особливостей і принципово відрізняється від аналогічних процесів в неорганічних матеріалах. В неорганічних НП атоми сильно зв'язані ковалентними зв'язками і поглинання фотона відбувається кристалом як єдиною системою. При цьому поглинута енергія фотона викликає перехід електрона з валентної зони в зону провідності. Таким чином утворюються носії заряду, які можуть вільно переміщуватись по об'єму кристала [5,6].

1.1.1. Нейтральні та збуджені стани. Ексітони Френкеля та СТ-стани

Органічні НП характеризуються слабкими міжмолекулярними взаємодіями типу Ван-дер-Ваальса, і фотон поглинається окремою молекулою, а не кристалом. При цьому відбувається електронний перехід між відповідними енергетичними рівнями молекули з утворенням не вільних носіїв заряду, а *нейтральних молекулярних збуджень*, які також можуть переміщуватись по кристалу. Необхідною умовою міграції молекулярних збуджень є наявність перекриття молекулярних хвильових функцій, що забезпечується регулярною та щільною упаковкою молекул в кристалі. Нейтральні молекулярні збудження, які можуть переміщуватись по кристалу називають молекулярним екситоном або *екситоном Френкеля* [6,7]. При поглинанні фотона в органічних НП зазвичай утворюються синглетні (S) екситони (сумарний спін електронів: s = 0). Триплетні (T) екситони (s = 1) практично не утворюються безпосередньо в процесі поглинання світла, і формуються переважно в результаті безвипромінювальної S→T конверсії [6].

Енергія екситона в молекулярних кристалах або органічних НП $\Delta E_e$ пов'язана з енергією збудження ізольованої молекули (мономера) $E_m$:

$$\Delta E_e = E_m + \Delta D + \gamma_e \qquad (1.1)$$

де $e$ – кількість молекул в елементарній комірці кристалу, $\Delta D$ – зміна енергії внаслідок взаємодії Ван-дер-Ваальса при переході від основного до збудженого стану та $\gamma_e$ – дисперсія енергії екситона внаслідок резонансної взаємодії нееквівалентних молекул в елементарній комірці, що призводить до „давидівського" розщеплення збуджених рівнів при $e > 1$. Величини $\Delta D$ та $\gamma_e$ невеликі, тому енергетичні рівні екситонів ізольованої молекули та кристала мають близькі значення [6,7].

Екситон Френкеля розглядається як сильно зв'язана система збудженого стану окремої молекули. З іншої сторони існує модель екситонів Ван'є-Мотта – слабозв'язаної електронно-діркової пари з малою кулонівською взаємодією. Ця модель справедлива для речовин з великими значеннями діелектричної проникності $\varepsilon$. Для більшості органічних речовин (за винятком сильно легованих) характерні невеликі значення $\varepsilon$ і, відповідно, сильна кулонівська взаємодія. Отже, на відміну від наближення екситонів Френкеля, модель Ван'є-Мотта вважається неприйнятною для органічних НП [5].

При переміщенні екситонів в молекулярних кристалах важливу роль відіграє взаємодія екситонів з коливаннями гратки – фононами. Тільки при дуже низьких температурах і малих коливаннях решітки, екситони в певній мірі делокалізовані і тому рухаються у вигляді когерентної хвилі, фактично не взаємодіючи з фононами. При температурах 30-40 K проявляється взаємодія екситона з фононами, при цьому відбувається перехід від когерентного хвилеподібного до некогерентного дифузійного переміщення екситона. Експериментально підтверджено, що при температурах більше 70 K рух екситонів некогерентний (екситони переміщуються в кристалі перескоками з однієї молекули на іншу) і рух екситонів можна описати в рамках макроскопічної дифузійної теорії [6]. Тобто внаслідок слабкої міжмолекулярної взаємодії в органічних кристалах екситони та окремі носії заряду мають тенденцію до локалізації на окремих молекулах з часом життя $10^{-14}$-$10^{-12}$ с. Таким чином, в органічному кристалі з'являється якісно новий ефект взаємодії носія заряду з оточуючою граткою, так звана електронна поляризація кристала носієм заряду. Електронна поляризація – процес дуже швидкий ($10^{-16}$-$10^{-15}$ с), тому за час локалізації носій встигає поляризувати електронні $\pi$-орбіталі сусідніх молекул і фактично переміщується разом зі своєю поляризаційною оболонкою. В неорганічних НП вільні носії заряду рухаються у вигляді електронної хвилі і не встигають поляризувати кристал. Електронна поляризація є багато-електронним явищем, тому електронні процеси в органічних НП в принципі не можна розглядати в рамках одно-електронного наближення зонної моделі. Оскільки, рівні провідності дірок та електронів органічних НП є багато-електронними рівнями власної енергії і не є аналогами краю валентної зони та зони провідності ковалентного кристалу, то оптичні переходи в органічних НП не можна описувати як міжзонні переходи. Але не зважаючи на це, на основі зонної теорії були розраховані рухливості носіїв струму в органічних НП, значення яких добре узгоджується з експериментом [6,7].

Окрім нейтральних екситонів Френкеля, коли електрон і дірка знаходяться на одній і тій же молекулі, існують збуджені стани, в яких електрон переходить на іншу (як правило, на сусідню або наступну за нею) молекулу, але залишається зв'язаний з діркою кулонівським полем взаємодії (рис. 1.1). Ці електронно-діркові пари називають екситонами або станами з переносом заряду (CT-стани – charge transfer states). CT-стани не є аналоги екситонів Ван'є-Мотта, оскільки електрон і дірка в CT-стані локалізовані на певних молекулах і можуть утворювати іонні стани. Вважається, що CT-стан

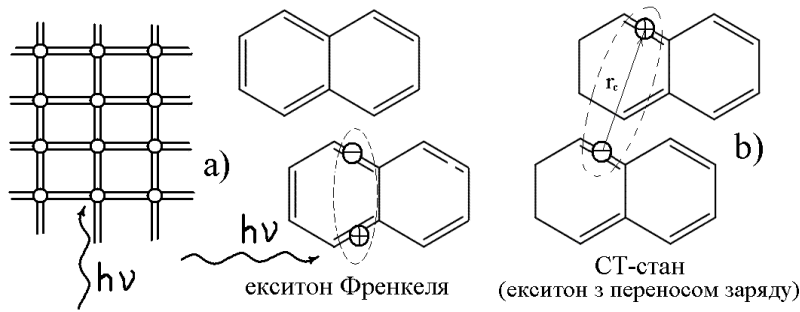

Рис. 1.1 Схематичне зображення взаємодії світла з ковалентним (a) та молекулярним (органічним) (b) НП

формується при відстані між електроном та діркою меншій за критичну відстань кулонівського захоплення: $r_c = e^2/\varepsilon\, kT$ (рис. 1.1). Електрон і дірка в СТ-стані мігрують некогерентно (перескакуючи з молекули на молекулу, змінюють відстань один відносно одного – модель Онзагера) [6,8].

Внаслідок слабкої міжмолекулярної взаємодії молекули органічних НП в твердому стані майже повністю зберігають свої властивості. Але в результаті колективної взаємодії молекул в молекулярному кристалі виникають властивості не притаманні молекулам з яких формується кристал (наприклад, фотопровідність). Так органічні молекули з поліспряженими зв'язками містять делокалізовані слабозв'язані π-електрони, які при освітленні є потенційними джерелами вільних носіїв заряду і виникнення фотопровідності. Тому часто органічні НП називають фотопровідниками [9].

### 1.1.2. Фотогенерація носіїв заряду в органічних фотопровідниках

Найбільш важливим для розуміння процесів фотопровідності є механізм власної багатоступінчастої фотогенерації носіїв заряду, на початковому етапі якого відбувається поглинання фотона з утворенням екситона Френкеля. На наступному етапі в результаті дисоціації екситона Френкеля утворюються квазівільний електрон та локалізована дірка, які віддаляються один від одного на деяку відстань $r_0$ – середню довжину термалізації. При $r_0 > r_c$ утворюються вільні носії, а при $r_0 < r_c$ – СТ-стани, які на останньому етапі розпадаються з утворенням вільних носіїв заряду [10-11].

Квантова ефективність фотогенерації ($\beta$) носіїв заряду для органічних фотопровідників сильно залежить від енергетичної структури нейтральних збуджених станів. Наприклад, антрацен є поганим фотопровідником внаслідок малих значень ефективності автоіонізації та термалізації екситонів Френкеля, при цьому основна частина енергії збудження втрачається на флуоресценцію, а Pn, навпаки, – хороший фотопровідник, через високі значення вищезгаданих ефективностей [10].

Для багатьох органічних НП експериментально встановлено, що при опроміненні світлом з $h\nu > E_g$, носії фотоструму генеруються, переважно в результаті власної фотогенерації, а при $h\nu < E_g$ – шляхом невласної фотогенерації [11]. Невласна фотогенерація носіїв заряду відбувається в результаті взаємодії екситонів з локальними центрами захоплення, домішковими молекулами, ГР, поверхнею кристалу, електродом, тощо. При цьому механізм фотогенерації носіїв заряду залежить не тільки від $h\nu$, а й від інтенсивності збуджуючого світла, типу дефектів, стану поверхні, або інжекційної здатності електродів [10-11].

Одновимірний розподіл концентрації фотогенерованих екситонів $\Delta n(x)$ в органічному шарі за дифузійним наближенням з врахуванням процесів фотогенерації та поверхневої рекомбінації має вигляд:

$$\Delta n(x) = n - n_0 = \beta F_0 \frac{\alpha L_e}{1 - \alpha^2 L_e^2}\left(exp(-\alpha x) - \frac{\alpha + S_e/D_e}{1/L_e + S_e/D_e} \cdot exp(-x/L_e)\right), (1.2)$$

де $F_0$ – кількість падаючих квантів на поверхню певної площі за секунду; $\alpha$ – коефіцієнт поглинання; $S_e$, $L_e$, $D_e$ – швидкість поверхневої анігіляції, довжина і коефіцієнт дифузії екситонів. [12,13]. Такий же вираз отримано для розподілу нерівноважних носіїв фотоструму, де параметри екситонів у формулі (1.2) слід замінити на параметри нерівноважних носіїв струму ($S$, $L$, $D$ – швидкість поверхневої рекомбінації, довжина та коефіцієнт дифузії нерівноважних носіїв струму) [14]. Інтегрування виразу (1.2) при певних граничних умовах дозволяє отримати залежність фотоструму від $\alpha$, $S$ та $L$. Так, відношення сили фотоструму до інтенсивності освітлення (спектральний відгук) сильно зменшується зі збільшенням $S$, особливо при великих $\alpha$, та зростає зі збільшенням $L$. В загальному випадку для збільшення спектрального відгуку слід зменшувати $S$ та збільшувати $L$ [15]. Аналізуючи окремі (в основному короткохвильові) області спектральної відгуку можна оцінити значення $L$ та відношення $S/D$ [16].

## 1.2. Фізичні властивості напівпровідникових двошарових гетероструктур
### 1.2.1. Енергетична структура анізотипних гетероструктур

В останні роки приділяється велика увага дослідженням ГС на основі органічних НП [3,4]. Це пов'язано з тим, що ГР ГС є ключовим елементом органічних СЕ [17-19] та світлодіодів [20-22]. В ГС розділення або рекомбінація електронно-діркових пар відбувається переважно на ГР шарів n- та p-типу провідності (анізотипні ГС), де в результаті вирівнювання енергетичних рівнів окремих шарів формується згин енергетичних зон, який відіграє важливу роль в роботі пристроїв на основі ГС [23]. Тому побудова діаграм вирівнювання енергетичних рівнів біля ГР компонент ГС є важливим елементом контролю та оптимізації пристроїв на основі ГС.

Модель ідеального різкого анізотипного гетеропереходу між двома НП з різними значеннями $E_g$, $\chi$ та $E_f$ (рис. 1.2a) була запропонована Андерсоном [3]. Після приведення в контакт і встановлення рівноваги, $E_f$ НП займають однакове положення (рис. 1.2b). Це досягається шляхом переміщення електронів з НП 2 (n-типу) в НП 1 (p-типу), внаслідок градієнта концентрацій носіїв заряду (рис. 1.2b). В результаті вирівнювання $E_f$ між двома компонентами ГС виникає контактна різниця потенціалів або потенціальний бар'єр рівний різниці $E_f$: $V_D = E_f(_p) - E_f(_n)$ та утворюються розриви зон провідності ($\Delta E_C = \chi_n - \chi_p$) та валентних зон ($\Delta E_V = I_n - I_p$) [3]. Після розділення полем p-n переходу неосновні носії заряду (дірки з n-області переходять в p-область, а електрони з p-області виявляються в n- області) стають вже основними носіями заряду у відповідних областях СЕ [16].

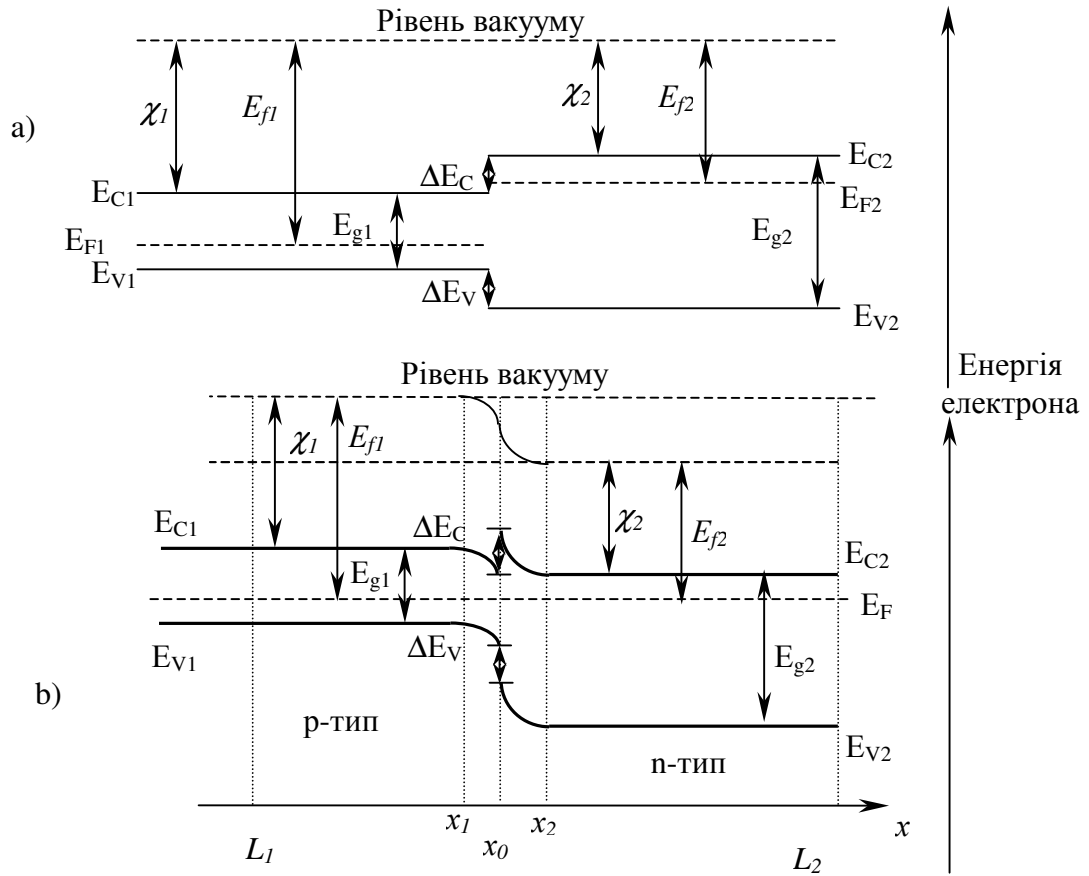

Рис. 1.2  Діаграми енергетичних зон двох НП p- (1) та n- (2) типу при умові нейтральності (a) та для гетеропереходу в умовах рівноваги (b).

В неорганічних ГС фотони з енергіями $E_{g2} > h\nu > E_{g1}$, поглинаються в першому (вузькозонному) НП, а – з $h\nu > E_{g2}$ в другому (широкозонному) НП (рис. 1.2) і гетероперехід буде розділювати носії, що фотогенеруються на відстані від переходу, яка не перевищує $L$, або ж безпосередньо в області просторового заряду переходу. Це явище називається „ефектом вікна" [15,24] і часто призводить до збільшення фотовідгуку в анізотипних ГС [24].

При $\Delta E_V < 0$ (або $\Delta E_C < 0$) на ГР компонент ГС утворюється бар'єр для фотоструму (рис. 1.2), який перешкоджає збиранню носіїв заряду. При цьому зі збільшенням $|\Delta E_V|$ (або $|\Delta E_C|$) відбувається обмеження інжекції носіїв заряду через перехід, що негативно впливає на фотовольтаїчні властивості ГС [4,15,25]. У випадку ж $\Delta E_C > 0$ створюється додаткове поле на ГР p-n переходу, яке не перешкоджає збиранню носіїв заряду. Але з іншої сторони, при великих $\Delta E_C > 0$ можлива рекомбінація носіїв заряду через стани на ГР [3]. При цьому у випадку $V_D < 0$ в ГС формується згин зон, що перешкоджає збиранню носіїв заряду, і такі ГС менш підходять для створення СЕ [26].

Для більш повного відображення реальної ГС Долега запропонував емісійно-рекомбінаційну модель (рис. 1.3), яка враховує наявність на ГР великої концентрації поверхневих станів, і як наслідок велику $S$ [3]. За цією моделлю вольт-амперні залежності при відсутності фотоструму можна представити у вигляді:

$$I = B \cdot \exp(-\frac{eV_D}{nkT}) \cdot [\exp(\frac{eU}{nkT}) - 1] \qquad (1.3)$$

де $B$ – коефіцієнт, слабо залежить від температури, значення $n$ залежить від відношення щільностей дефектів в двох НП [3].

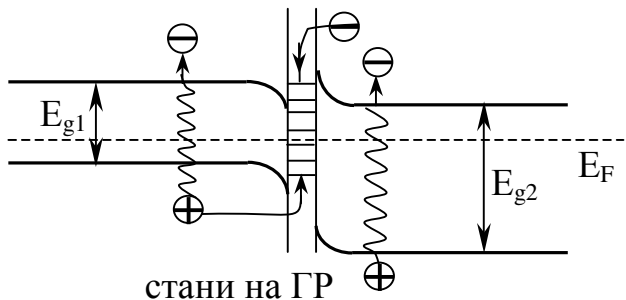

стани на ГР

Рис. 1.3 Схематичне зображення емісійно-рекомбінаційної моделі різкого p-n гетеропереходу

В ГС проходять два основних процеси фотогенерації носіїв заряду: утворення вільних дірок чи електронів внаслідок фотозбудження домішкових станів або станів на ГР та формування електронно-діркових пар внаслідок переходу електрона з валентної зони в зону провідності. При цьому фотострум через гетероперехід з'являється, якщо фотогенерація відбувається біля ГР в межах $L$ та при наявності внутрішнього електричного поля [24].

Загальний вираз вольт-амперної характеристики має вигляд:

$$I = I_S [exp(eU/kT) - 1] - I_{ph}, \qquad (1.4)$$

де $I_S$ та $I_{ph}$ – сила "темнового" струму насичення та фотоструму, відповідно. $I_{ph}$ залежить від геометричних розмірів ГС, напрямку освітлення та фізичних параметрів НП шарів. Коли $I_{ph}$ не залежить від $U$, тобто при відсутності фотогенерації та рекомбінації носіїв струмі в збідненому шарі: $I_{ph} = I_{кз}$. У випадку розімкнутого кола ($I = 0$) вираз для ФЕ або $U_{xx}$ має вигляд [14-16,24]:

$$U_{xx} = \frac{kT}{e} \cdot \ln(1 + \frac{I_{ph}}{I_S}) . \tag{1.5}$$

Ефективність фотоперетворення (ККД) визначається добутком $U_{xx}$, $I_{кз}$ та FF поділеному на інтенсивність освітлення ГС.

При освітленні ГС перпендикулярно до площини переходу та зі сторони НП з більшою $E_g$ (на рис. 1.2 $E_{g2}$) потоком фотонів $F_0$ та в результаті поглинання фотонів в ГС генеруються дві компоненти $I_{ph}$: у вузькозонному ($I_{ph1}$) та, відповідно, широкозонному ($I_{ph2}$) НП [24,27]:

$$I_{ph1} = eF_0\beta_1(h\nu)\,exp(-\alpha_2 d_2)\{1 - exp(-\alpha_1 l_1) + \frac{\alpha_1 L_1}{1-(\alpha_1 L_1)^2} \times$$

$$\left[ \left( \frac{sh\left(\frac{d_1-l_1}{L_1}\right) + \frac{S_1 L_1}{D_1} ch\left(\frac{d_1-l_1}{L_1}\right)}{ch\left(\frac{d_1-l_1}{L_1}\right) + \frac{S_1 L_1}{D_1} sh\left(\frac{d_1-l_1}{L_1}\right)} \right) - \alpha_1 L_1 \right] exp(-\alpha_1 l_1) + \frac{\left(\alpha_1 L - \frac{S_1 L_1}{D_1}\right) exp(-\alpha_1 d_1)}{ch\left(\frac{d_1-l_1}{L_1}\right) + \frac{S_1 L_1}{D_1} sh\left(\frac{d_1-l_1}{L_1}\right)} \right] \} \tag{1.6}$$

$$I_{ph2} = eF_0\beta_2(h\nu)\,exp(-\alpha_2 d_2)\{exp(\alpha_2 l_2) - 1 + \frac{\alpha_2 L_2}{1-(\alpha_2 L_2)^2} \times$$

$$\left[ \left( \frac{sh\left(\frac{d_2-l_2}{L_2}\right) + \frac{S_2 L_2}{D_2} ch\left(\frac{d_1-l_1}{L_1}\right)}{ch\left(\frac{d_2-l_2}{L_2}\right) + \frac{S_2 L_2}{D_2} sh\left(\frac{d_2-l_2}{L_2}\right)} \right) + \alpha_2 L_2 \right] exp(\alpha_2 l_2) - \frac{\left(\alpha_2 L_2 + \frac{S_2 L_2}{D_2}\right) exp(\alpha_2 d_2)}{ch\left(\frac{d_2-l_2}{L_2}\right) + \frac{S_2 L_2}{D_2} sh\left(\frac{d_2-l_2}{L_2}\right)} \right] \} \tag{1.7}$$

де $\beta$, $\alpha$, $d$, $l$, $L$, $D$, $S$ – квантова ефективність фотогенерації, коефіцієнт поглинання, товщина шару, область просторового заряду, довжина та коефіцієнт дифузії, швидкість поверхневої рекомбінації (індекси 1 та 2 відносяться до нижнього та верхнього шару НП, відповідно).

Одним з найбільш важливих параметрів ГС є спектральна залежність фотовідгуку – залежність фотоструму ($I_{кз}$ або $U_{xx}$) від $h\nu$. В наближенні малих сигналів [24,28,29] $I_{кз}$ анізотпних ГС прямо пропорційний $U_{xx}$. Тому спектр фотовідгуку не залежить від способу вимірювання ($I_{кз}$ чи $U_{xx}$) і на практиці визначається зовнішнім імпедансом [24].

Біля p-n переходу відбувається ефективне розділення як нерівноважних носіїв заряду, так і екситонів. У випадку коли екситони не мають достатньої енергії щоб дисоціювати в об'ємі органічних НП, то при наявності на ГР згину зон більшого, ніж енергія зв'язку екситона, можливе розділення екситонів полем гетеропереходу [30]. Найбільш ефективне розділення екситонів в органічних НП відбувається на ГР НП p- і n-типу. Особливо у випадку, коли формується ГС на основі НП p-типу провідності з малим значенням $I_C$ і n-типу провідності з великим $\chi$. При цьому ефективність фотогенерації носіїв заряду в двошарових органічних СЕ на основі p-n переходу може бути на кілька порядків більше, в порівнянні з одношаровими СЕ [31].

### 1.2.2. Електронні процеси на границі розділу ізотипних гетероструктур

Використання ізотипної та анізотипної ГС (напр. $p^+$-p-$n^+$), дозволяє збільшувати ККД СЕ. Так, введення ізотипного ($p^+$-p або $n^+$-n) переходу в конструкцію СЕ дозволяє збільшити коефіцієнт збирання неосновних носіїв заряду [32]. Ізотипні ГС можуть мати випрямляючі властивості. А при наявності поверхневих станів з захопленням електронів на ГР можуть існувати збіднені шари, які значно збільшують $V_D$ ГС. Переважну роль в ізотипних ГС, на відміну від анізотипних, відіграють основні носії заряду [4]. Для опису ізотипних ГС Ван Опдорпом запропоновано різні моделі як без врахування, так і з врахуванням станів на ГР, наявність яких залежить від узгодження граток і методу виготовлення структур. Так, за відсутності станів на ГР поява $I_{ph}$ пов'язана з фотоемісією носіїв через перехід вузькозонного шару та генерацією електронно-діркових пар в широкозонному матеріалі поблизу ГР. Згідно з цією моделлю $I_{ph}$ не залежить від прикладеної напруги і еквівалентний $I_{кз}$, а $U_{xx}$ виражається формулою (1.5). При наявності ж станів на ГР, з'являється $I_{ph}$ пов'язаний з фотоемісією захоплених носіїв з вищезгаданих станів. Стани на ГР призводять до утворення двох збіднених

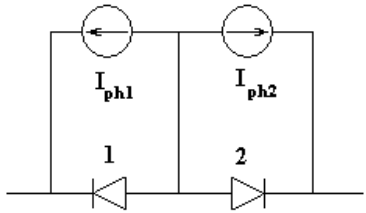

Рис. 1.4 Еквівалентна схема ізотипної ГС з великою концентрацією станів на ГР.

шарів, і тому гетероперехід можна представити у вигляді двох діодів Шотткі, з'єднаних назустріч один одному. А згідно з еквівалентною схемою (рис. 1.4), $U_{xx}$ (або $\varphi$) можна представити наступним чином:

$$\varphi = \varphi_1 \cdot ln\left(1 + \frac{I_{ph1}}{I_{S1}}\right) - \varphi_2 \cdot ln\left(1 + \frac{I_{ph2}}{I_{S2}}\right). \qquad (1.9)$$

Оскільки $I_{ph1}$ та $I_{ph2}$ сильно залежить від hv, то і фотовідгук ізотипних ГС також буде залежати від hv (при цьому можлива переміна знаку) [24].

Ізотипні ГС з подвійним збідненням (при наявності станів на ГР) можуть призводити до випрямлення або насичення в обох напрямках. Схожий ефект спостерігається у випадку послідовного з'єднання металічними контактами двох діодів метал – НП. Структуру, в якій носії струму переходять спочатку з одного НП на стани біля ГР, а потім в інший НП, важко відрізнити від послідовно з'єднаних контактів метал – НП [33].

### 1.3. Переваги та недоліки, проблеми та перспективи широкого застосування органічних сонячних елементів

Дослідження та розробка органічних СЕ викликані пошуком дешевих та гнучких модулів для виробництва електроенергії. Останнім часом спостерігається суттєве зростання ККД органічних СЕ (від 1 до 5%) [17,34] на основі багатошарових та "об'ємних" тонкоплівкових ГС. Кращі результати отримані для органічних СЕ наведені в табл. 1.2 (без врахування результатів для гібридних СЕ).

До кінця 70-их років XX ст. досліджувались лише одношарові органічні СЕ (органічний шар розміщується між двома різними електродами), ККД

яких не перевищував 1% [5,35,36]. Лише після отримання Тангом в 1986 р. [17] ККД порядку 1% для тонкоплівкових двошарових органічних СЕ на основі CuPc та бензімідної похідної перилену (PTCBI), розпочалися активні дослідження та пошук нових органічних матеріалів та ГС з метою фотовольтаїчного застосування [37,38]. $U_{xx}$ одношарових СЕ визначається

Таблиця 1.1

Основні характеристики та параметри СЕ на основі органічних НП

| Тип структури | $U_{xx}$, В | $I_{кз}$, мА/см$^2$ | $P$, мВт/см$^2$ | FF | ККД, % | рік |
|---|---|---|---|---|---|---|
| Ag/мероціанін/Al | 1.2 | 0.18 | 78 | 0.25 | **0.7** | 1978[35,36] |
| ITO/CuPc/PTCBI/Ag | 0.45 | 2.3 | 75 | 0.65 | **0.9** | 1986 [17] |
| ITO/MPP/H$_2$Pc/Au | 0.4 | 2.7 | 78 | 0.56 | **0.76** | 1990 [39] |
| ITO/MPP/CuPc/Au | 0.42 | 1.9 | 100 | 0.41 | **0.33** | 1995 [40] |
| ITO/MPP/TiOPc/Au | 0.34 | 3 | 108 | 0.36 | **0.34** | 1996 [41] |
| ITO/TiOPc/MPP/Au | ~0.4 | ~0.6 | ~108 | ~0.38 | **~0.08** | |
| ITO/MPP/C$_{60}$:TiOPc/TiOPc/Au | 0.53 | 3.4 | 94 | 0.32 | **0.61** | 2000 [42] |
| ITO/MPP/ZnPc/Au | 0.37 | 2.5 | 82 | 0.385 | **0.41** | 2000 [43] |
| ITO/MPP/ZnPc:C60(1:1)/ZnPc/Au | 0.4 | 5.2 | 86 | 0.5 | **1.04** | 2000 [44] |
| ITO/PEDOT:PSS/.../ ZnPc:C$_{60}$(1:2)/MPP/LiF/Al | 0.5 | 6.3 | 100 | 0.33 | **1.04** | 2003 [45] |
| ITO/PEDOT:PSS/.../ ZnPc:C$_{60}$(1:2)/MPP/LiF/Al | 0.46 | 7.6 | 100 | 0.5 | **1.75** | 2004 [46] |
| ITO/PEDOT:PSS/CuPc/C$_{60}$/.../Al | 0.56 | 50 | ~400 | 0.6 | **4.2** | 2004[47] |
| нанокристалічна суміш CuPc/C$_{60}$ | – | – | ~85 | – | **6.2** | 2007 [48] |
| ITO/PEDOT:PSS/MDMO–PPV/PCBM/Al | 0.78 | 0.96 | 78 | ~0.5 | **0.5** | 2000 [49] |
| ITO/PEDOT:PSS/MDMO–PPV:PCBM/LiF/Al [1] | 0.82 | 5.25 | 80 | 0.61 | **2.5** | 2001[18,50] |
| ITO/PEDOT:PSS/P3HT:PCBM/LiF/Al[1] | 0.55 | 8.5 | 80 | 0.6 | **3.5** | 2003 [51] |
| ITO/PEDOT:PSS/P3HT:PCBM/LiF/Al[1] | 0.6 | 11.1 | 80 | 0.54 | **4.5** | 2005 [52] |

Примітка: 1. об'ємні ГС;

різницею $E_F$ електродів або бар'єром Шотткі між електродом та органічним шаром, тому фотовольтаїчні властивості одношарових структур сильно залежать від природи електродів. Недоліком одношарових СЕ є малий FF, що зазвичай пов'язано з великим послідовним опором, який мають органічні НП та залежністю $\beta$ від прикладеної напруги. Для двошарових СЕ $\beta$ практично не залежить від прикладеної напруги, тому ці структури мають більше значення FF [17]. Також ККД органічних СЕ сильно залежать від $\beta$ та ефективності транспорту носіїв заряду [53].

Так, для СЕ ITO/MPP/ZnPc/Au зі збільшенням інтенсивності освітлення зменшуєшся FF, зростає $I_{кз}$, а $U_{хх}$ не змінюється. При цьому спектри $I_{кз}$ корелюють з СП шару, який освітлюється останнім, що свідчить про генерацію фотоструму в тонкій області поблизу ГР СЕ. Адже, якщо фотон поглинається далеко від ГР, то спектр $I_{кз}$ корелює з СП обох шарів, а при освітленні з різних сторін спектри $I_{кз}$ були б однаковими [43]. Оскільки процеси фотогенерації в органічних СЕ відбуваються переважно в тонкому шарі біля ГР, то для органічних НП характерні великі сили розділення носіїв заряду, яких немає в неорганічних НП. Густина струму електронів або дірок через СЕ буде визначатися сумою просторових градієнтів електричної ($\nabla U$) та хімічної ($\nabla \mu$) потенціальних енергій. В неорганічних НП переважну роль відіграє електрична складова $\nabla U$, оскільки фотогенерація вільних носіїв заряду відбувається у всьому об'ємі і завдяки великій рухливості носіїв струму просторовий розподіл заряду "швидко" врівноважується, тому вплив $\nabla \mu$ є мінімальним. Внаслідок домінуючої ролі $\nabla U$ в неорганічних СЕ $U_{хх}$ обмежується внутрішнім електричним полем чи згином зон. В органічних НП майже всі носії генеруються у вузькій області поблизу ГР шляхом дисоціації екситонів. Це призводить до утворення великого градієнту концентрації фотогенерованих носіїв заряду, пропорційного $\nabla \mu$, який значно більше, ніж в неорганічних СЕ. Тому в органічних СЕ слід враховувати вклад як $\nabla U$, так і $\nabla \mu$, в результаті чого $U_{хх}$ в органічних СЕ може бути більше внутрішнього електричного поля [30,54,55].

При впровадженні органічних СЕ не ставиться задача замінити високоефективні технології на основі неорганічних НП [37]. Органічні СЕ мають ряд переваг над звичайними неорганічними СЕ:

☼ використання простих технологій виготовлення (нанесення з розчину на підкладку, яка обертається, або термічне напилення) тонкоплівкових структур великої площі [37];

☼ витрати невеликої кількості речовини: для органічних НП $\alpha \sim 10^5$ см$^{-1}$, що дозволяє використовувати дуже тонкі плівки (товщиною $\sim 100$ нм) для поглинання основної частини сонячного світла [53].

☼ можливість підбору та зміни хімічним шляхом різних параметрів ($E_g$, $\chi$, рухливості носіїв заряду, розчинності, тощо) [37].

З іншої сторони, оскільки фотогенерація носіїв заряду відбувається в тонкому шарі поблизу ГР органічних ГС, то тільки невелика частина сонячного випромінювання поглинається в області ГР ГС. Це обмежує ефективність органічних ГС і є недоліком таких систем. Одним із шляхів збільшення області ефективної фотогенерації носіїв заряду є одночасне нанесення двох органічних речовин (їх суміші) і утворення так званої об'ємної ГС [18,19,53]. Перевагою об'ємних ГС є велика площа ГР, де відбувається ефективне розділення носіїв заряду, але головною проблемою таких систем є розчинність компонент ГС та сильна залежність ККД від наноморфології ГР. На практиці (див. табл. 1.1) в об'ємних ГС одна компонента ГС (барвники, наприклад, РСВМ) фотогенерує носії заряду, а за допомогою іншої (спряжені полімери, наприклад, MDMO–PPV чи P3HT) забезпечується ефективний транспорт носіїв заряду до електроду [18,19]. Морфологія об'ємної та шаруватої ГР може обмежувати ККД органічних СЕ, в той час як структура з частковим розділенням фаз може стати ідеальною, якби екситони могли дифундувати до ГР, а розділені електрон та дірка переміщувались в окремих областях (доменах) p- та n-типу провідності [44,45,53,56] і переходити в зовнішнє коло навантаження.

Отже, ключовими перевагами органічних СЕ є простий та дешевий спосіб нанесення шаруватих структур великої площі. При цьому можна використовувати недорогі, напівпрозорі та гнучкі підкладки. Проте для комерційного застосування СЕ повинні мати достатню ефективність (ККД), час роботи та собівартість (відносно до виготовленої енергії) [57]. Так, ККД перших органічних СЕ в атмосфері повітря деградував приблизно на 50% за пів доби [58], тоді як для сучасних органічних СЕ повідомляється про стабільність роботи до 6000 [59] годин. Крім того показана можливість покращення параметрів органічних СЕ в атмосфері $O_2$ [60].

Також ККД органічних СЕ обмежений слабким поглинанням в ближній ІЧ області, малою рухливістю носіїв заряду та низькою стабільністю органічних матеріалів. Перераховані вище проблеми намагаються вирішити шляхом синтезу нових речовин, використання різних компонент та оптимізації морфології органічних ГС [61]. Адже завдяки використанню Pc та периленів відбулися революційні зміни в ксерографії та інтенсивний розвиток копіювальної техніки [53]. Органічні світлодіоди на основі ГС вже завоювали значний сектор світового ринку світлодіодної техніки [62], органічні елементи успішно використовуються в сенсорних системах [63], а тонкоплівкові транзистори на основі Pn знаходяться на стадії практичного впровадження [64,65]. Практичне застосування органічних СЕ можливе при досягненні ККД 10%, що прогнозується реалізувати найближчим часом [66].

1.4. Основні класи фоточутливих органічних напівпровідників

1.4.1. Вплив замісників на властивості лінійних аценів

Лінійні ацени та їх похідні є одними з найбільш досліджених органічних НП. Будівельною ланкою аценів є бензольна молекула, яка має плоску структуру: всі атоми С та Н жорстко з'єднані Ω-зв'язками в площині молекули. А 2 $P_z$-орбіталі вуглецю спрямовані перпендикулярно молекулярній площині і утворюють колективну π-орбіталь делокалізованих електронів, яка визначає основні електронно-оптичні властивості аценів [67].

Встановлено, що з ростом числа бензольних кілець, а отже і кількості π-електронів, в лінійних аценів спостерігається зміщення краю власного поглинання в сторону менших $h\nu$ (батохромне зміщення), що пов'язано зі зменшенням енергії електронного збудження та іонізації цих молекул [68]. Збільшення кількості π-електронів призводить до зростання фоточутливості, наприклад, фоточутливість плівок Pn значно більша, ніж антрацену та тетрацену [67]. Також приєднання атомів сірки до молекул аценів призводить до суттєвого покращення їх властивостей. Наприклад, утворення молекули ТТТ, внаслідок приєднання 4-ох атомів сірки до молекули тетрацену, призводить до підвищення електропровідності на 9 порядків та розширення СП в довгохвильову область [69]. Але як наслідок великої провідності, фоточутливість ТТТ невелика. При заміщенні шести атомів сірки в молекулі Pn і утворенні молекули НТР (рис. 1.5), також спостерігається довгохвильовий зсув краю власного поглинання (рис. 1.6). При цьому спостерігається помітна фоточутливість плівок НТР при $h\nu > 1.55$ еВ [69,70], в той час як Pn фоточутливий при $h\nu > 1.8$ еВ [71]. Слід відмітити, що фоточутливість структур на основі Pn зростає (збільшується $I_{кз}$) зі збільшенням концентрації кисню адсорбованого в плівках Pn [72]. Утворення плівки НТР призводить до розширення смуг поглинання та невеликого батохромного зсуву максимумів СП на 0.08 еВ, що в три рази менше, ніж для Pn (рис. 1.6) і свідчить про зменшення міжмолекулярної взаємодії з приєднанням атомів сірки до молекули Pn [69].

Лінійні ацени (в.т.ч. ТТТ і НТР) є НП p-типу провідності, при цьому питомий опір плівок суміші тіопохідних Pn з переважним вмістом НТР складає приблизно $10^{10}$ Ом·см [69], тоді як для плівок Pn має порядок $10^{14}$-$10^{13}$ Ом·см [73] (хоча згідно з більш новими даними – $10^{7}$-$10^{8}$ Ом·см [74]). Але слід відмітити, що в польових транзисторах опір плівок НТР зменшується приблизно на сім порядків [75], а для Pn – на п'ять порядків [76], що говорить про перспективність практичного застосування цих шарів.

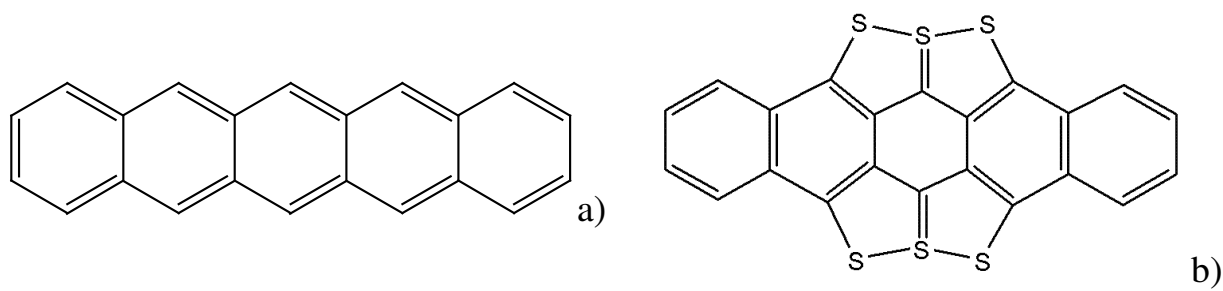

Рис. 1.5  Структурна формула молекул Pn (a) [77] та HTP (b) [75].

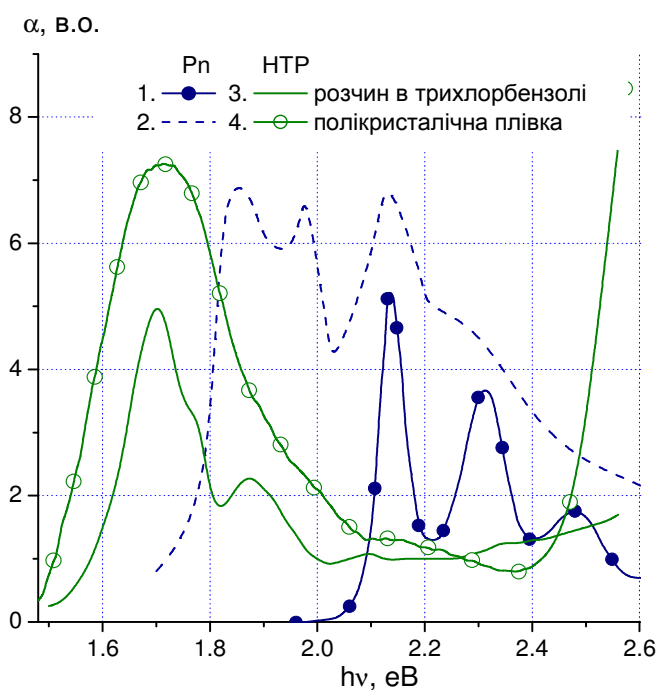

Рис. 1.6  СП розчинів Pn [78] (1) та HTP [70] (3) в трихлорбензолі та полікристалічних плівок Pn [68] (2) та HTP [70] (4) отриманих при кімнатній температурі ($T_S$ = 293 K).

Оскільки органічні НП в принципі не можна описувати в рамках зонної моделі одноелектронного наближення, то енергетичну схему іонізованих станів органічних НП часто будують за феноменологічною моделлю Лайноса [79], в якій враховуються енергетичні параметри кристалу, ефекти електронної поляризації та утворення СТ-станів. Енергетичне положення зон іонізованих станів органічного НП визначається емпіричними параметрами $I_C$, $\chi$ та $E_g$, які залежать від молекулярної структури кристалів. Наприклад, для молекул лінійних аценів з ростом числа $\pi$-електронів значення $I_C$ та $E_g$ зменшуються, а – $\chi$ та $\beta$ збільшуються (див. табл. 1.2).

Таблиця 1.2

Значення енергетичних параметрів монокристалічних органічних НП

аценового ряду

| Параметри | Антрацен | Тетрацен | ТТТ | Pn |
|---|---|---|---|---|
| $I_c$, еВ [77] | **5.93** | **5.40** | **4.75** | **5.07** |
| $\chi$, еВ [77] | **2.03** | **2.5** | **2.75** | **2.87** |
| $E_g$, еВ [77] | **3.90** | **2.9** | **2.0** | **2.20** |
| $\beta$, електрон/ фотон [77,80] | **$10^{-4}$** [77] | **$4\cdot10^{-2}$** [77] | **$10^{-2}$-$10^{-1}$** [80] | **0.3** [77] |

Також введення атомів сірки в молекулу тетрацену та утворення ТТТ призводить до збільшення $\chi$ і зменшення $I_C$ та $E_g$ (див. табл. 1.2). Таким чином, введення гетероатомів у молекули лінійних аценів призводить до покращення їх фотоелектричних властивостей, розширення фоточутливості та оптичного діапазону поглинання світла і зміни енергетичних параметрів.

### 1.4.2. Залежність властивостей фталоціанінів від структури молекули

Серед органічних НП особливе місце посідають гетероциклічні сполуки, які, на відміну від карбоциклічних, містять гетероатоми в вуглеводневому циклі спряження $\pi$-електронів. Гетероатоми N, O та S мають пари n-електронів, які можуть брати участь в як в міжмолекулярних зв'язках, так і в утворенні вільних носіїв заряду. Тому гетероциклічні сполуки (особливо Pc з центральними атомами металів: PbPc, CuPc, SnPc) характеризуються сильно вираженими НП властивостями. Молекула Pc (рис. 1.7а) близька за структурою до порфірину – основи фотосинтетичного хлорофілу та гему, за допомогою яких забезпечується транспорт електронів в цитохромних ферментах, мітохондріях та хлоропластах [67]. Перевагами Pc є утворення достатньо чистих кристалів (тільки деякі з аценів можуть бути очищені до такого рівня), висока термічна та хімічна стабільність: не розкладаються при температурах до 670 К в повітрі та до 1170 К у вакуумі, а також на них не впливають сильні кислоти та основи [81].

Структура молекули Рс є плоскою (в межах 0.3Å), якщо іонний радіус їх центрального атома не перевищує 0.7-0.8 Å і цей атом може розміщуватись в порожнині Рс кільця. У випадку коли розмір центрального металічного іону або групи перевищує розміри порожнини Рс кільця, утворюються неплоскі молекули: наприклад, іонний радіус атома свинцю $Pb^{2+}$ складає 1.2 Å, що призводить до утворення неплоскої молекули PbPc (рис. 1.7 a) [81].

Введення гетероатомів в цикл спряження змінює не тільки електронні властивості, а і кристалічну структуру органічних НП: в яких можуть з'явитися кілька поліморфних модифікацій кристалічної структури в залежності від умов виготовлення [67]. Так PbPc може формувати моноклінну (молекули складаються в стопку, так що атоми Pb утворюють одновимірний ланцюжок), триклинну (молекули PbPc формують дві незалежні колонки) та третю мало вивчену кристалічну модифікації. При цьому електропровідність триклинної форми PbPc приблизно на 8 порядків нижче моноклінної [81]. Це пов'язують з тим, що в кристалах PbPc моноклінної модифікації при пакуванні в „стовпчик" утворюється лінійний ланцюжок іонів $Pb^{2+}$, відстань між якими складає 0.373 нм, що трохи більше, ніж в металічній решітці свинцю (0.348 нм) [82]. При цьому провідність структур PbPc моноклінної модифікації складає $10^{-4}$ Ом$^{-1}$·см$^{-1}$ при кімнатній температурі [83,84], що свідчить про перспективність оптоелектронного застосування шарів PbPc.

СП Рс складається з двох інтенсивних смуг $\pi$-$\pi$* переходів в області 300-400 нм (В-смуга або смуга Соре) і в області 650-700 нм (Q-смуга). Для Рс з неплоскою структурою молекул спостерігаються додаткові смуги поглинання, які пов'язуються з появою заборонених синглет – триплетних переходів, внаслідок порушення спін-орбітального зв'язку при введенні в молекулу Рс важких атомів металу [81], або формуванням СТ-станів в результаті зменшення міжмолекулярної відстані та збільшення взаємодії між центральними та периферійними атомами сусідніх молекул [85]. Це пов'язано з тим, що важкий атом чи група атомів виступає з площини молекули Рс кільця і його взаємодія з π-електронною системою Рс кільця сусідньої молекули суттєво зростає. Поява додаткових смуг поглинання призводить до довгохвильового зсуву краю СП плівок Рс з неплоскою структурою молекул (наприклад, PbPc, TiOPc, SnCl$_2$Pc, і.т.д.). Так, наприклад, в плівках PbPc спостерігається поява додаткових смуг поглинання в області 1.1-1.6 та 2.0-2.3 еВ (рис. 1.7) [85,87], для плівок SnCl$_2$Pc – в області 1.25-1.45 еВ, для SnPc – 1.4-1.5 еВ [86]. Ці додаткові смуги поглинання пов'язують з формуванням в плівках СТ-станів (по два для кожної з поліморфних модифікацій, які спостерігались в плівках Рс [85]). При цьому кристалічна структура шарів забезпечує сприятливе розміщення молекул для здійснення переносу заряду і утворення СТ-станів [86,88]. При введенні в молекулу Рс атомів більш легких елементів (Cu, Zn) її планарність не порушується, і довгохвильовий зсув СП практично не проявляється [81].

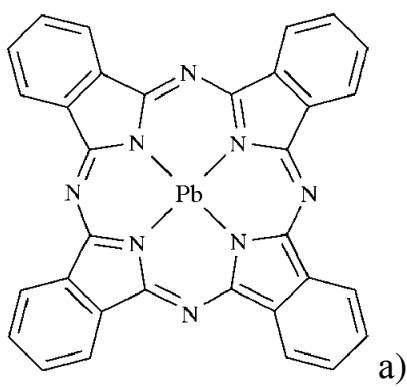

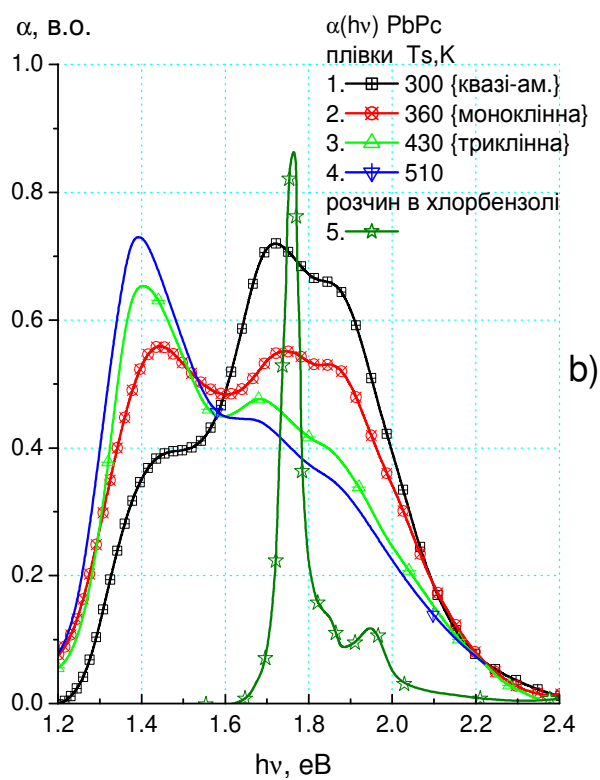

Рис. 1.7  Структурна формула молекули PbPc (a) та СП (b) плівок PbPc при $T_S$ = 300 К (1), 360 К (2), 430 К (3) та 510 К (4) [85] та розчину PbPc в хлорбензолі (5) [86].

Також показано, що для плівок Pc ефективність фотогенерації носіїв заряду високоенергетичних CT-станів більша, ніж низькоенергетичних CT-станів та екситонів Френкеля [85,89]. Значення $\beta$ та $S$ в тонкоплівкових структурах PbPc залежать від природи смуг поглинання (типу переходів) [90] та від матеріалу електрода, біля якого може формуватися велика концентрація поверхневих станів. Зі збільшенням $T_S$ від 297 до 433 K при виготовленні структур Ni/PbPc зростає $I_{кз}$ та зменшується $U_{xx}$ (внаслідок зменшення послідовного опору: провідність плівок PbPc отриманих при $T_S = 433$ K співпадає з провідністю монокристалів в напрямку осі $c$), в результаті максимальний ККД досягається при $T_S = 333$ K [91].

Встановлено, що вплив кисню ($O_2$) на фотоелектричні властивості Pc значно більший, ніж на темнові властивості. Наприклад, $I_{ph}$ монокристалів PbPc в атмосфері $O_2$ збільшується більше, ніж на два порядки [92]. Зростання $I_{ph}$ в атмосфері $O_2$ обумовлено ефективною фотогенерацією вільних носіїв заряду за допомогою слабких комплексів з переносом заряду (Pc$^+$,$O_2^-$) [81].

### 1.4.3. Агрегація метилзаміщених похідних перилену

Однією з найбільш перспективних НП компонент n-типу [93] провідності є МРР, який активно використовується при розробці органічних СЕ (див табл. 1.2) та світлодіодів [93]. Метилові замісники розміщені на кінцях молекули МРР і виходять за межі площини периленової основи, тому молекула МРР не повністю плоска (рис. 1.8a) [94]. При утворенні монокристалу МРР утворюється структура з дуже малою відстанню між сусідніми шарами (~ 0.32 нм), що призводить до значного перекриття молекулярних орбіталей сусідніх шарів (рис. 1.8b) [95].

СП слабоконцентрованого розчину МРР (5 мкмоль) [93] з електронними переходами 2.37 еВ та відповідними внутрішньо-молекулярними коливаннями (2.54, 2.70 та 2.84 еВ) батохромно зміщений відносно СП розчину перилену [96] на 0.5 еВ (рис. 1.9, кр. 3). Це пов'язується з більшою

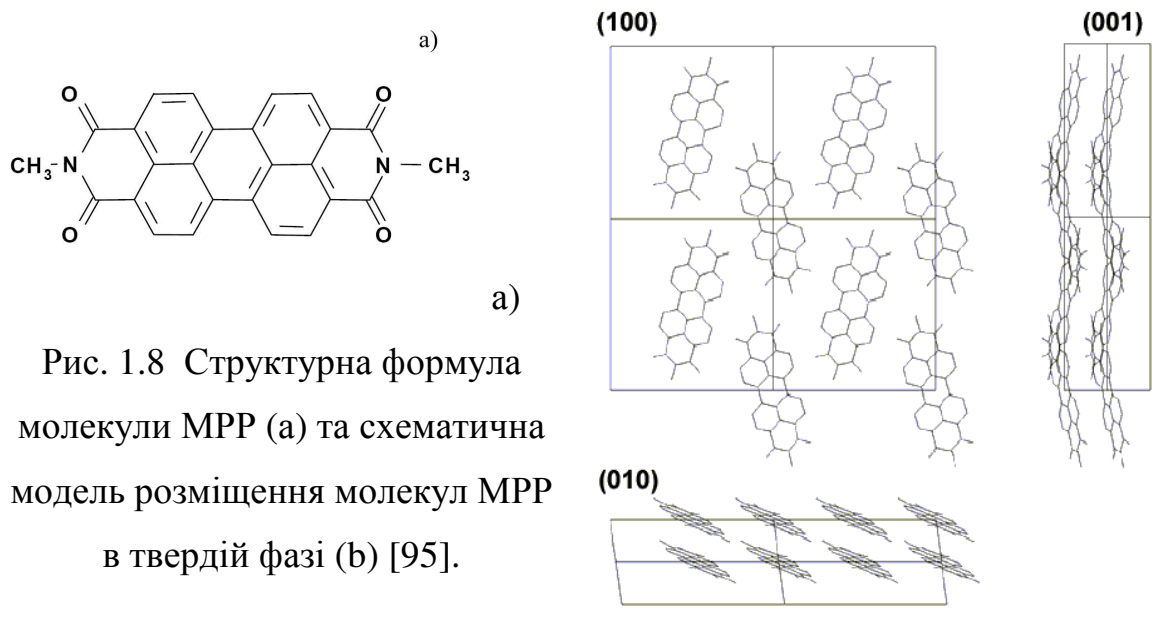

a)

Рис. 1.8  Структурна формула молекули МРР (a) та схематична модель розміщення молекул МРР в твердій фазі (b) [95].

b)

кількістю π-електронів у молекулі МРР, ніж в молекулі перилену. Відстань між смугами СП розчину МРР складає 0.16 еВ, що відповідає енергії внутрішньомолекулярних коливань перилену 0.15 еВ [93,96]. В СП більш концентрованих розчинів МРР (15-30 μмоль) спостерігається широка смуга в області 1.8-2.3 еВ, якої немає в СП слабоконцентрованого розчину МРР (рис. 1.9). Зі збільшенням концентрації розчину інтенсивність цієї смуги зростає сильніше, ніж інших смуг, що свідчить про утворення димерів або більших агрегатів, внаслідок збільшення міжмолекулярної взаємодії зі зростанням концентрації розчину МРР [93].

В СП плівок МРР при температурі 1.4К спостерігаються 2 підгрупи піків, які починаються зі смуг при 2.12 та 2.52 еВ і супроводжується коливальними смугами віддаленими ~ на 0.15 еВ (рис. 1.10). Зсув цих смуг відносно переходу при 2.37 еВ в розчині пов'язується з розщепленням збуджених станів в результаті утворення агрегатів в плівках МРР. Величина розщеплення залежить від орієнтації молекул в агрегаті та напрямку їх диполів: для паралельно розміщених диполів молекул характерний зсув у високо-енергетичну область, для диполів "хвіст до голови" – у низько-енергетичну (довгохвильову) область, а для нахилених димерів смуги розщеплюються в обидві сторони. Тому припускається, що в плівках МРР існують димери в яких молекули нахилені одна відносно одної [97].

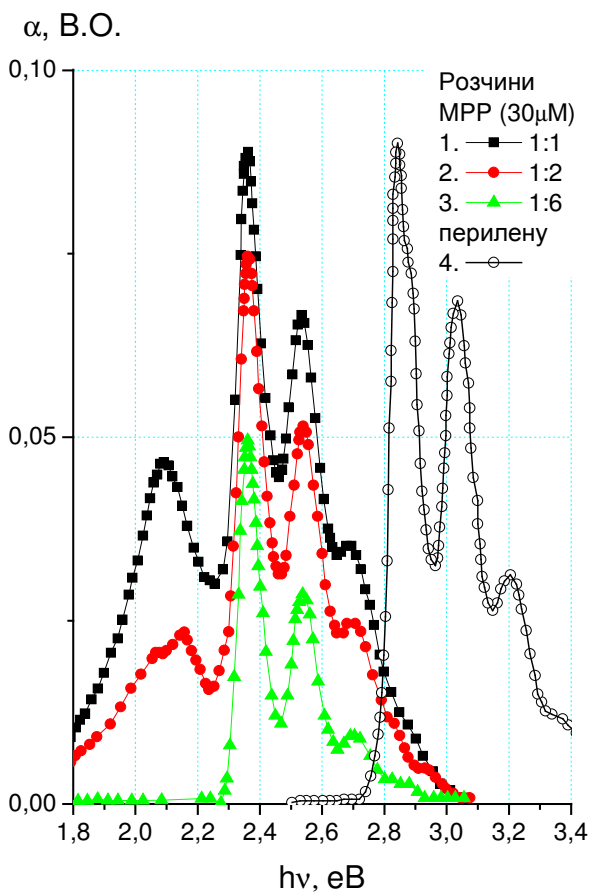

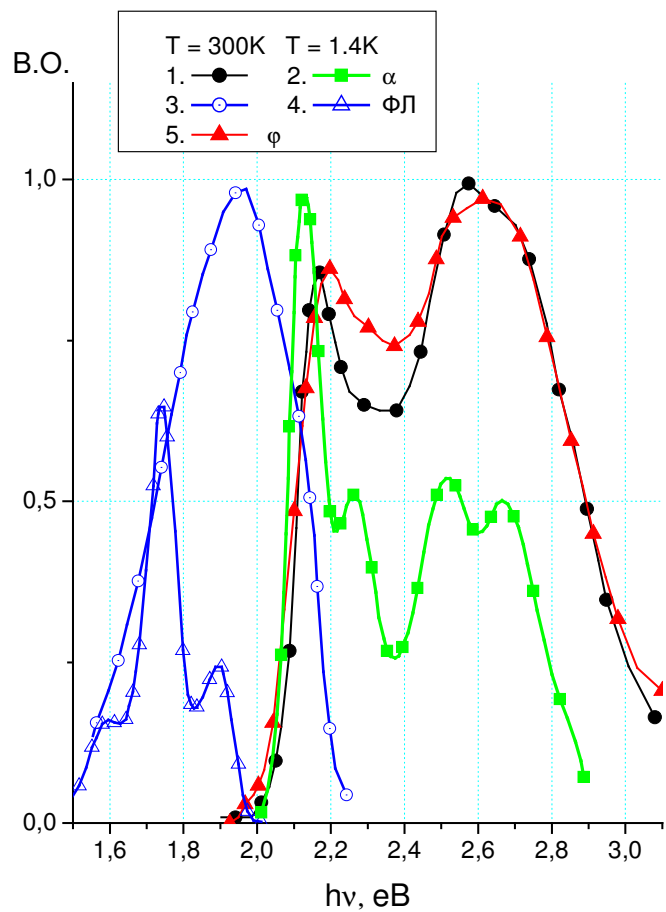

Рис. 1.9 СП розчинів МРР з концентраціями 30 (1), 15 (2) та 5 (3) ммоль в хлороформі [93] та СП розчинів перилену [96].

Рис. 1.10 СП (1,2), спектри ФЛ (3,4) та ФЕ (5) плівок МРР при температурах 300 К (1,3,5) [93] та 1.4 К (2,4) [97].

Квантова ефективність випромінювання ФЛ та електролюмінісценції при кімнатних температурах в плівках МРР значно менша, ніж для інших похідних перилену. Це пояснюється низькою квантовою ефективністю випромінювання димерної електронної структури, яка домінує в плівках МРР [59]. При температурі 1.4 К спостерігається тонка структура спектру ФЛ (рис. 1.10, кр.4), яка також підтверджує відсутність мономерів та утворення димерів чи більших агрегатів в плівках МРР [97].

Висновки до розділу 1 та їх зв'язок з постановкою задачі

Органічні НП мають високу технологічність, можливість зміни спектрального діапазону фоточутливості в залежності від молекулярної структури та ряд інших переваг над неорганічними НП. При цьому одними з найбільш перспективних органічних НП p-типу провідності є Pn та PbPc, а n-типу – МРР. На даний час добре вивчені властивості Pn та PbPc, досліджено структурні та оптичні властивості МРР [59,65,67]. Проте практично не досліджено властивості фоточутливих шарів n-типу – $SnCl_2Pc$ [98] та p-типу – HTP [75]. Як приклад, не вивчено ні вплив $T_S$ під час напилення плівок, ні вплив відпалу на енергетичну структуру, ФЕ, $L$ чи $S$, що дуже важливо для оптимізації характеристик та покращення фотовольтаїчних властивостей ГС та СЕ на основі органічних НП.

Також досліджено властивості ряду анізотипних ГС на основі органічних НП, при цьому ізотипні органічні ГС досліджені дуже мало. Слід відмітити, що спектральний діапазон фоточутливості досліджуваних ГС обмежений, а параметри отримання цих ГС не оптимізовані, що є причинами низького ККД СЕ отриманих на їх основі. При цьому практично не вивчено залежність фотовольтаїчних властивостей від діаграм енергетичних зон біля ГР двох органічних НП.

Отже, необхідність цілеспрямованих досліджень нових органічних НП та визначення оптимальних умов створення фоточутливих в широкій області спектру ГС на основі органічних НП є актуальною задачею, оскільки розкриває нові можливості практичного застосування органічних матеріалів.

# РОЗДІЛ 2
## МЕТОДИКА ДОСЛІДЖЕНЬ ТОНКОПЛІВКОВИХ СТРУКТУР

2.1. Вибір досліджуваних речовин для створення фоточутливих тонкоплівкових структур

Вибір речовин для успішної розробки ефективних органічних СЕ є дуже актуальною задачею. Ціль – вибрати оптимально ефективні компоненти ГС, які в подальшому можна використовувати при розробці та виготовлені СЕ. Для цього повинні виконуватися, по крайній мірі, дві необхідних умови: ГС повинна поглинати більшу частину сонячного випромінювання у видимій, ближній ІЧ та ультрафіолетовій (УФ) області (~ 1.2-3.2 еВ) і поглинання цього світла повинно призводити до ефективного утворення носіїв заряду. Таким чином обидві компоненти ГС повинні бути фоточутливими в спектральній області ~ 1.2-3.2 еВ. Але цього недостатньо. Третьою важливою умовою є необхідність створення на ГР двох компонент ГС значного запірного бар'єру з невеликою концентрацією центрів поверхневої рекомбінації носіїв заряду. Для перевірки можливості створення запірного бар'єру можна використовувати енергетичну схему ГС за моделлю Андерсона, яка була запропонована для неорганічних НП (п. 1.2). А для оцінки концентрації центрів поверхневої рекомбінації носіїв заряду на ГР компонент ГС необхідно провести детальне дослідження оптичних та фотовольтаїчних властивостей ГС.

Отже, отримання перспективних для фотовольтаїчного перетворення сонячного світла ГС відбувається при умові фоточутливості компонент ГС в області ~ 1.2-3.2 еВ та формування значного згину зон на ГР ГС. На відміну від неорганічних НП, більшість фоточутливих органічних НП p-типу провідності слабо поглинають сонячне світло в області 2.0-3.0 еВ. Тому для створення ГС нами вибраний фоточутливий органічний НП n-типу провідності MPP (рис. 1.8а), який ефективно поглинає сонячне світло ($\alpha > 10^5$ см$^{-1}$) з утворенням нерівноважних носіїв заряду в області 2.0-3.0 еВ. Також перевагами МРР є термічна стабільність та впорядкована структура тонких шарів. Тому шари МРР активно використовуються при розробці органічних СЕ (див. табл. 1.1).

Для створення анізотипних ГС на основі шарів МРР було підібрано ряд НП компонент р-типу провідності: Pn (рис. 1.5а), НТР (рис. 1.5b), PbPc (рис. 1.7а) та CuI, які поглинають сонячне світло в спектральній області прозорості шарів МРР (рис. 1.10). Так, тонкоплівкові структури на основі Pn ефективно поглинають сонячне світло тільки в області 1.8-2.4 еВ (рис. 1.6), при цьому Pn є найбільш фоточутливим НП серед лінійних аценів (див. табл. 1.2). Плівки похідної Pn – НТР поглинають світло та фоточутливі в спектральній області 1.6-2.0 еВ (рис. 1.6), а плівки PbPc – в області 1.3-2.2 еВ (рис. 1.7). Створення анізотипних ГС на основі шарів МРР, які фоточутливі в області 2.0-3.0 еВ вирішує проблему слабкого поглинання сонячного світла та незначної фоточутливості шарів р-типу провідності в області 2.2-2.7 еВ. Плівки CuI ефективно поглинають світло в ближній УФ області (hv > 3.0 еВ). Саме тому було вибрано компоненти р-типу провідності Pn, НТР, PbPc та CuI, спектральна фоточутливість яких зміщена в ближню ІЧ або УФ область в порівнянні з областю фоточутливості шарів МРР. Це може призвести до збільшення, як частини поглинутого сонячного випромінювання, так і фотогенерованих носіїв заряду в ГС на основі запропонованих компонент n- та р-типу. Отже, такий вибір компонент р-типу провідності при створенні анізотипних ГС на основі МРР, дозволяє розширити діапазон фоточутливості майбутніх ГС в порівнянні з областю фоточутливості окремих компонент.

Як видно з табл. 2.1 вибрані НП р-типу провідності (Pn, НТР, PbPc та CuI) мають різні значеннями роботи виходу носіїв або рівня Фермі ($E_f$), що дозволяє проаналізувати залежність фоточутливості досліджуваних анізотипних ГС на основі МРР від енергетичної структури компонент р-типу провідності.

Таблиця 2.1

Значення параметрів енергетичної структури компонент ГС.

| | $I_C$, еВ | $E_g$, еВ | $\chi$, еВ | $E_f$, еВ |
|---|---|---|---|---|
| МРР | **6.47** [98,99] | **2.6** [98,99] | **3.87** [98,99] | **4.38** [98,99] |
| PbPc | **4.9** [100] | **1.6** [101] | **3.3** | **4.2-4.6** [102] |
| Pn | **5.07** [77] | **2.2** [77] | **2.87** [77] | **4.3-4.8** [103-106] |
| ТТТ | **4.75** [77] | **2.0** [77] | **2.75** [77] | **4.4** [107-108] |
| НТР [1] | **4.7-4.8** | **1.95-2.05** | **2.7-2.8** | **4.3-4.5** |
| CuI | **6.1** [109] | **3.0** [109] | **3.1** [109] | **6.0** [109] |

[1] (Параметри енергетичної схеми НТР на даний момент не встановлені. Але оскільки молекулярна структура молекул НТР і ТТТ близька. СП у видимій та ближній ІЧ області плівок НТР і ТТТ практично співпадають, то припускається, що значення енергетичних параметрів НТР і ТТТ будуть близькими.)

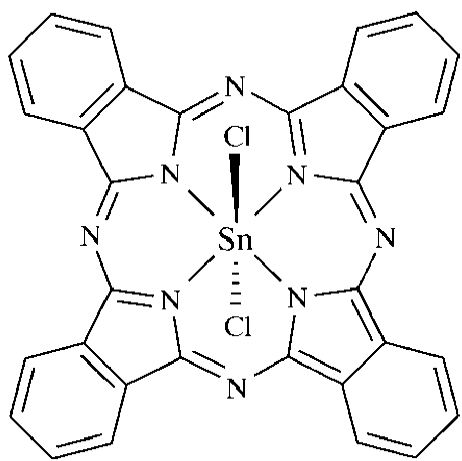

Рис. 2.1  Структурна формула
молекули SnCl$_2$Pc

В якості альтернативної компоненти n-типу використані також шари SnCl$_2$Pc (рис. 2.1), які є фоточутливими у видимій та ближній ІЧ області, і можуть бути використані для експериментальної перевірки можливості виго-товлення органічних ГС перспек-тивних для створення ефективних багатошарових органічних СЕ.

## 2.2. Методика виготовлення тонкоплівкових структур

При дослідженні об'ємних властивостей тонких плівок необхідно створювати тонкоплівкові структури товщиною більше довжини вільного пробігу носіїв заряду або області просторового заряду біля поверхні (в залежності, що більше). При цьому плівка повинна бути суцільною, достатньо однорідною і мати омічні контакти. В протилежному випадку отримані результати дослідження властивостей тонкоплівкових структур будуть сильно залежати від властивостей поверхні плівок [110].

Виготовлення тонкоплівкових ГС, тобто приведення в контакт двох тонких НП шарів, є складним технологічним процесом. Так, компоненти ГС (тонкі плівки) повинні задовольняти ряду важливих умов (напр. однорідність по товщині, велика адгезія, незмінність інших властивостей по товщині плівки) [110]. Також для отримання якісної ГР двох НП шарів потрібно перешкодити взаємодифузії компонент, що можна досягнути при достатньо низьких температурах виготовлення ГС [111]. З іншої сторони для отримання більш ефективних ГС, потрібно підвищувати температуру виготовлення компонент з метою отримання тонких шарів з більш впорядкованою кристалічною структурою молекул.

Компоненти досліджуваних ГС та самі ГС були виготовлені методом термічного напилення у вакуумі ($2 \cdot 10^{-4}$ Па) з очищеного вакуумною сублімацією порошку.

Метод термічного вакуумного напилення, через свою простоту та технологічність дозволяє змінювати та контролювати параметри компонент (плівок) і є одним з найбільш зручних для виготовлення ГС. При термічному напиленні можна створити чисті технологічні умови. Крім цього для отримання рівномірних та однорідних плівок використовуються випаровувачі, задачею яких є створення рівномірного потоку частинок, що випаровуються і осідають на підкладку. Швидкість осідання залежить від взаємного положення випаровувача та підкладки, а також від коефіцієнта конденсації [111]. При виготовленні тонкоплівкових структур в якості випаровувача нами використовувався танталовий човник.

При напиленні верхнього шару ГС використовувались маски заданої форми. Адже для отримання якісних зразків верхній шар ГС має напилятись лише на поверхню нижнього шару і не повинен наноситись на нижній електрод. При напиленні верхнього шару на електрод буде відбуватися часткова закоротка ГР двох компонент ГС і отриманий зразок стане непридатним для фотовольтаїчних досліджень. Використання масок дозволяє обмежувати потік частинок при напиленні верхнього шару і отримувати якісні зразки без закоротки.

Досліджувані плівки та тонкоплівкові ГС наносились на скляні круглі підкладки діаметром 14 мм. Для дослідження фотовольтаїчних властивостей тонкоплівкових структур на підкладку спочатку напилювали провідний шар ІТО, який у вигляді тонких плівок має високу електропровідність (поверхневий опір $\approx$ 100 Ом см$^{-2}$) та оптичну прозорість (пропускання $\approx$ 80%). Оскільки в нашому випадку вимірювання ФЕ проводились безконтактним конденсаторним методом, то другий електрод нам був непотрібен. Для дослідження оптичних властивостей використовувались підкладки без шару ІТО.

Температура утворення тонкоплівкових структур визначається температурою підкладок ($T_S$), яка задається в процесі термічного напилення. $T_S$ при виготовленні досліджуваних тонких плівок та ГС задавалась в діапазоні від 300 до 430 К. Стабілізація $T_S$ з точністю до 2 К проводилась за допомогою спеціального термостатуючого пристрою. Вимірювання $T_S$ відбувалось мідь-константовою термопарою. Швидкість напилення плівок складала ~ 1 нм/с.

Розміщення атомів та молекул в плівці сильно залежить від $T_S$ в процесі термічного напилення. При низьких $T_S$ утворюються квазіаморфні плівки, в той час як при більш високих $T_S$ в плівках формується полікристалічна структура. При цьому молекулярні площини $ab$ поліциклічних ароматичних молекул при утворенні тонких плівок намагаються орієнтуватися паралельно поверхні підкладки. Кристалічну структуру плівок можна покращити відпалом при температурах ($T_A$) вищих, ніж температура виготовлення плівок [110]. Так як для ряду органічних НП p-типу провідності (в т.ч. для Pn) кисень є допантом і збільшення концентрації кисню адсорбованого в плівках призводить до збільшення фоточутливості цих тонкоплівкових структур (див. п. 1.4 [72,81,92]). Тому відпал досліджуваних тонких плівок НТР (p-тип) проводився (протягом години) на повітрі. Досліджувані плівки MPP та $SnCl_2Pc$ (n-тип) відпалювались протягом години у вакуумі з метою зменшення концентрації кисню в плівках НП n-типу провідності. Температура відпалу ($T_A$) задавалась в діапазоні від 330-475 К.

Необхідною умовою отримання ефективних ГС є оптимальний підбір товщини шарів, адже для того щоб більша частина світла поглинулась в НП, товщина шару повинна перевищувати значення $\alpha^{-1}$, при цьому щоб більша частина генерованих світлом носіїв розділялась гетеропереходом, товщина плівки повинна бути близькою до $L$ носіїв заряду або екситонів. Досліджувані плівки були різних товщин від 100 до 800 нм в залежності від задачі досліджень. Товщина компонент ГС була близькою до $L$ екситонів в цих шарах і задавалась від 50 до 200 нм в залежності від $\alpha$ використовуваної речовини. Товщина шарів під час напилення контролювалась за вимірюванням частоти кварцового датчика, а після виготовлення – за допомогою атомно-силового мікроскопа (AFM). Дослідження структури поверхні та морфології плівок проводилось за допомогою AFM «Nanoscope IIІа» (DI, США) в режимі періодичного контакту (Tapping mode[TM]) кремнієвим зондом з радіусом заокруглення 10 нм.

2.3. Вимірювання спектрів поглинання, фотомодульованого відбивання та люмінесценції плівок

В даній роботі СП вимірювались на двопроменевих спектрофотометрах «Hitachi М356» та «Shimadzu SP 2101 UV-Vis».

Поглинання світла при проходженні крізь шар певної речовини описується законом Бугера – Ламберта – Бера:

$$I = I_0 \cdot e^{-\alpha d} \qquad (2.1)$$

де $I_0$ – інтенсивність падаючого світла, $I$ – інтенсивність світла, яке пропустив зразок (пройденого світла), $\alpha$ – коефіцієнт поглинання речовини, $d$ – відстань, що проходить світло в даній речовині, або в нашому випадку товщина плівки. Вираз (2.1) описує лише поглинання окремих молекул і не враховує відбивання та розсіювання світла речовиною [112].

При досліджені оптичних характеристик плівок (монокристалів) слід враховувати, що

$$I_0 = I_n + I_в + I + I_p \qquad (2.2)$$

де $I_n$ – інтенсивність поглинутого світла, $I_в$ – інтенсивність відбитого світла, $I$– інтенсивність пройденого світла, $I_p$ – інтенсивність розсіяного світла. В тонких та однорідних плівках $I_p << I, I_n, I_в$. При цьому конструкція спектрофотометру «Hitachi М356» дозволяє забезпечувати інтенсивність розсіяного світла менше 0.02% при довжині хвилі 600 нм та близько 0.1% при 250 нм, що значно менше інших складових інтенсивності світла ($I, I_n, I_в$). Тому складовою розсіювання світла ($I_p$) у виразі (2.2) можна знехтувати. В цьому випадку значення $I$ з врахуванням багатократного відбивання описується за формулою:

$$I = I_0 \frac{(1-R) \cdot e^{-\alpha d}}{1 - R^2 \cdot e^{-2\alpha d}}, \qquad (2.3)$$

де $R$ – коефіцієнт відбивання.

При умові $R^2 \cdot exp(-2\alpha d) << 1$ рівняння (2.3) спрощується:

$$I = I_0 \cdot (1-R) \cdot e^{-\alpha d}. \qquad (2.4)$$

В більшості спектрофотометрів автоматично вимірюється оптична густина ($D$) зразків, яка згідно з виразом (2.4) рівна:

$$D = \ln (I_o/I) = \alpha d - \ln(1 - R). \tag{2.5}$$

В області прозорості (при $\alpha = 0$):

$$D\,(\alpha = 0) = D_o = -\ln(1 - R)\,. \tag{2.6}$$

Тому в багатьох випадках для вимірювання $\alpha$ допустимо користуватись спрощеним виразом:

$$\alpha = (D - D_0)\,/\,d. \tag{2.7}$$

Отримане значення $\alpha$ є характеристикою поглинаючого середовища і залежить від довжини хвилі або енергії квантів (hv) падаючого світла. Залежність коефіцієнта поглинання світла ($\alpha$) від енергії фотонів (hv), що падають на речовину $\alpha$(hv) називають спектром поглинання (СП) [112].

Для тонких плівок використовувався режим безпосереднього вимірювання $D$ та $D_0$, але для плівок з оптичною густиною більше одиниці СП доцільніше вимірювати в режимі пропускання:

$$D - D_0 = \ln\left(\frac{I^{100} - I^{0}}{I - I^{0}}\right), \tag{2.8}$$

де $I^{100}$ та $I^{0}$ інтенсивності пройденого світла за відсутності зразка та при повному відбиванні, відповідно [112]. Відносна похибка вимірювання СП складає не більше 3%.

Вимірювання фотомодульованого відбивання ($\Delta R/R$) проводились за оригінальною методикою, детальний опис якої наведено в роботах [113,114]. Для отримання спектральної залежності фотомодульованого відбивання зразок, крім немодульованого монохроматичного світла галогенної лампи, освітлювався модульованим випромінюванням He-Ne лазера з hv = 1.96 еВ (рис. 2.2). Відбите від зразка світло реєструвалося фотодіодом. Для уникнення потрапляння відбитого від зразка лазерного випромінювання перед фотодетектором ставиться відповідний світлофільтр. Система детектування за допомогою комп'ютера розділює сигнал на постійну та

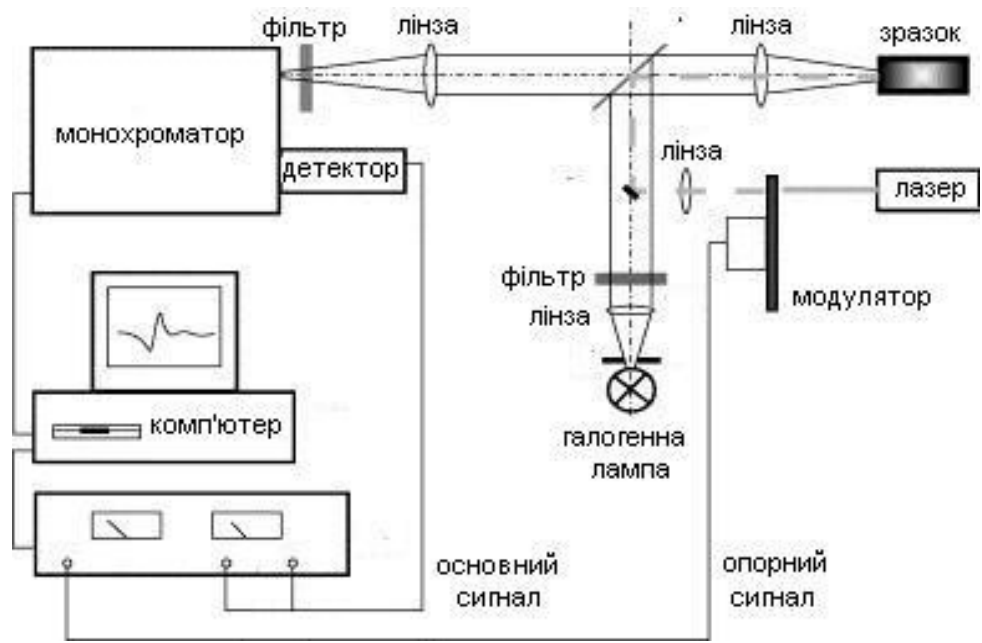

Рис. 2.2 Схема експериментальної установки призначеної для вимірювання спектральних залежностей фотомодульованого відбивання досліджуваних тонкоплівкових структур.

модульовану складові. Де постійна складова пропорційна $R$, а перемінна (модульована) складова – $\Delta R$. Після розділення сигналів автоматизована система ділить модульований сигнал на постійний, і в результаті на виході реєструється спектральна залежність фотомодульованого відбивання – $\Delta R/R$.

Спектри ФЛ та збудження ФЛ вимірювались за допомогою спектрофлюориметра (spectrofluorimetr) «Hitachi F-4500».

### 2.4. Методика досліджень фотовольтаїчних властивостей

ФЕ досліджуваних тонкоплівкових структур вимірювалась за допомогою методу Бергмана або статичного конденсатора. Метод статичного конденсатора був запропонований в Бергманом та Хенслером [115], а потім був вдосконалений Пуцейком та Акімовим, отримавши при цьому широке застосування в різних дослідженнях пов'язаних з фотовольтаїчними ефектами [116-126].

Для правильного використання методу та інтерпретації результатів вимірювань важливо визначити природу конденсаторної ФЕ. Досліджувана ФЕ в більшості випадків складається з кількох компонент, які важко розділити одна від одної. Головними компонентами ФЕ є дифузійна та бар'єрна компоненти.

Дифузійна ФЕ ($\varphi_d$ – ФЕ Дембера) виникає при різній рухливості електронів ($\mu_n$) та дірок ($\mu_p$), внаслідок градієнта концентрацій надлишкових носіїв струму і в загальному випадку має вигляд [126,127]:

$$\varphi_d = \frac{kT}{e} \frac{\mu_n - \mu_p}{\mu_n + \mu_p} \ln \frac{\sigma_i}{\sigma_n},$$ (2.9)

де $\sigma_i$ і $\sigma_n$ – провідність біля освітленої та неосвітленої поверхні зразка, відповідно. З рівняння (2.9) видно, що величина $\varphi_d$ буде збільшуватись зі зростанням різниці рухливостей електронів та дірок: $|\mu_n - \mu_p|$, а знак $\varphi_d$ буде від'ємним у НП p-типу і додатнім для n-типу. Відношення $\sigma_i$ до $\sigma_n$, а, отже, і $\varphi_d$ зростає зі збільшенням $\alpha$ та фоточутливості зразка. Тому спектральна залежність $\varphi_d$ повинна корелювати зі спектральною залежністю $\alpha$.

Бар'єрна ФЕ ($\varphi_\delta$) виникає внаслідок розділення нерівноважних фотогенерованих носіїв струму внутрішнім електричним полем, яке обумовлене згином енергетичних зон біля вільної поверхні, на ГР метал-НП, анізотипного чи ізотипного гетеропереходу в НП, тобто наявністю бар'єру. Нерівноважні електронно-діркові пари, що підходять до бар'єру, розділяються його полем, внаслідок чого зростає концентрація носіїв заряду. Знак $\varphi_\delta$ співпадає зі знаком $\varphi_d$ при наявності запірного згину зон і $\varphi_\delta$ має протилежний знак $\varphi_d$ у випадку антизапірного згину зон. Довгохвильовий край спектральної залежності $\varphi_\delta$ зазвичай зміщений в сторону менших hν відносно СП. Величина дифузійної компоненти $\varphi_d$ пропорційна не концентрації нерівноважних носіїв заряду, а їх градієнту. Тому спектральна залежність $\varphi_d$ близька до СП досліджуваних зразків і зміщена в сторону сильного поглинання в порівнянні з спектром $\varphi_\delta$ [126].

Порівняння експериментальних та розрахункових залежностей $\varphi_\delta$ та $\varphi_d$ для германію та кремнію показало наявність суттєвих розходжень між теорією та експериментом. Встановлено, що головною причиною розходжень є виникнення компоненти ФЕ, обумовленої захопленням і накопиченням нерівноважних носіїв заряду одного знаку на поверхні тонкоплівкової структури. Знак цієї компоненти ФЕ може бути протилежним знаку дифузійної [128]. Отже, при аналізі отриманих експериментальних даних необхідно враховувати вклад різних компонент ФЕ (дифузійної, бар'єрної та захоплювальної) та вплив рекомбінації носіїв струму на ГР.

Перевагою конденсаторного методу є можливість вимірювання ФЕ без нанесення верхнього контакту, молекули якого можуть дифундувати в приповерхневу область і суттєво змінювати властивості плівок. Проте відсутність контактів не дозволяє визначати ККД досліджуваних нами структур, що є недоліком цієї методики. Схема установки яка використовувалась для дослідження фотовольтаїчних властивостей зображена на рис. 2.3. Згідно з методикою статичного конденсатора досліджуваний зразок розміщується між напівпрозорими обкладками конденсатора в спеціальній діелектричній комірці. В якості електродів (обкладок) конденсатора використовувались круглі плоско-паралельні пластинки діаметром 14 мм, які використовувались як підкладки для напилення тонкоплівкових структур. Пластинки виготовлені з кварцового скла, по всій поверхні вкриті провідним шаром ІТО. На одній з пластинок наносилася досліджувана тонкоплівкова структура. Діелектричною прокладкою конденсатора була прозора тефлонова плівка, товщиною 20 мкм. При освітленні комірки між поверхнями зразка виникає різниця потенціалів – конденсаторна ФЕ. В результаті цього з'являється змінний струм через опір навантаження, який за допомогою схеми синхронізації реєструється нановольтметром. Тонкоплівкові структури освітлювались з обох сторін: зі сторони ІТО-електрода та зі сторони вільної поверхні (ГР з навколишнім середовищем, тобто повітрям). Освітлення з обох сторін дозволяє враховувати властивості вільної поверхні плівок та ГР з ITO контактом.

Світло від йодованої лампи розжарювання («Hitachi» потужністю 120 Вт) через фокусуючу систему двох кварцових лінз, модулятор з частотою модуляції 80 Гц та відповідний світлофільтр потрапляє на вхідну щілину монохроматора «МДР-4» (ЛОМО). Монохроматичне світло з вихідної щілини потрапляє на вікно вимірювальної комірки, всередині якої розміщується зразок (рис. 2.3). Слід відмітити, що вимірювання ФЕ при модульованому освітленні з основною частотою приблизно 80 Гц дозволяє зменшити шуми в порівнянні з немодульованим світлом, чи світлом модульованим в іншому діапазоні частот.

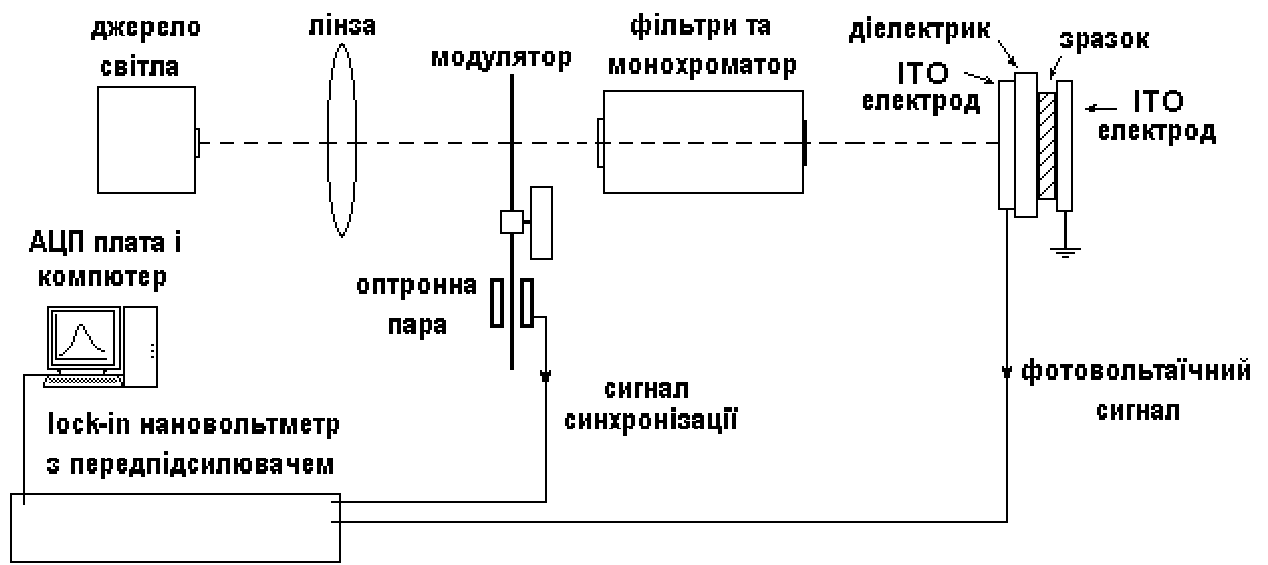

Рис. 2.3 Схема експериментальної установки призначеної для вимірювання спектральних залежностей ФЕ досліджуваних тонкоплівкових структур.

ФЕ вимірюється фазочутливим селективним (lock-in) нановольтметром «Unipan-232B» з передпідсилювачем моделі 233-7. Для синхронізації вимірювань використовувавсь опорний сигнал оптронної пари фотодіод-світлодіод, яка розміщувалась на модуляторі (рис. 2.3). З виходу нановольтметра сигнал подається через інтерфейсну плату АЦП-ЦАП (ЕТ-1050) на комп'ютер.

Використовується також схема, яка дозволяє швидко і якісно визначити фоточутливість досліджуваних зразків. За даною схемою ФЕ збуджувалась випромінюванням серійних світлодіодів, перелік яких наведено в табл. 2.2. Напруга на світлодіоди подавалась з генератора «ГЗ-109». Світлодіоди вмонтовувались в спеціальні касети з метою розміщення безпосередньо поблизу зразка для найменшого розходження пучка світла.

Таблиця 2.2

Параметри серійних світлодіодів, що використовувались для дослідження фотовольтаїчних властивостей

| Колір випромінювання світлодіода | $\lambda$, нм | h$\nu$, еВ |
|---|---|---|
| синій | 470 | 2.6 |
| зелений | 570 | 2.2 |
| червоний | 650 | 1.9 |
| ІЧ | 940 | 1.3 |

Також за допомогою випромінювання світлодіодів створювалась постійна (немодульована) тильна підсвітка при дослідженні ГР ізотипних ГС. При створенні підсвітки на світлодіоди подавалось постійна напруга від джерела каліброваних напруг П4108. Потужність підсвітки вимірювалась каліброваним радіометром ППТН-02 на основі кремнієвого фотодіода.

Після кожної серії вимірювань ФЕ спектральний розподіл інтенсивності випромінювання лампи та інтенсивність випромінювання світлодіодів вимірювався за допомогою піроприймача (СКТБ ІФ НАНУ), який встановлюється на місце зразка. Після цього спектри ФЕ перераховуються на однакову кількість падаючих квантів світла.

Відносна похибка вимірювання спектрів ФЕ становить 5%. При цьому точність експериментального вимірювання слабких фотовольтаїчних сигналів забезпечувалась використанням модуляційної методики. Так в схемах вимірювань (особливо при збудженні ФЕ випромінюванням світлодіодів) використовувався селективний нановольтметр «Unipan-237», який має систему селективного підлаштовування частоти (з точністю до 0.1 Гц). А фазочутливий селективний (lock-in) нановольтметр «Unipan-232В» забезпечує більш високу точність за рахунок використання опорного сигналу (рис. 2.3). Обидва прилади «Unipan-237» та «Unipan-232В» мають вхідний опір 100 МОм, а при використанні передпідсилювача «Unipan 233-7» вхідний опір збільшується до 1 ГОм. Використання спеціальних екранованих кабелів дозволяє суттєво підвищити співвідношення «сигнал-шум» та покращити точність експериментальних вимірювань. Крім того при вимірюванні ФЕ застосовувалась система запису та накопичення експериментальних даних з використанням комп'ютера, що також дозволяє підвищувати точність та достовірність вимірювань.

## 2.5. Аналіз експериментальних даних за допомогою гаусових кривих

Для більш адекватного опису та інтерпретації спектральних особливостей поглинання та ФЛ досліджуваних плівок використовувався розклад спектрів на гаусові компоненти. Тобто спектральні залежності поглинання чи ФЛ розкладалися на декілька елементарних гаусових кривих, які описують окремі електронні переходи в молекулах чи внутрішньо-молекулярні коливання. Елементарні смуги спектрів поглинання чи ФЛ мають контур "дзвоноподібної" форми, яка апроксимується в першому наближенні нормальним (гаусовим) розподілом випадкової неперервної величини:

$$f_i(h\nu) = A \cdot \exp[-\frac{(h\nu - Ei)^2}{2 \cdot \sigma_i^2}], \qquad (2.11)$$

де $A$ – інтенсивність елементарної смуги спектру в точці hν = $E_i$, $E_i$ – енергетичне положення максимуму гаусової кривої, $\sigma$ – середньоквадратичне відхилення від положення максимуму або півширина елементарної смуги на рівні половини інтенсивності ($A/2$).

Опис спектрів за допомогою гаусових кривих можна проводити при відомій кількості та приблизного положення елементарних смуг, з яких складається спектр. Гаусовий розклад спектрів поглинання та ФЛ досліджуваних плівок дозволяє більш точно встановити положення максимумів ($E_i$) елементарних смуг в широких мультиплетних смугах, що складаються з декількох елементарних смуг. Особливо у випадку часткового перекриття елементарних смуг і слабкого візуального розділення широких мультиплетних смуг на окремі компоненти: електронні переходи чи внутрішньо-молекулярні коливання. Визначивши значення $E_i$ можна побудувати енергетичну структуру досліджуваної речовини при наявності необхідних для цього даних. Також за допомогою параметрів гаусового розкладу $\sigma$ та $A$ можна отримати величину площі під елементарною гаусовою кривою, що дозволяє оцінити вклад окремих елементарних смуг в широкому мультиплетному спектрі, який складається з декількох окремих елементарних смуг.

Висновки до розділу 2

Отримання органічних ГС перспективних для фотовольтаїчного перетворення сонячного світла відбувається при умові фоточутливості компонент ГС в області ~ 1.2-3.2 еВ та формування ефективного згину зон на ГР ГС. Копмонентами для створення органічних ГС були вибрані: МРР, SnCl₂Pc (n-тип провідності) та Pn, HTP, PbPc (p-тип провідності).

Неорганічний НП CuI використовувався в якості модельного обєкта для дослідження електронних процесів на ГР органічних НП.

Для детального дослідження енергетичної структури шарів МРР та $SnCl_2Pc$ вимірювались спектри поглинання, фотомодульованого поглинання та ФЛ плівок МРР та $SnCl_2Pc$. А для адекватної інтерпретації енергетичної структури спектри поглинання та ФЛ розкладались на гаусові компоненти. Також для дослідження фотовольтаїчних властивостей окремих компонент ГС та самих ГС використовувався метод статичного конденсатора. Перевагою вищезгаданого методу є відсутність потреби нанесення верхнього електроду. І як наслідок неможливість визначення ККД досліджуваних структур.

# РОЗДІЛ 3

# ДОСЛІДЖЕННЯ ВЛАСТИВОСТЕЙ ОРГАНІЧНИХ ФОТОПРОВІДНИКІВ, ЯК КОМПОНЕНТ ГЕТЕРОСТРУКТУР

3.1. Морфологія та структура поверхні тонких плівок органічних напівпровідників

Досліджувані плівки MPP, $SnCl_2Pc$ та HTP отримані термічним вакуумним напиленням були полікристалічними, а форма та розміри кристалітів залежали від температури підкладки під час напилення ($T_S$).

Для плівок MPP при $T_S = 300$ K формуються овальні кристаліти величиною 40-70 нм, орієнтовані переважно перпендикулярно до підкладки (рис. 3.1 a), а при $T_S \geq 370$ K домінують стрічкоподібні хаотично орієнтовані кристаліти, шириною 100-200 нм, довжина яких збільшується зі зростанням $T_S$ до 1000 нм (рис. 3.1 b,c). Перехід від овальних до стрічкоподібних кристалітів починається приблизно при 330 K (рис. 3.1 b). Морфологія (форма та розміри кристалітів) практично не залежить від типу підкладок, хоча мікрорельєф (шорсткість) плівок MPP нанесених на шар ITO більш однорідний, ніж на кварцовому склі без шару ITO.

Морфологію плівок MPP, в яких спостерігаються два типи кристалітів в залежності від $T_S$, можна пояснити враховуючи подібність молекулярної структури досліджуваного матеріалу та вихідного перилену. Молекули перилену кристалізуються в двох поліморфних кристалічних модифікаціях [129]. Крім того дослідження кристалічної структури ряду диімід-карбоксильних похідних перилену [130-131] вказує на існування різних молекулярних упаковок в залежності від групи заміщення. Це дозволяє припускати існування в плівках MPP поліморфних структур, які відповідають двом типам кристалітів, що спостерігались експериментально. На даний час тільки одна кристалічна модифікація MPP описана в прямих експериментах [95,130-132], але зміна форми домінуючих кристалітів, що спостерігалась для плівок МРР отриманих при $T_S$ = 330 К (рис. 3.1), вказує на існування іншої фази [133].

Мікроструктура отриманих нами плівок якісно не відрізняється від структури плівок МРР, отриманих в [134-136] при $T_S$ = 300 та 420 К, відповідно.

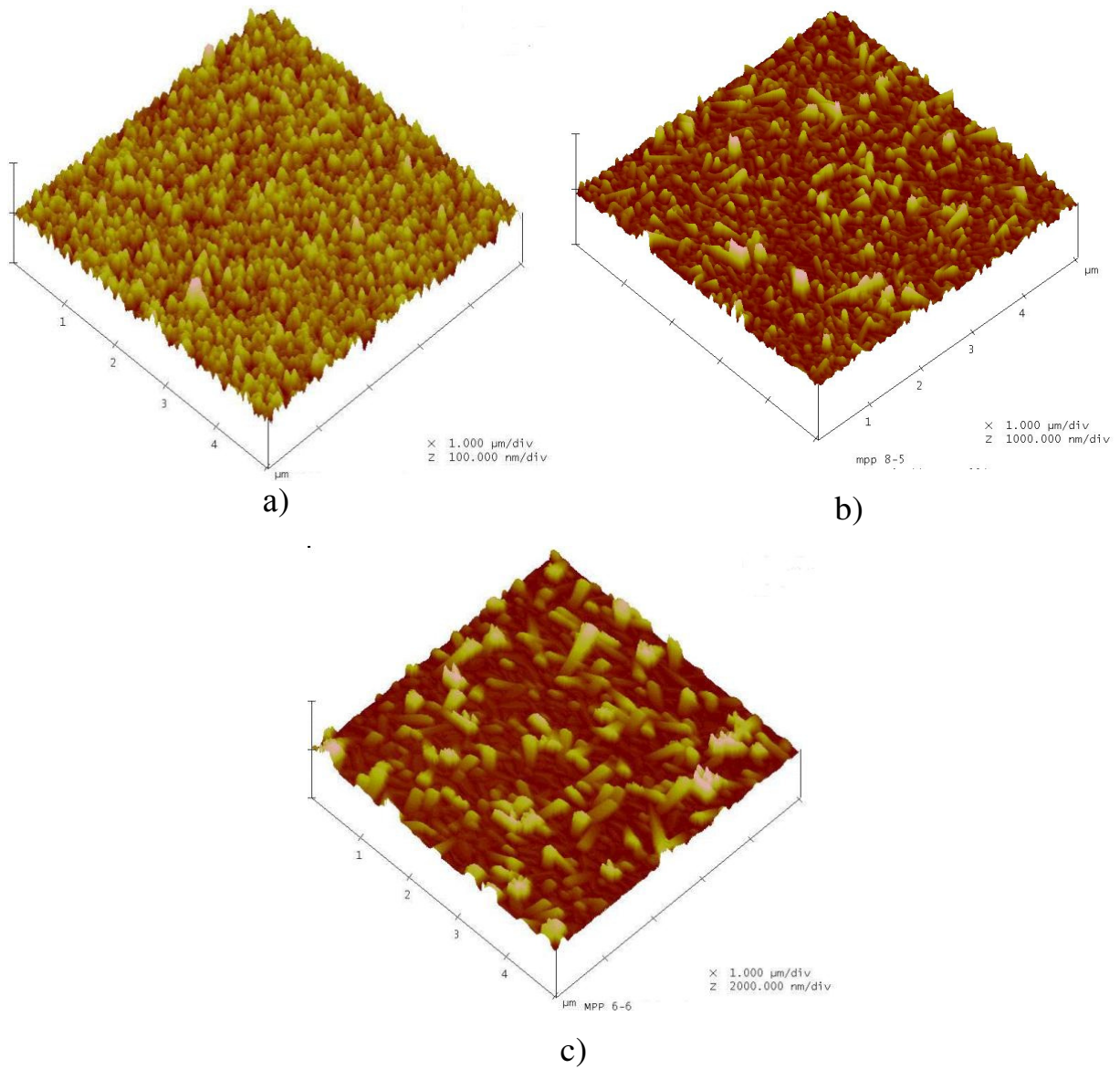

a)

b)

c)

Рис. 3.1  Зображення поверхні розміром 5 × 5 мкм, отримане за допомогою AFM, для плівок МРР виготовлених при $T_S$ = 300 (a), 330 (b) і 370 К (c). Масштаб вертикальних поділок складає 100 (a), 1000 (b) та 2000 (c) нм.

Плівки SnCl₂Pc при $T_S$ = 300 К (рис. 3.2 а) формуються з невеликих кристалітів круглої форми діаметром приблизно 100 нм. Зі збільшенням $T_S$ до 410 К в плівці SnCl₂Pc спостерігаються скупчення декількох круглих кристалітів (конгломерати складної форми). В наслідок чого утворюється відносно гладка поверхня з шорсткістю приблизно 30 нм (рис. 3.2 b). При цьому в плівці виникають великі кристаліти довжиною більше 1 мкм і висотою приблизно 200 нм над поверхнею (рис. 3.2 b).

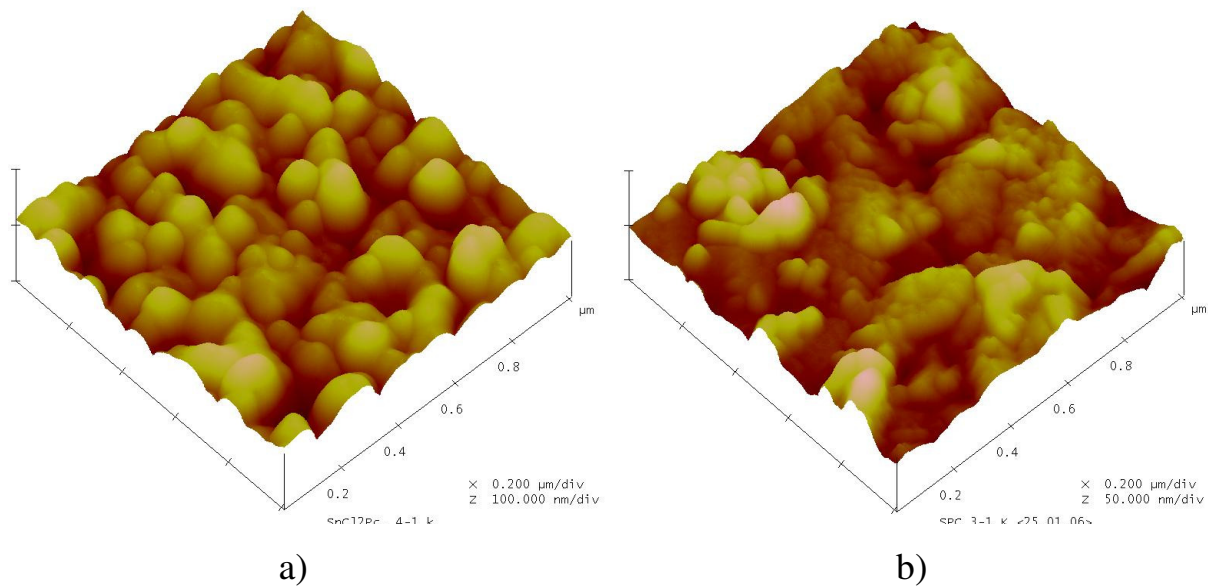

a)                       b)

Рис. 3.2 Зображення поверхні розміром 1 × 1 мкм, отримане за допомогою AFM, для плівок SnCl₂Pc виготовлених при $T_S$ = 300 (а) і 410 К (b). Масштаб вертикальних поділок складає 100 (а) і 50 (b) нм.

Плівки НТР при $T_S$ = 300 К формуються з паличкоподібних кристалітів з округленими гранями величиною приблизно 100 нм. В результаті відпалу кристаліти паличкоподібної форми в плівках НТР витягуються і мають ширину приблизно 100 нм та довжину до 300 нм (рис. 3.3 а). При $T_S$ = 370 К в плівках НТР спостерігаються скупчення кристалітів (конгломерати) складної форми, довжина і ширина яких складає приблизно 100-300 нм, а висота – менше 25 нм (рис. 3.3 b). Отже, зі збільшенням $T_S$ від 300 до 370 К заокруглені паличкоподібні кристаліти в плівках HTP також рекристалізуються в гладкі конгломерати, як це відбувалось в плівках SnCl$_2$Pc.

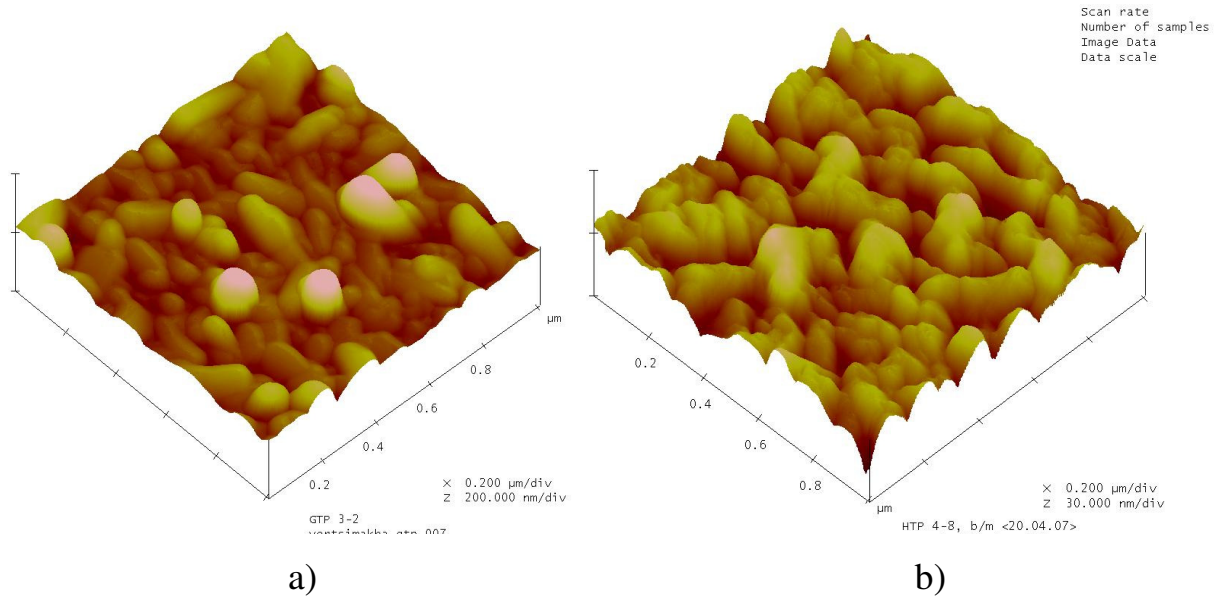

a)                 b)

Рис. 3.3  Зображення поверхні розміром 1 × 1 мкм, отримане за допомогою AFM, для плівок HTP виготовлених при $T_S$ = 300 К та відпалених при $T_A$ = 350 К (a) та виготовлених при $T_S$ = 370 К (b). Масштаб вертикальних поділок складає 200 (a) та 30 (b) нм.

За допомогою функції AFM "Image Statistics: Roughness Analysis" визначені параметри поверхні: розмах висот ("Z range") – $Z$, середньоарифметична шорсткість ("Mean roughness") – $R_a$ та середньоквадратична шорсткість („RMS" – Root-Mean-Square) – $R_q$ для досліджуваних плівок MPP, HTP та SnCl$_2$Pc отриманих при різних $T_S$.

Як бачимо з табл. 3.1, для плівок MPP при переході від овальних ($T_S$ = 300 К) до стрічкоподібних ($T_S$ = 330 К) кристалітів параметри поверхні ($Z$, $R_a$, $R_q$) збільшуються майже на порядок, а зі збільшенням $T_S$ від 330 до 370 К в кілька разів (в 1.5 разів для $Z$ та в 2.5 разів для $R_a$ та $R_q$). Для плівок MPP виготовлених при $T_S$ = 410 К не вдалось отримати зображення структури поверхні через велику шорсткість.

Таблиця 3.1

Параметри поверхні досліджуваних плівок визначені за допомогою функції

AFM "Image Statistics: Roughness Analysis"

| Площа поверхні | $T_S$, К | $Z$, нм – розмах висот поверхні ("Z range") | $R_a$, нм – середньоарифме-тична шорсткість ("Mean roughness") | $R_q$, нм – середньоквадра-тична шорсткість („RMS") |
|---|---|---|---|---|
| MPP (5 × 5 мкм) | 300 | 77.9 | 3.6 | 5.3 |
| | 330 | 553.9 | 37.4 | 51.9 |
| | 370 | 865.3 | 98.0 | 126.4 |
| SnCl$_2$Pc (5 × 5 мкм) | 300 | 180.6 | 11.2 | 15.0 |
| | 410 | 212.4 | 23.2 | 32.5 |
| SnCl$_2$Pc (1 × 1 мкм) | 300 | 99.7 | 10.6 | 13.3 |
| | 410 | 83.2 | 10.1 | 13.3 |
| НТР (5 × 5 мкм) | 300 | 651.7 | 57.6 | 95.0 |
| | 300 [1] | 164.8 [1] | 10.4 [1] | 14.6 [1] |
| | 300 [2] | 166.7 [2] | 12.6 [2] | 16.9 [2] |
| | 370 | 77.3 | 3.4 | 4.8 |
| НТР (1 × 1 мкм) | 300 | 22.4 | 2.3 | 3.0 |
| | 370 | 27.6 | 2.7 | 3.4 |

Примітки:

1. – відпал плівок при $T_A$ = 350К;   2. – відпал плівок при $T_A$ = 370К;

При статистичному аналізі поверхні плівок SnCl$_2$Pc розміром 5 × 5 мкм враховувалось утворення великих кристалітів, а для плівок розміром 1 × 1 мкм, відповідно, великі кристаліти не враховувалось. Так, на поверхні розміром 1 × 1 мкм параметр $Z$ зменшується від 99.7 до 83.2 нм при збільшенні $T_S$ від 300 до 410 К, а при врахуванні великих кристалітів на поверхні розміром $5 \times 5$ мкм $Z$ зростає від 180.6 до 212.4 нм зі збільшенням $T_S$ від 300 до 410 K. Теж саме і з $R_a$ та $R_q$, які на поверхні розміром $5 \times 5$ мкм зростають майже в два рази, а на поверхні розміром $1 \times 1$ мкм практично не змінюються зі збільшенням $T_S$ від 300 до 410 K (див. табл. 3.1).

Аналіз поверхні плівок НТР розміром $1 \times 1$ мкм показує зменшення параметрів $Z$, $R_a$, $R_q$ в результаті відпалу плівок при $T_A = 350$ та 370 K приблизно в 5 разів, а зі збільшенням $T_S$ від 300 до 370 K – практично на порядок (див. табл. 3.1). При цьому шорсткість ($R_a$, $R_q$) поверхні плівок НТР зменшується майже в 20 разів. В результаті чого плівки НТР отримані при $T_S = 370$ K мають найбільш гладку поверхню в порівнянні з МРР та $SnCl_2Pc$.

Отримані дані показують, що зі збільшенням $T_S$ відбувається покращення локальної шорсткості і одночасне формування великих неструктурованих скупчень кристалітів на поверхні плівок $SnCl_2Pc$ та НТР. Утворення таких скупчень кристалітів відбувається за рахунок збираючої рекристалізації (процесу утворення і росту одних кристалічних зерен полікристалу за рахунок інших, тої ж фази) круглих кристалітів. В результаті рекристалізації кристалітів зменшується кількість структурних дефектів в плівках [137], що пояснює покращення параметрів поверхні плівок з підвищенням $T_S$. Так збільшення $T_S$ може призвести до переміни вкладу існуючих поліморфних модифікацій в полікристалічних плівках $SnCl_2Pc$ та НТР, але утворення нової поліморфної модифікації не проявляється, як це спостерігалось для плівок PbPc [81]. При цьому для монокристалічної форми НТР експериментально спостерігається лише одна (триклинна) кристалічна модифікація [75].

## 3.2. Оптичні властивості тонких плівок органічних напівпровідників

### 3.2.1. Екситони з переносом заряду в плівках $SnCl_2$ фталоціаніну

У випадку коли іонний радіус центрального атома (Pb, Sn) чи атомної групи (VO, TiO) молекули Pc великий, цей атом (чи атомна група) виступає за межі площини Рс кільця [82,138], і в результаті чого утворюється неплоска молекула (п. 1.4.1). При формуванні плівки з неплоских молекул Рс, термодинамічно вигідною стає структура, в якій центральний атом однієї молекули знаходиться навпроти атомів С та Н периферії Рс кільця сусідньої молекули. В такому випадку між молекулами Рс виникає електростатична взаємодія, енергія якої порівнянна або більше енергії Ван-дер-ваальсівської взаємодії. Обидві ці причини призводять до збільшення ймовірності переходів за участю СТ-станів і ці переходи уже можна зареєструвати в СП та спектрах ФЕ плівок Рс [85]. У Рс з плоскою структурою молекул (CuPc, ZnPc) переходи за участю СТ-станів заборонені по симетрії і ці переходи можна зареєструвати тільки в спектрах електропоглинання [139,140].

Нелінійна оптична сприйнятливість ($\chi^{(3)}$) плівок $SnCl_2Pc$ на 3 порядки вища, ніж плівок інших Рс [141,142], що може бути обумовлено резонансним співпадінням енергії лазерного збудження з енергією одного із СТ-станів або з енергією подвійних синглет-триплетних переходів в плівках $SnCl_2Pc$. Однак, енергетична структура переходів для плівок $SnCl_2Pc$ на даний час не визначена, а вклад СТ-станів у електронні процеси в (плівках) $SnCl_2Pc$ практично не досліджувався, при тому, що дані про розміри атомів Sn і Cl дозволяють передбачати, що молекули $SnCl_2Pc$ мають неплоску структуру (рис. 2.1).

Тому було досліджено особливості оптичних властивостей плівок $SnCl_2Pc$. Так, СП розчину $SnCl_2Pc$ в диметилформаміді (ДМФА) практично співпадає з СП інших розчинів $SnCl_2Pc$ [143] і добре описується електронним переходом з енергією 1.79 еВ та більш слабкими смугами, пов'язаними з внутрішньо-молекулярними коливаннями величиною 0.08 та 0.19 еВ (рис. 3.4 а). СП плівок $SnCl_2Pc$ схожий на СП розчину, зміщеного в область менших hv на 0.12 еВ (рис. 3.4 b), що обумовлено виникненням слабкої Ван-дер-Ваальсівської взаємодії молекул в плівці.

Півширина смуг поглинання плівок $SnCl_2Pc$ зменшується в результаті відпалу на повітрі при $T_A$ від 330 до 475 К. Отримані дані по впливу відпалу на СП плівок SnCl₂Pc якісно узгоджуються з експериментальними даними в роботі [144].

СП розчину (при зміщенні на 0.12 еВ в сторону менших hν) та плівок SnCl₂Pc (рис. 3.4 b) узгоджуються тільки в області сильного поглинання

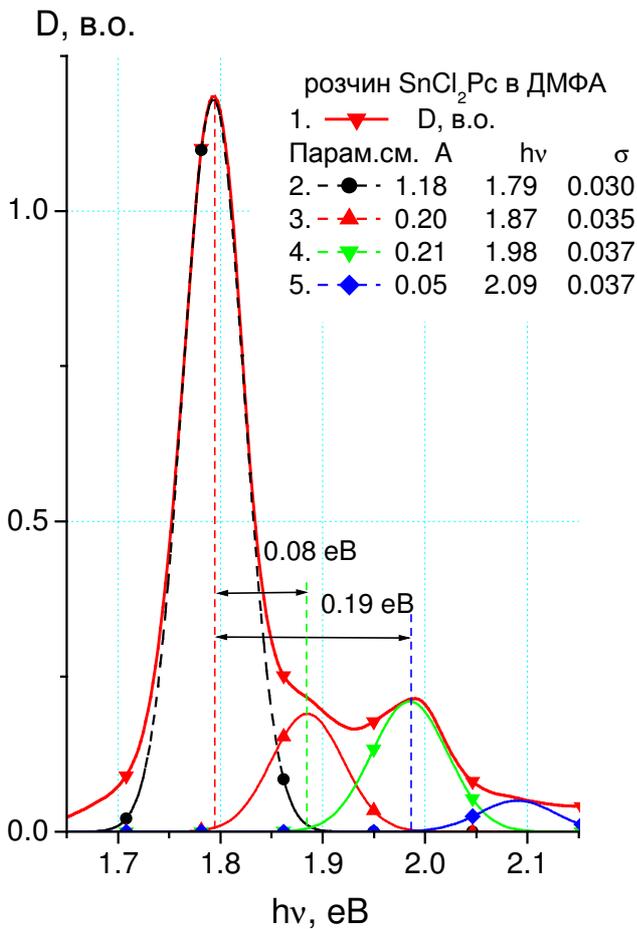 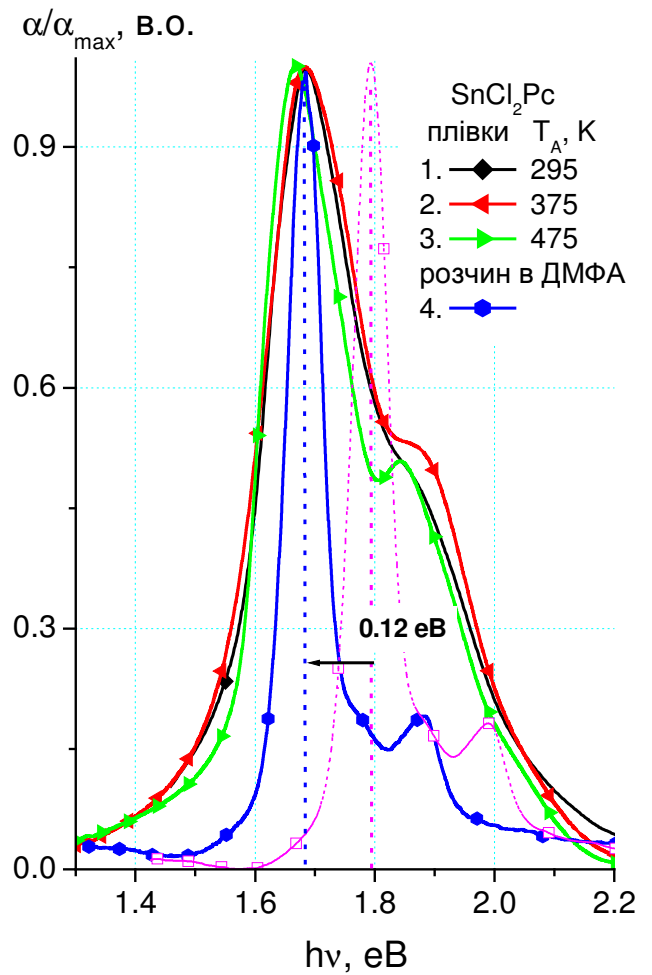

Рис. 3.4a  Спектральна залежність оптичної густини (*D*) розчину SnCl₂Pc в ДМФА та її гаусові складові.

Рис. 3.4b  СП плівок SnCl₂Pc товщиною 180 нм, напилених при *T_S* = 300 К, до відпалу (1) та після відпалу при *T_A* = 375 (2) та 475 К (3) і СП розчину SnCl₂Pc (4), зміщеного на 0.12 еВ в сторону менших енергій.

плівок (1.6-1.9 еВ). В області слабкого поглинання плівок (1.2-1.6 та 1.9-2.2 еВ) інтенсивність поглинання плівок значно більше, ніж поглинання розчину, що вказує на виникнення в плівках $SnCl_2Pc$ додаткового поглинання, аналогів якого немає в молекул $SnCl_2Pc$.

Суттєве розширення смуг СП плівок $SnCl_2Pc$ (в порівнянні зі СП розчину $SnCl_2Pc$) може бути обумовлено як статистичним розупорядкуванням кристалітів в плівці, так і появою додаткового поглинання. Додаткове поглинання в плівках $SnCl_2Pc$ може проявляться при утворенні домішок в процесі термічного напилення. Проте, утворення домішок в процесі напилення малойймовірне, так як в цій області температур молекули Рс стабільні та не розкладаються. Навпаки, при термічному напиленні у вакуумі відбувається додаткове очищення Рс. Також додаткове поглинання в плівках $SnCl_2Pc$ може бути обумовлено утворенням в плівках СТ-станів, які утворюються внаслідок виникнення взаємодії між центральними атомами Cl та периферійними С та Н атомами Рс кільця сусідніх молекул.

Для визначення енергій смуг додаткового поглинання в плівках $SnCl_2Pc$, були виміряні спектри фотомодульованого відбивання – $\Delta R/R$(hν). В області 1.15-1.65 еВ на спектрах $\Delta R/R$(hν) при $T_A$ = 375 та 475 К (рис. 3.5 а) проявляються піки при 1.34±0.02 та 1.55±0.02 еВ. При цьому інтенсивність піку при 1.34±0.02 еВ зростає зі збільшенням $T_A$, а піку 1.55±0.02 еВ, навпаки, зменшується.

Для підтвердження наявності цих смуг були проведені вимірювання СП плівок товщиною 800 нм відпалених при різних температурах. Оптична густина ($D$) плівок $SnCl_2Pc$ після відпалу збільшується в області 1.30-1.38 еВ та зменшується в області 1.40-1.60 еВ зі збільшенням $T_A$. При цьому зміна оптичної густини плівок $SnCl_2Pc$ – $\Delta D$(hν) зі збільшенням $T_A$ має максимум при 1.34±0.02 еВ та мінімуми при 1.55±0.02 еВ та 2.05±0.02 еВ (рис. 3.5 а, кр. 3,4; рис. 3.5 b, кр. 3). Інтенсивність поглинання в максимумі зростає зі збільшенням $T_A$, а в мінімумі навпаки – зменшується.

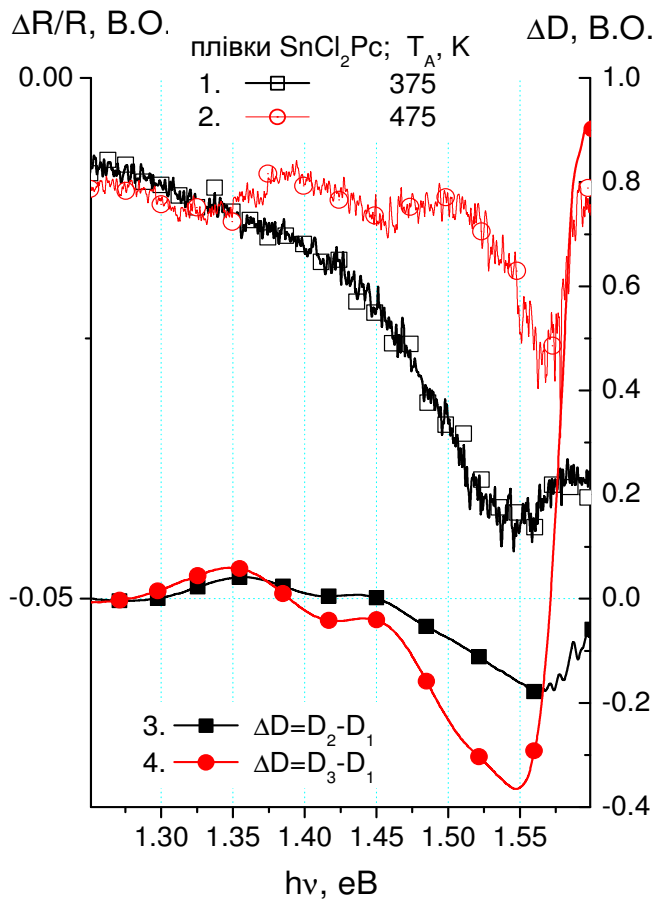
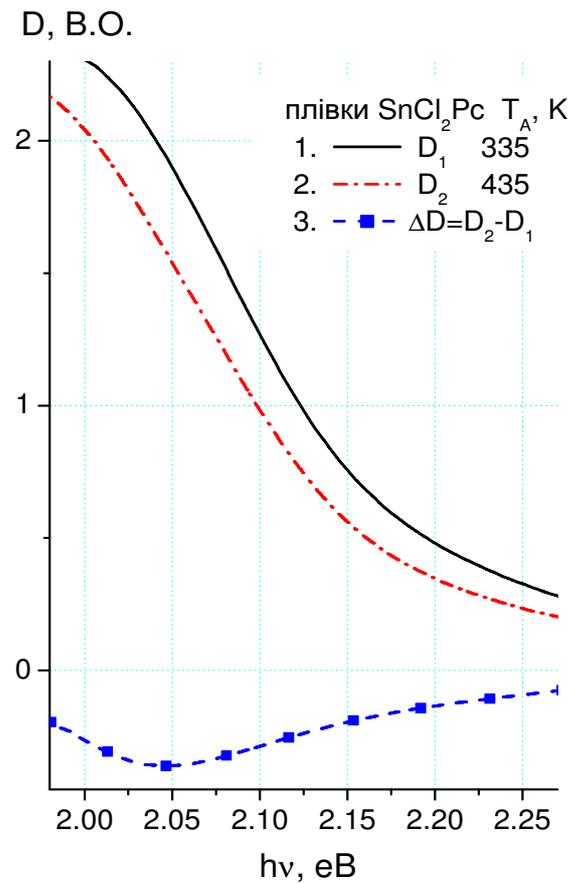

Рис. 3.5a Спектри фотомодульованого відбивання ($\Delta R/R$) після відпалу при $T_A = 375$ К (1), 475 К (2) плівок SnCl$_2$Pc товщиною 180 нм та зміни оптичної густини ($\Delta D$) зі збільшенням $T_A$ від 335 до 435 К (3) і 475 К (4) для плівок товщиною 800 нм.

Рис. 3.5b Спектри оптичної густини ($D$) плівок SnCl$_2$Pc товщиною 800 нм після відпалу при 335 (1) та 435К (2) та її зміна – $\Delta D$(hv) (3) зі збільшенням $T_A$ від 335 до 435 К.

Оптична густина плівок SnCl$_2$Pc відпалених при $T_A = 335$ К – $D_1$, 435 К – $D_2$ та 475 К – $D_3$.

Утворення СТ-станів сильно залежить від кристалічної модифікації плівок. В плівках Рс з неплоскою структурою молекул (PbPc, VOPc, TiOPc) чітко проявляються по два СТ-стани кожної з модифікацій цих сполук. Причому енергія одних СТ-станів менше енергії екситонів Френкеля, а інших, навпаки, – більше [85]. Структура плівок SnCl$_2$Pc має дві (триклінну та моноклінну) кристалічні модифікації [88,145]. Тому в плівках SnCl$_2$Pc також повинні проявлятися два СТ-стани з енергіями менше енергії екситонів Френкеля та два СТ-стани з енергіями більше енергії екситонів Френкеля.

Отже, СП плівок SnCl$_2$Pc формується зі смуг поглинання ізольованих молекул (зсунутих в сторону менших енергій) з максимумами при 1.65, 1.73, 1.84, 1.92 еВ та додаткових смуг поглинання, пов'язаних з утворенням в плівці SnCl$_2$Pc СТ-станів з максимумами при 1.35, 1.52 та 2.05 еВ, що спостерігалися в спектрах $\Delta R/R$(hv) і $\Delta D$(hv). Ймовірно, що поглинання одного з СТ-станів при hv > 1.65 еВ мале і його важко ідентифікувати на фоні інтенсивних смуг поглинання екситонів Френкеля.

Оскільки визначено енергетичне положення та число смуг СП, то можна оцінити їх інтенсивність за допомогою розкладу спектральної залежності на гаусові складові (п. 2.5). В табл. 3.2 наведені результати розкладу на сім смуг гаусової форми з максимумами при 1.35, 1.52, 1.65, 1.73, 1.84, 1.92 та 2.05 еВ для СП плівок SnCl$_2$Pc без відпалу та відпалених при різних $T_A$. Співставлення інтенсивностей ($A$) і площ ($\Theta$) гаусових компонент СТ-станів в плівці SnCl$_2$Pc (див. табл. 3.2) показує, що інтенсивність смуги з максимумом при 1.35 еВ зростає зі збільшенням $T_A$, а інтенсивність смуг з максимумом при 1.52 та 2.05 еВ, навпаки, зменшується з ростом $T_A$. Це підтверджує, що смуга з максимумом при 1.35 еВ обумовлена поглинанням СТ-станів однієї з поліморфних модифікацій, а смуги з максимумами при 1.52 та 2.05 еВ – іншої модифікації.

Вклад смуг поглинання СТ-станів в СП плівки SnCl$_2$Pc визначається сумою параметрів $\Theta_i / \Sigma\Theta_i$ для смуг з максимумами при 1.35, 1.52 та 2.05 еВ (див. табл. 3.2). При порівнянні цих сум для смуг поглинання СТ-станів в СП плівок SnCl$_2$Pc відпалених при різних $T_A$ видно, що загальний вклад смуг поглинання в СП плівки SnCl$_2$Pc дещо зменшується приблизно від 20 до 15% зі збільшенням $T_A$ (див. табл. 3.2). Це пояснюється тим, що ймовірність

Таблиця 3.2

Параметри гаусових складових СП плівок SnCl$_2$Pc без відпалу та відпалених при різних $T_A$.

| $T_A$, К | № смуги | 1 | 2 | 3 | 4 | 5 | 6 | 7 |
|---|---|---|---|---|---|---|---|---|
| без від-палу | hv$_i$, еВ | 1.35 | 1.52 | 1.653 | 1.73 | 1.84 | 1.922 | 2.05 |
| | $A_i$, в.о. | 0.06 | 0.185 | 0.83 | 0.80 | 0.41 | 0.33 | 0.14 |
| | $\Theta_i=A_i\cdot\sigma_i$, в.о. | 0.005 | 0.014 | 0.043 | 0.042 | 0.022 | 0.021 | 0.011 |
| | $\Theta_i$ / $\Sigma\Theta_i$, % | 3.2 | 8.7 | 27.1 | 26.7 | 13.9 | 13.5 | 6.9 |
| 375 | hv$_i$, еВ | 1.35 | 1.52 | 1.65 | 1.73 | 1.84 | 1.924 | 2.05 |
| | $A_i$, в.о. | 0.066 | 0.21 | 0.95 | 0.78 | 0.48 | 0.37 | 0.135 |
| | $\Theta_i=A_i\cdot\sigma_i$, в.о. | 0.006 | 0.014 | 0.048 | 0.04 | 0.025 | 0.022 | 0.01 |
| | $\Theta_i$ / $\Sigma\Theta_i$, % | 3.4 | 8.3 | 29.5 | 24.2 | 15.5 | 13.3 | 5.8 |
| 475 | hv$_i$, еВ | 1.35 | 1.52 | 1.65 | 1.731 | 1.84 | 1.925 | 2.05 |
| | $A_i$, в.о. | 0.09 | 0.19 | 1.38 | 0.75 | 0.64 | 0.37 | 0.13 |
| | $\Theta_i=A_i\cdot\sigma_i$, в.о. | 0.007 | 0.012 | 0.065 | 0.037 | 0.032 | 0.02 | 0.009 |
| | $\Theta_i$ / $\Sigma\Theta_i$, % | 4.0 | 6.5 | 35.7 | 20.2 | 17.6 | 11.2 | 4.8 |

переходів за участю СТ-станів в високотемпературній поліморфній модифікації плівок SnCl$_2$Pc менша, ніж в низькотемпературній модифікації.

Порівняння отриманих даних для досліджуваних плівок SnCl$_2$Pc з даними для плівок інших Pc з неплоскою структурою молекул (PbPc, VOPc та TiOPc [85]) показує, що вклад СТ-станів в СП плівок SnCl$_2$Pc характерний вмісту СТ-станів в СП моноклінних модифікацій цих Pc (15-16%), але значно менше, ніж у триклинних модифікацій VOPc та TiOPc (21-29%). Останнє може бути обумовлено утворенням в плівках VOPc та TiOPc водневоподібних зв'язків між атомами О та Н сусідніх молекул [85].

3.2.2. Спектральні особливості поглинання тонких шарів метил-заміщеного периленового барвника (МРР)

Спектральні особливості агрегації молекул МРР в розчинах та плівках детально описано в літературі (п. 1.4.3), але при цьому не достатньо проаналізовано енергетичну структуру збуджених станів в тонких шарах МРР. Детальне порівняння спектральних залежностей для плівок МРР отриманих при різних $T_S$ повинно дозволити глибше проаналізувати властивості та енергетичну структуру збуджених станів в плівках МРР.

СП в області 1.8-3.1 еВ для плівок МРР напилених на кварцові підкладки як без ITO, так і з ITO при різних $T_S$ складаються з двох мультиплетних смуг з максимумами при (2.17±0.01) еВ та (2.58±0.05) еВ. Форма цих смуг слабо залежить від $T_S$ (рис. 3.6, кр. 1-3), але значно відрізняється від СП молекул МРР в розчині [93] (рис. 1.9; рис. 3.6), що пов'язано з агрегацією молекул МРР в плівці (п. 1.4.3).

При детальному розгляді СП плівок МРР, видно, що зі збільшенням $T_S$ відбувається відносне зменшення інтенсивності смуг поглинання в області 2.5-2.9 еВ та збільшення в області 2.9-3.1 еВ. Крім того, в області 1.75-2.0 еВ спостерігається суттєве збільшення поглинання в зразках виготовлених при $T_S$ = 410 К. Ці результати показують, що кожна з широких мультиплетних смуг поглинання, складаються з більш вузьких смуг, інтенсивності яких залежать від $T_S$.

В спектрах ФЛ отриманих при температурі: $T_m$ = 300 К для плівок МРР спостерігаються по крайній мірі дві смуги з максимумами (1.81±0.01) еВ та (1.98±0.02) еВ (рис. 3.7). Форма цих смуг залежить від $T_S$. Розклад спектрів ФЛ на гаусові компоненти (п. 2.5) вказує на наявність слабкої смуги з максимумом при (1.64±0.02) еВ в спектрах ФЛ виміряних при $T_m$ = 300 К для плівок МРР. Цей висновок узгоджується з результатами досліджень спектрів ФЛ плівок МРР виміряних при низьких температурах ($T_m$ = 1.4 К) [97,146] в

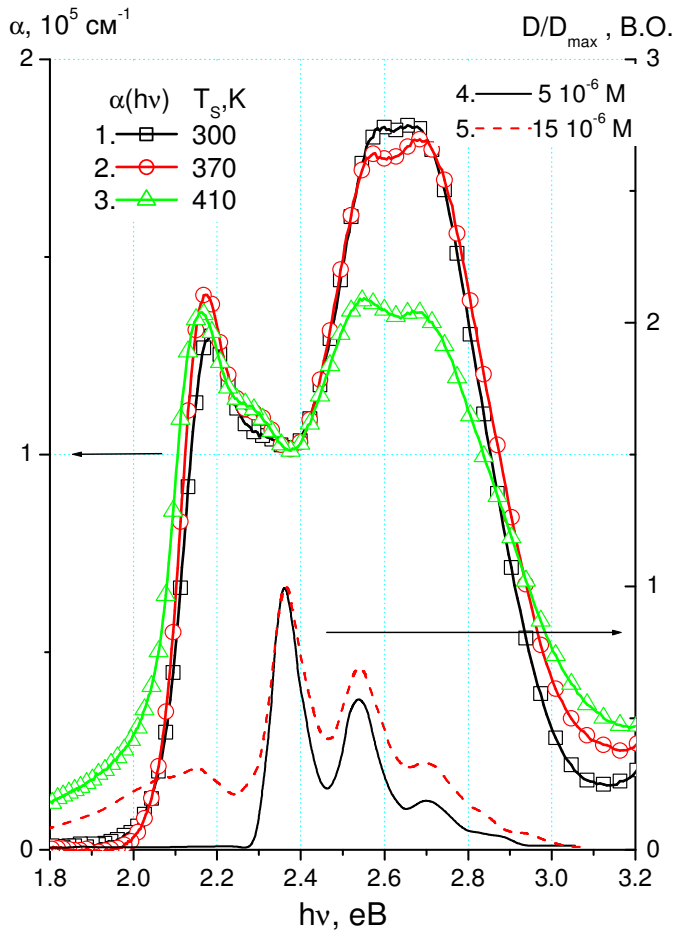

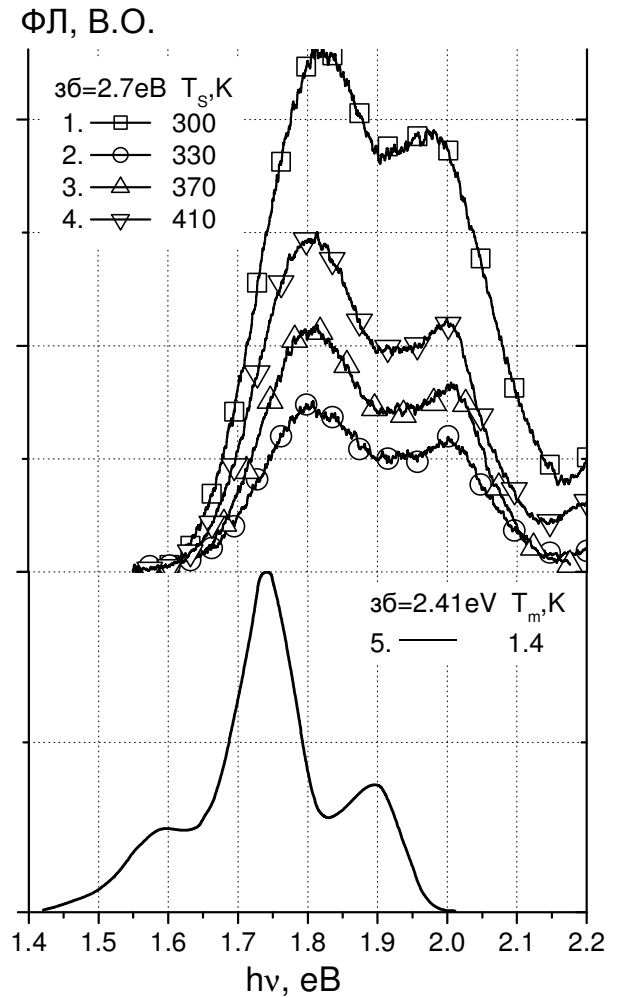

Рис. 3.6  СП плівок MPP нанесених при різних $T_S$ (1-3) та розчинів MPP в хлороформі [93] (4,5). Спектри розчинів нормовані на максимум.

Рис. 3.7  Спектри ФЛ плівок MPP виготовлених при різних $T_S$ і виміряні при $T_m = 300$ К (1-4) і виміряних при низькій температурі ($T_m = 1.4$ К) [97] (5).

яких спостерігались 3 піки при 1.58 еВ, 1.74 еВ та 1.90 еВ (рис. 3.7, кр. 5). Крім того інтенсивності смуг при 1.74 еВ та 1.90 еВ зростають зі збільшенням $T_m$, а смуги при 1.58 еВ, навпаки, зменшується [97,146]. Наші вимірювання проведені при $T_m = 300$ К показують, що інтенсивність ФЛ дійсно залежать від $T_S$: повна інтенсивність смуг різко зменшується зі збільшенням $T_S$ до 330 К, а при більш високих $T_S$ – зростає (рис. 3.7).

Спектри збудження ФЛ плівок МРР (рис. 3.8) мають такі ж особливості як і СП. Піки в спектрах збудження ФЛ є вужчими, що дозволяє визначити положення максимумів смуг 2.17 еВ, 2.55 еВ та 2.66 еВ.

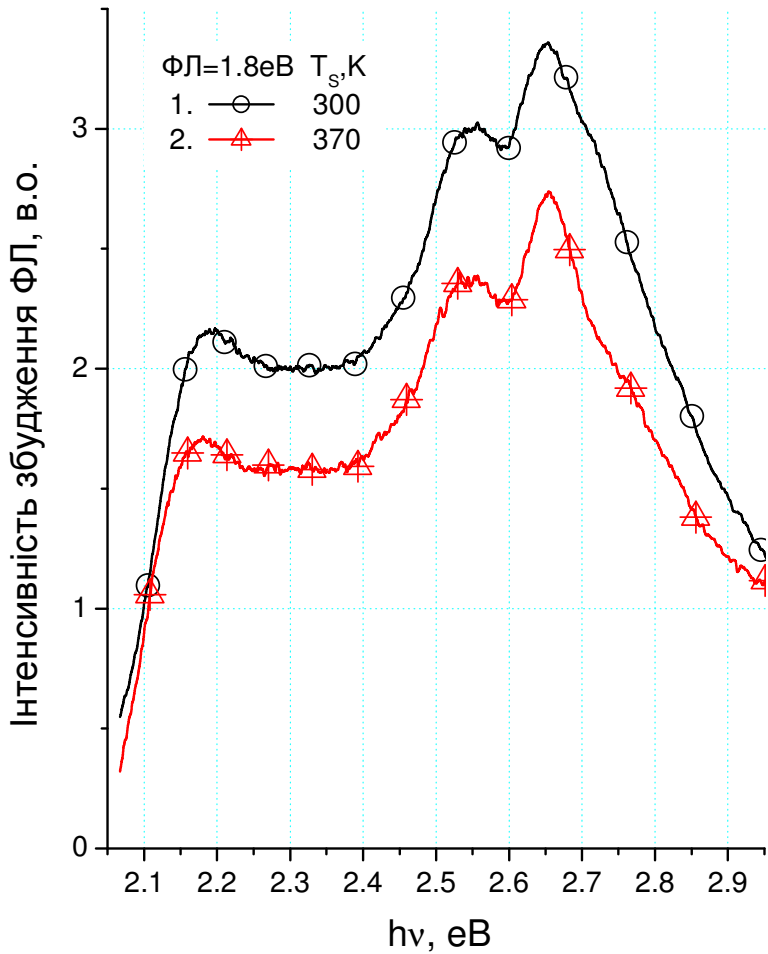

Рис. 3.8  Спектри збудження ФЛ плівок МРР. Інтенсивність випромінювання вимірювалась при 1.8 еВ, $T_S$ при напиленні плівок була 300 (1) та 370 К (2).

Різниця між СП плівок та слабоконцентрованого розчину МРР, а також зміна СП розчинів від концентрації підтверджує існування агрегації молекул МРР [93,97,146]. Слід відмітити, що зростання концентрації розчину призводить до зростання інтенсивності головних смуг з $h\nu > E_m$ ($E_m = 2.37$ еВ), а в більш концентрованих розчинах з'являється група смуг поглинання при $h\nu < E_m$ (рис. 3.6, кр. 5). Структуру цих смуг можна краще побачити в диференційних СП, отриманих відніманням спектру слабоконцентрованого розчину від спектру розчину МРР з більшою концентрацією, де крім смуг при $h\nu < E_m$ проявляються смуги при $h\nu > E_m$.

Інтенсивність смуг з максимумами в області 2.4-2.9 eB зростає зі збільшенням $T_m$ значно сильніше, ніж смуг при hv < $E_m$.

При розкладі СП розчинів та тонких плівок MPP на гаусові складові з максимумами (згідно з експериментом), як показано в табл. 3.3, спостерігається хороше узгодження положень максимумів СП розчинів та плівок MPP. На основі цього узгодження хотілось приписати смуги з максимумами більше $E_m$ до переходів в ізольованих молекулах MPP, а смуги з максимумами менше $E_m$ – до переходів в їх агрегатах ("фізичних димерах"). Але існують аргументи, що заперечують такий висновок. По-перше, найбільш інтенсивна смуга в СП розчину з максимумом при $E_m$ = 2.37 eB, пов'язана з чисто електронним переходом в молекулі MPP, відсутня (чи дуже слабка) в СП плівок. По-друге, на наших AFM зображеннях поверхні плівок MPP не проявляється аморфна фаза (рис. 3.1), наявність якої було б підтвердженням існування "мономерних" піків. Таким чином потрібно шукати інше пояснення.

Особливості СП плівок MPP в літературі [97,146] пояснюються існуванням одного типу молекулярних димерів (чи агрегатів), молекули в яких нахилені одна відносно одної з кутом нахилу ±36°. Через наявність

Таблиця 3.3

Енергії смуг в СП розчинів та тонких плівок MPP.

| Зразок | Енергія смуг (eB) | | | | | | |
|---|---|---|---|---|---|---|---|
| Розчин в хлороформі [93] концентрацією | | | | | | | |
| $5 \times 10^{-6}$ M | | | | 2.37 | 2.54 | 2.70 | 2.86 |
| $15 \times 10^{-6}$ M | 2.03 | 2.16 | 2.28 | 2.37 | 2.54 | 2.70 | 2.86 |
| Тонка плівка | | | | | | | |
| $T_m = 1.4$K [97] | 2.01 | 2.13 | 2.26 | 2.38 | 2.51 | 2.67 | 2.82 |
| $T_S = 300$ K чи 370 K | **2.03** | **2.16** | **2.29** | **2.42** | **2.54** | **2.70** | **2.85** |

агрегатів електронні рівні розщеплюються на два підрівні, відповідно виразу (1.1). Тому, один чистий електронний перехід повинен спостерігатись в кожній групі смуг (2.03 еВ та 2.54 еВ в наших експериментах) і супроводжуватись коливальними смугами (2.16 еВ та 2.29 еВ в низько-енергетичній групі смуг і 2.70 еВ та 2.85 еВ у високо-енергетичній групі смуг) (рис. 3.9 а).

При такій схемі величина розщеплення – $\gamma_e$ за формулою (1.1), є надто великою для типових молекулярних кристалів, оскільки відношення $\gamma_e / \Delta E_m$ перевищує 0.1. Тобто міжмолекулярні взаємодії в димерах не можна розглядати в рамках наближення слабкого зв'язку.

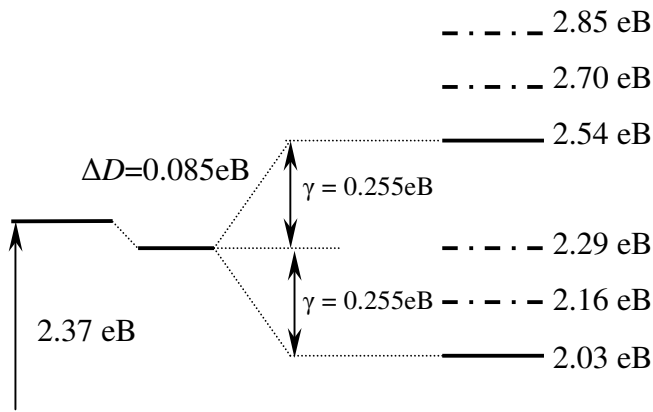
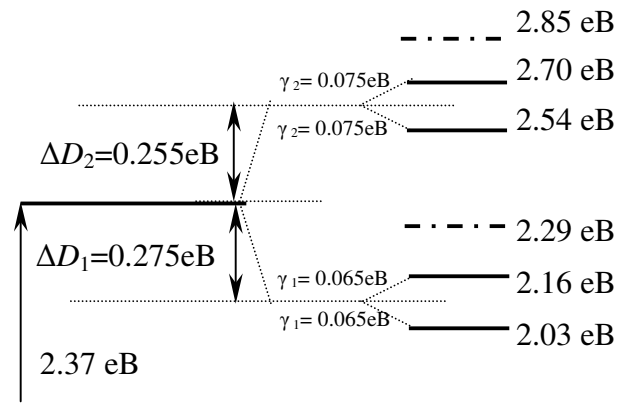

Рис. 3.9a  Діаграма енергетичних рівнів для плівок МРР за моделлю однотипних димерів [97,146].

Рис. 3.9b  Діаграма енергетичних рівнів для плівок МРР за моделлю двох типів взаємодій [147].

Тому було запропоновано інший механізм з врахуванням двох основних типів взаємодій. Оскільки для монокристалу МРР [95,131] є характерним існування молекулярних шарів з високим ступенем перекриття сусідніх молекул (рис. 3.10), то між молекулами МРР існують переважно два основних типи взаємодій: між молекулами в межах шару («в площині», що відповідає нахиленій геометрії «хвіст до голови»), та між молекулами в сусідніх шарах («в стопці», тобто з домінуванням паралельної геометрії). Ці два типи взаємодій призводять до різних значень $\Delta D$ та $\gamma_e$ в формулі (1.1); і в обох випадках електронні рівні розщеплюються на два підрівні відповідно до принципу заборони Паулі (рис. 3.9 b). Тому в кожній групі смуг повинні спостерігатись два електронні переходи з різними поляризаціями (в наших спектрах це повинні бути смуги з максимумами при 2.03 еВ та 2.16 еВ в низько-енергетичній групі і з максимумами при 2.54 еВ та 2.70 еВ у високо-енергетичній групі), які супроводжуються смугами вібронної природи (в нашому випадку 2.29 та 2.85 еВ) (рис. 3.9 b).

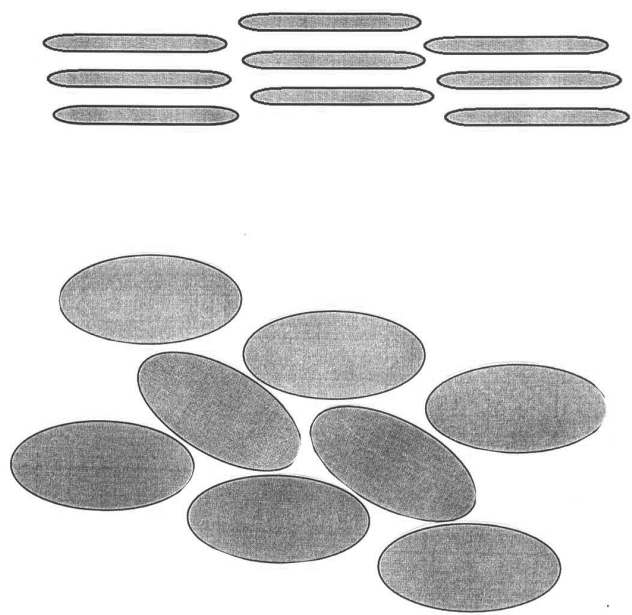

Рис. 3.10  Схематичне зображення впорядкування молекул в монокристалі МРР отримане на основі літературних даних [95].

За цією моделлю, формування агрегатів «в площині» призводить до утворення двох електронних рівнів, що знаходяться на $(0.275 \pm 0.01)$ еВ нижче рівня $E_m$ ізольованої молекули і розщеплені на $(0.065 \pm 0.01)$ еВ, а формування агрегатів «в стопці» призводить до утворення двох електронних рівнів розміщених на $(0.255 \pm 0.01)$ еВ вище рівня $E_m$ і розщеплені на $(0.075 \pm 0.01)$ еВ. Таким чином, в обох випадках переважає слабка міжмолекулярна взаємодія типова для молекулярних кристалів ($\gamma_e/E_m \approx 0.03$). Порівняння результатів отриманих в цій роботі і літературних даних з передбаченнями обох моделей, показує, що більшість результатів можна просто пояснити за допомогою моделі, що припускає існування двох типів агрегатів:

1. Відповідно до [148-149], переважний вклад в формування смуг при hv < $E_m$ дають екситони Френкеля, а в формування смуг при hv > $E_m$ – СТ-стани. При цьому утворення СТ-станів найбільш ймовірне для найближчих сусідніх молекул. А для кристалічної структури МРР характерна дуже мала відстань між сусідніми шарами (~ 0.32 нм) [95], тобто між молекулами «в стопці»;

2. В СП та ФЛ інтенсивності смуг з енергіями менше, ніж $E_m$ мають інші залежності від $T_m$ та $T_S$, ніж смуги з енергіями більше $E_m$;

3. Інтенсивність смуги при 2.16 еВ значно більше, ніж смуги при 2.03 еВ в СП плівок МРР. В спектрах ФЛ МРР інтенсивності відповідних смуг випромінювання мають таку ж закономірність: смуга при 1.81 еВ є більш інтенсивною, ніж при 1.98 еВ.

Ці співвідношення добре узгоджуються з закономірностями, що спостерігаються в органічних кристалах, в яких формування кристалічної структури починається з утворення агрегатів (напр. димерів). Всі ці співвідношення було б набагато важче пояснити припускаючи, що смуги 2.03 еВ та 2.54 еВ є компонентами одного і того ж електронного переходу.

### 3.3. Фотовольтаїчні властивості органічних тонкоплівкових структур

В області слабкого поглинання ($\alpha$ < 3-6·$10^4$ см$^{-1}$) ФЕ структури ITO/SnCl$_2$Pc практично не змінюється в досліджуваній області $T_S$. корелює з СП плівки SnCl$_2$Pc (рис. 3.11). При цьому ФЕ при освітленні вільної поверхні – $\varphi_S$(hv) та контакту з ITO – $\varphi_C$ (hv) практично однакова. В області сильного поглинання ($\alpha$ > 6·$10^4$ см$^{-1}$) плівок SnCl$_2$Pc $\varphi_C$ менше $\varphi_S$ і спостерігається антикореляція $\varphi_C$(hv) та $\alpha$(hv) плівок SnCl$_2$Pc. Так, при $\alpha$ < 3·$10^4$ см$^{-1}$ залежність $\varphi(\alpha)$ лінійна при освітленні з різних сторін плівки SnCl$_2$Pc, а при $\alpha$ $> 4\cdot10^4$ см$^{-1}$ $\varphi_S(\alpha)$ прямує до насичення, в той час як $\varphi_C$ зменшується зі збільшенням $\alpha$ (рис. 3.12).

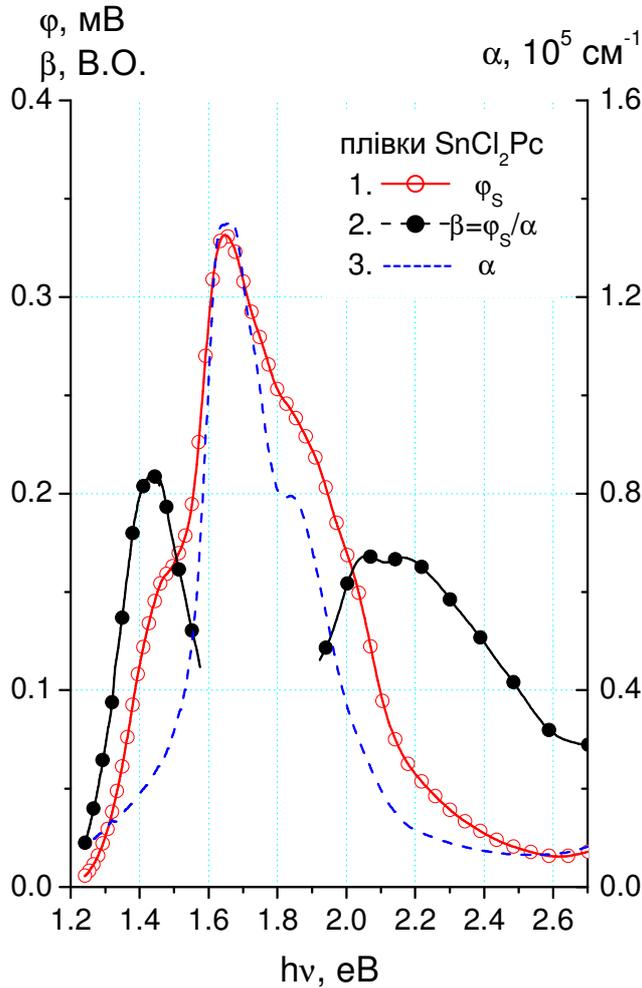 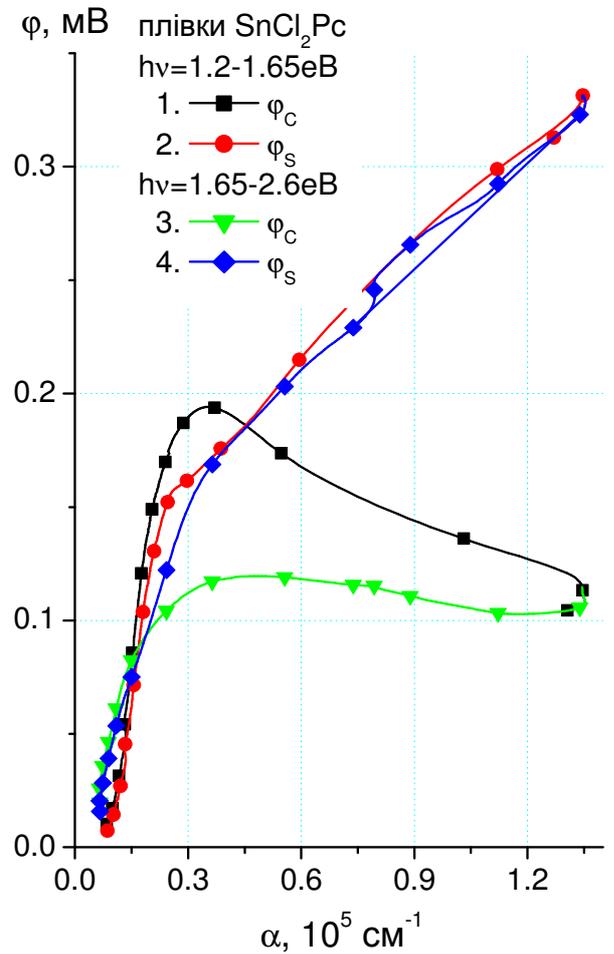

Рис. 3.11 Спектральні залежності ФЕ $\varphi_S$ (1), коефіцієнта поглинання $\alpha$ (2) та відношення $\beta = \varphi_S/\alpha$ (3) для структур ITO/SnCl$_2$Pc, напилених при $T_S$ = 415 К.

Рис. 3.12 Залежності ФЕ від коефіцієнта поглинання $\varphi(\alpha)$ при освітленні контакту (1,3), вільної поверхні (2,4) в області 1.20-1.65 еВ (1,2) та області 1.65-2.60 еВ (3,4) плівок SnCl$_2$Pc.

Однакове значення ФЕ плівок SnCl$_2$Pc при освітленні з обох сторін і кореляція ФЕ зі СП в області слабкого поглинання (рис. 3.12) свідчить, що згин зон біля вільної поверхні та ГР з ITO контактом для плівки SnCl$_2$Pc однаковий. А прямування до насичення залежності $\varphi(\alpha)$ зі збільшенням $\alpha$ та антикореляція $\varphi_C$(hv) та $\alpha$(hv) для плівок SnCl$_2$Pc в області сильного поглинання $(\alpha > 6 \cdot 10^4$ см$^{-1})$ пояснюється наявністю одно – та, відповідно, двоступінчатої рекомбінації в плівках SnCl$_2$Pc. Пряма (одноступінчата) рекомбінація носіїв заряду відбувається при відсутності захоплення носіїв заряду домішковими центрами, а двоступінчата – навпаки (нерівноважні носії спочатку захоплюються домішковими центрами і після захоплення цим же зарядженим центром носія протилежного знаку відбувається рекомбінація) [150-152].

Пряма рекомбінація носіїв заряду переважає при освітленні вільної поверхні плівки SnCl$_2$Pc і може бути обумовлена наявністю центрів рекомбінації носіїв заряду зі сторони вільної поверхні. Наявність антикореляції $\varphi_C$(hv) та $\alpha$(hv) для структури ITO/SnCl$_2$Pc свідчить про двоступінчату рекомбінацію через центри захоплення нерівноважних носіїв заряду, які утворюються на ГР з ITO-електродом. Різниця між залежностями $\varphi(\alpha)$ в різних спектральних областях (рис. 3.12) невелика і може бути обумовлена різним вкладом СТ-станів в цих спектральних областях.

Отримані данні не дозволяють визначити природу центрів рекомбінації чи захоплення носіїв заряду. Проте можна припустити, що центри рекомбінації утворюються при адсорбції на вільній поверхні плівки SnCl$_2$Pc активних молекул з повітря (наприклад, кисню), а центри захоплення утворюються внаслідок дифузії атомів з ITO-електроду в процесі формування плівки SnCl$_2$Pc.

Порівняння ФЕ плівок SnCl$_2$Pc, MPP та фулерену C$_{60}$ (всі n-типу провідності) показує, що фоточутливість плівок SnCl$_2$Pc (у видимій та ближній ІЧ області) порівняна з фоточутливістю шарів MPP і приблизно в 5 разів більше термічно напилених плівок фулерену C$_{60}$ [153], які широко використовуються для розробки CE. Це говорить про перспективність НП шарів n-типу провідності SnCl$_2$Pc для створення фоточутливих ГС на їх основі.

Згідно з [150-152] величина ФЕ може бути представлена за допомогою наближеного виразу:

$$\varphi = A\beta\alpha/(1+\alpha L), \tag{3.1}$$

де $A$ – параметр не залежний від hν і визначається величиною згину зон на освітленій поверхні. Звідси витікає, що в області невеликого поглинання ($\alpha L < 1$) спектральна залежність $\beta$(hν) ~ $\varphi$(hν)/$\alpha$(hν).

При збудженні CT-станів відбувається часткове розділення носіїв заряду і тому для утворення вільних носіїв заряду необхідно затратити менше енергії, ніж для іонізації екситонів Френкеля. Тому $\beta$ при збудженні CT-станів може бути більше, ніж при збудженні екситонів Френкеля, якщо різниця енергій між рівнем CT-станів і зоною провідності невелика. В такому випадку на залежності $\beta$(hν) будуть з'являтися максимуми при енергіях, де суттєвий вклад в поглинання дають CT-стани.

На спектральній залежності $\varphi$(hν)/$\alpha$(hν) в області слабкого поглинання (рис. 3.11, кр. 2) для плівок $SnCl_2Pc$, проявляються максимуми при 1.5 та 2.0 еВ, енергетичне положення яких практично співпадає з максимумами поглинання 2-х CT-станів (рис. 3.5 a,b). В області сильного поглинання (1.6-1.95 еВ) спектральна залежність $\varphi$(hν)/$\alpha$(hν) не корелює з $\beta$(hν), тому значення $\varphi$(hν)/$\alpha$(hν) не наведені на рис. 3.11. CT-стани з енергією 1.35 еВ в спектрах $\varphi$(hν) та відношення $\varphi$(hν)/$\alpha$(hν) не проявляються, тобто $\beta$ при його збудженні менше, ніж при збудженні екситонів Френкеля та інших CT-станів. Невеликі значення $\beta$ для CT-станів при енергії 1.35 еВ можуть обумовлюватись більшою різницею енергії цього стану і зони провідності, ніж для екситонів Френкеля та CT-станів 1.52 та 2.05, і тому ймовірність утворення вільних носіїв для CT-станів при енергії 1.35 еВ дуже мала.

Незважаючи на невелику зміну СП плівок МРР (рис. 3.6), збільшення $T_S$ від 300 до 370 К призводить до суттєвого зростання ФЕ – як $\varphi_C$, так і $\varphi_S$ (рис. 3.13 та рис. 3.14). Слід відмітити, що збільшення ФЕ співпадає зі зменшенням інтенсивності ФЛ (рис. 3.7). При цьому зміни величини ФЕ залежать як від hν, так і від напрямку освітлення. Це обумовлено складною залежністю ФЕ від hν та $\alpha$, які суттєво відрізняються для першої та другої груп смуг поглинання (рис. 3.6). В спектральній області першої групи (1.9-2.3 еВ), $\varphi$ корелює з $\alpha$. При цьому в області слабкого поглинання (1.9-2.1 еВ) $\varphi$(hν) практично не залежить від напрямку освітлення плівки ($\varphi_C \approx \varphi_S$). Що свідчить про утворення однакового згину зон біля ГР з ITO контактом та на вільній поверхні плівок МРР. А в області сильного поглинання (2.1-2.3 еВ) $\varphi_C$(hν) значно більше $\varphi_S$(hν), що підтверджується прямуванням залежності $\varphi$($\alpha$) до насичення зі збільшенням $\alpha$, яке для $\varphi_S$ досягається при менших значеннях $\alpha$ (рис. 3.14).

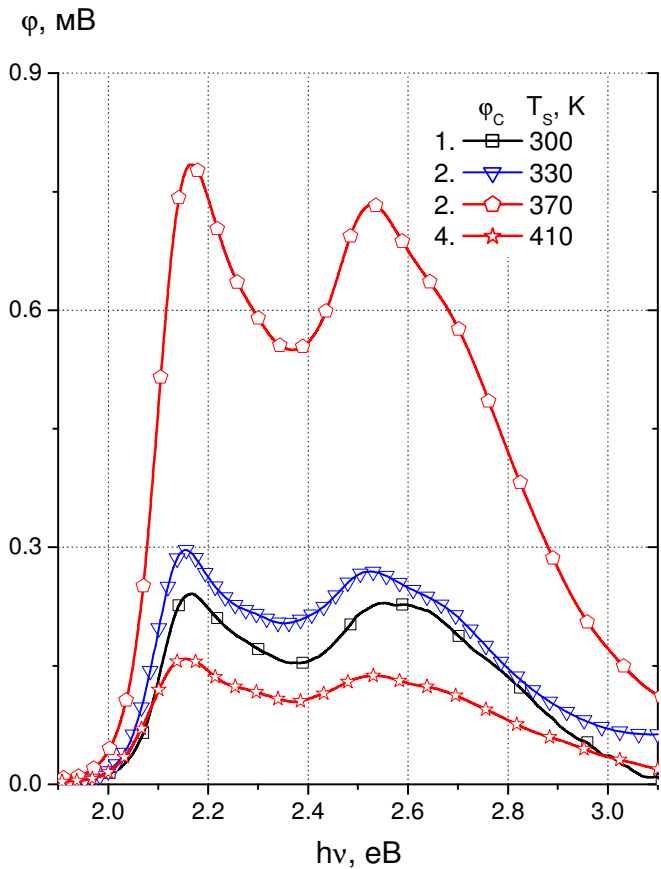

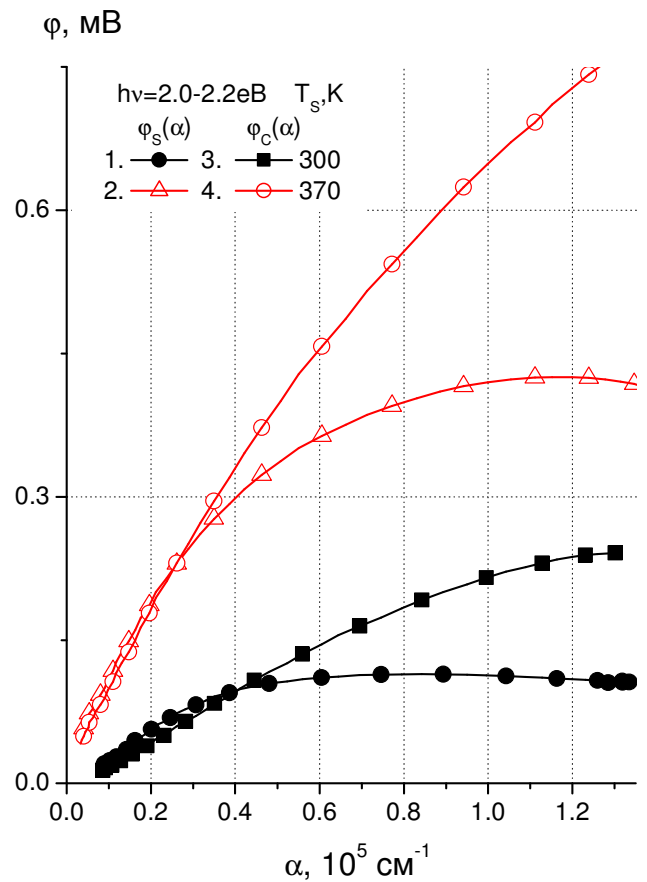

Рис. 3.13  Спектри ФЕ при освітленні через ITO контакт ($\varphi_C$) плівок МРР нанесених при різних $T_S$.

Рис. 3.14  Залежності $\varphi_S$ (1,2) та $\varphi_C$ (3,4) від $\alpha$ в спектральній області 2.0-2.2 еВ для плівок МРР отриманих при $T_S$ = 300 K (1,3) та 370 K (2,4).

В області другої групи смуг поглинання (2.3-3.1 еВ) залежності ФЕ для плівок МРР від hν та $\alpha$ більш складні. А порівняння $\varphi$(hν) для плівок отриманих при різних $T_S$ показує, що зміни в цій групі смуг обумовлені за рахунок зростання $\varphi$ в області 2.40-2.55 еВ, тобто в області переходу 2.54 еВ.

В загальному випадку, $\varphi$(hν) залежить від кількох параметрів, найбільш важливими з яких є $L$, $S$, товщина області просторового заряду та кінетичні коефіцієнти захоплення носіїв заряду в області просторового заряду. В літературі приводяться числові розрахунки залежностей величини $\varphi$ від цих параметрів. Для пояснення наших результатів можна скористатися результатами розрахунків для аморфного кремнію Si [150-152], для якого $L$ та концентрація вільних носіїв заряду є близькими до таких же значень в плівках органічних НП [72,154]. Відповідно до згаданих вище обчислень при відсутності ефективного поверхневого захоплення носіїв заряду залежність $\varphi(\alpha)$ повинна бути лінійною в області слабкого поглинання і прямувати до насичення в області сильного поглинання. При цьому значення $\varphi$ зростає зі збільшенням $L$ і навпаки зменшується зі збільшенням $S$ (див. п. 1.1).

Нахил залежностей $\varphi_S(\alpha)$ та $\varphi_C(\alpha)$ в області слабкого поглинання і їх значення в області насичення в кілька разів ($\approx 4$) більше для плівок МРР отриманих при $T_S = 370$ K, ніж для плівок отриманих при $T_S = 300$ K. Це може бути обумовлено як збільшенням $L$, так і зменшенням $S$. В плівках МРР отриманих при заданих $T_S$, початкові нахили залежностей $\varphi_S(\alpha)$ та $\varphi_C(\alpha)$ в області слабкого поглинання однакові, але при виході на насичення: $\varphi_S < \varphi_C$. Так як концентрація кисню на вільній поверхні плівок МРР більше, ніж біля ГР з ITO, то можна припустити, що цей ефект відбувається внаслідок збільшення $S$ на вільній поверхні за рахунок адсорбованого кисню. Найбільш ймовірно, що плівки нанесені при $T_S = 370$ K мають оптимальне структурне впорядкування молекул та кристалітів і адсорбують менше кисню, в наслідок цього формується менше центрів поверхневої рекомбінації.

При відпалі плівок МРР у вакуумі тривалістю 1 година при різних температурах 330-410 K спостерігається незначне (до 40 %) зростання ФЕ.

Тобто, відпал (зміна $T_A$) менш ефективно впливає на фотовольтаїчні властивості досліджуваних плівок, ніж зміна $T_S$ при напиленні плівок.

Дослідження оптичних властивостей плівок НТР показує, що СП плівок НТР практично на змінюються, як з відпалом при $T_A$ = 350-370К, так і зі збільшенням $T_S$ від 300 К до 370 К при напиленні плівок. А оскільки НТР мало досліджений в літературі, то встановити тонку структуру СП плівок НТР, або вклад можливих тіопохідних Pn іншої структури [70,75] нам не вдалося. Зі спектральних залежностей ФЕ для плівок НТР (рис. 3.15 та 3.16)

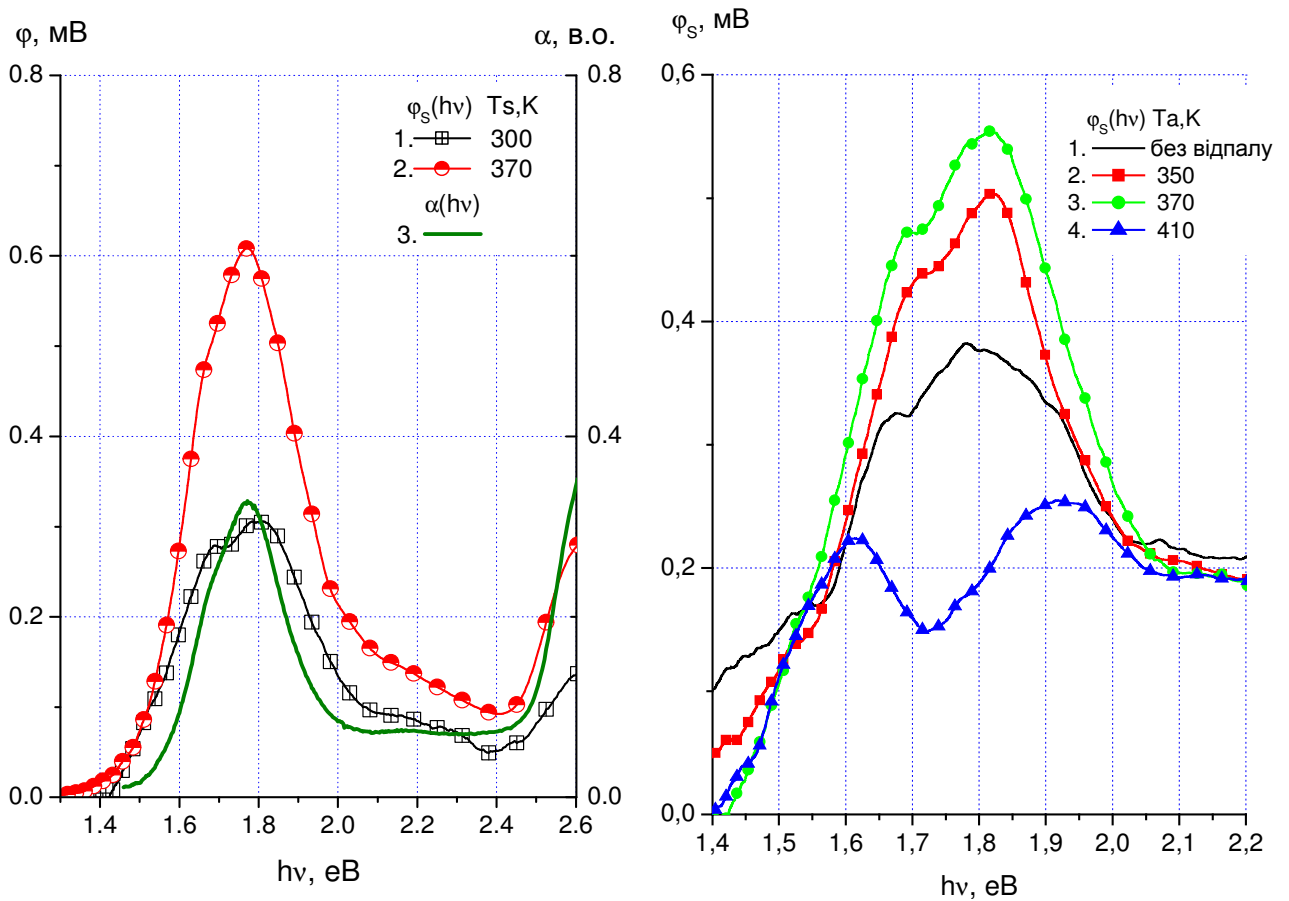

Рис. 3.15 Спектри ФЕ при освітленні вільної поверхні ($\varphi_S$) плівок НТР отриманих при $T_S$ = 300 (1) та 370 (2) К та СП плівок НТР (3).

Рис. 3.16 Спектри ФЕ при освітленні вільної поверхні ($\varphi_S$) плівок НТР отриманих при $T_S$ = 300 К без відпалу (1) та відпалених при різних $T_A$ (2-4).

видно, що в області першого електронного переходу (1.45-1.75 еВ) при $\alpha < 4 \cdot 10^4$ см$^{-1}$, $\varphi_C \approx \varphi_S$ і корелює з $\alpha$ (рис. 3.16), що також свідчить про однаковий згин зон с обох сторін плівки. При подальшому збільшенні $\alpha$ ($\alpha > 4 \cdot 10^4$ см$^{-1}$) $\varphi_C(\alpha)$ та $\varphi_S(\alpha)$ прямують до насичення, при цьому $\varphi_S < \varphi_C$ (рис. 3.16), що вказує на наявність суттєвої $S$, яка більше зі сторони вільної поверхні плівки НТР.

В результаті відпалу на повітрі при $T_A = 350$-$370$ К в області першого електронного переходу діапазон кореляції спектрів ФЕ та $\alpha$ плівок НТР збільшується до $\alpha < 7 \cdot 10^4$ см$^{-1}$ (рис. 3.16, кр 1, 2), а при $T_A = 390$ К – зменшується до $\alpha < 3 \cdot 10^4$ см$^{-1}$ і навіть виникає антикореляція $\varphi_S$ та $\alpha$ (рис. 3.16, кр. 3). Оскільки згин зон з обох сторін плівки НТР однаковий, то це свідчить про зменшення $S$ при $T_A = 370$ К і, навпаки, суттєве зростання $S$ в результаті відпалу при більш високих $T_A$.

### 3.4. Визначення довжини дифузії екситонів в плівках органічних напівпровідників

Як відомо з літератури, екситони Френкеля та CT-стани можна описати дифузійним рівнянням, враховуючи кулонівську взаємодію між електронно-дірковою парою в CT-станах, так зване наближення Онзагера [155]. При цьому згідно з розв'язком цього рівняння, або згідно з виразами (1.2) та (3.1) чи відповідно до [151,152], при невеликих $S$ можна визначити $L$ екситонів за допомогою інтерполяції залежності $\varphi^{-1}(\alpha^{-1})$ в області сильного поглинання:

$$\frac{1}{\varphi} = A\left(1 + \frac{1}{\alpha L}\right),$$
(3.2)

де $A$ не залежить від $\alpha$. Таким чином лінійна екстраполяція залежності $\varphi^{-1}(\alpha^{-1})$ в граничній області дає значення $L$ екситонів в точці $\varphi^{-1} = 0$.

Використовуючи цю процедуру для плівок МРР до смуг в спектральній області 2.05-2.15 еВ, де домінуючий вклад в СП дають екситони Френкеля, отримуємо значення $L = (25 \pm 5)$ нм для плівок виготовлених при 300 К, і $L=$ (55 ± 5) нм для плівок виготовлених при 370 K (рис. 3.17). В спектральній області 2.6-2.95 еВ, де головний вклад дають СТ-стани [148,149], екстраполяція залежності $\varphi^{-1}(\alpha^{-1})$ дає $L = (10 \pm 5)$ нм для плівок нанесених при 300 K, і $L = (25 \pm 5)$ нм для плівок нанесених при 370 K. Таким чином $L$ при збудженні СТ-станів приблизно в два рази менше, ніж при збудженні екситонів Френкеля в однакових умовах експерименту.

В плівках $SnCl_2Pc$ $L = (130 \pm 30)$ нм і однакова, як при $T_S = 300$ K, так і при $T_S = 410K$ (рис. 3.18). На жаль, цим способом не можна достовірно

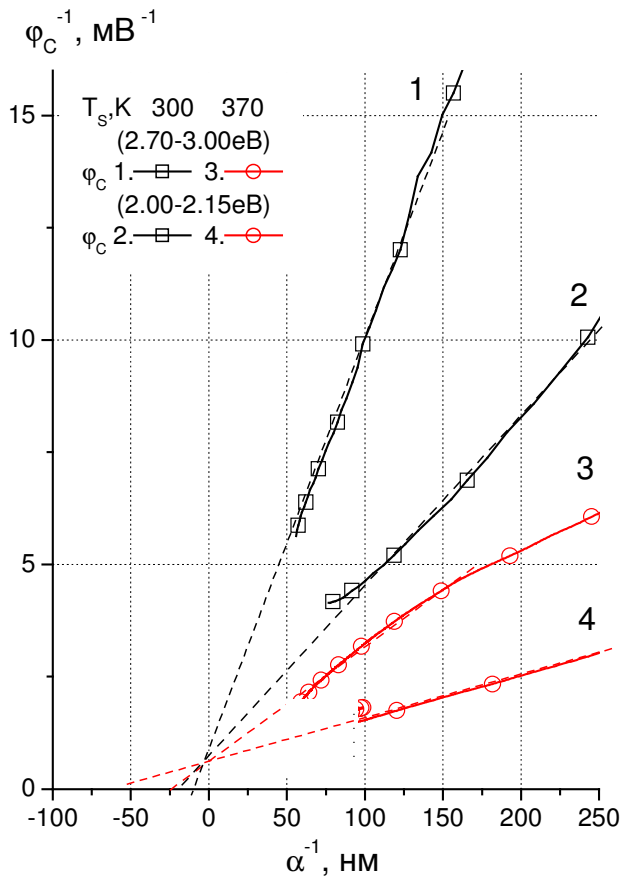
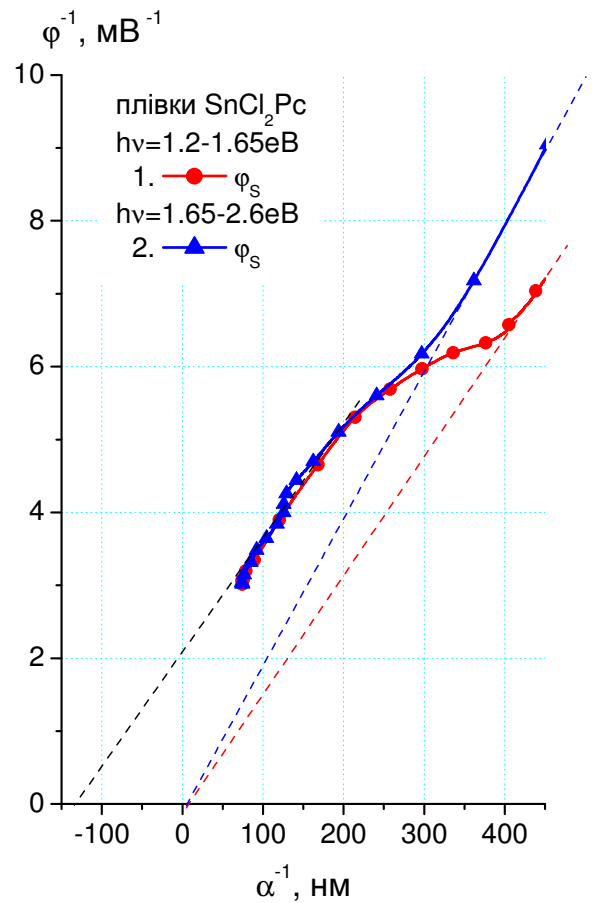

Рис. 3.17 Прямолінійна екстраполяція залежностей $\varphi_C^{-1}(\alpha^{-1})$ в областях 2.7-3.0 еВ (1,3) та 2.0-2.15 еВ (2,4)для плівок MPP нанесених при $T_S = 300$ (1,2) та 370 K (3,4).

Рис. 3.18 Прямолінійна екстраполяція залежностей $\varphi_S^{-1}(\alpha^{-1})$ в областях 1.20-1.65 еВ (1) та 1.65-2.60 еВ (2) для плівок $SnCl_2Pc$ напилених при $T_S = 415$ K.

визначити $L$ при збудженні СТ-станів в досліджуваних плівках SnCl$_2$Pc, так як вони проявляються в СП і ФЕ в області слабкого поглинання ($\alpha L < 1$), де $\varphi^{-1}$ практично не залежить від $L$. В плівках НТР $L = (200 \pm 50)$ нм і практично не змінюється, як в результаті відпалу, так і зі збільшенням $T_S$ від 300 до 370 К (рис. 3.19).

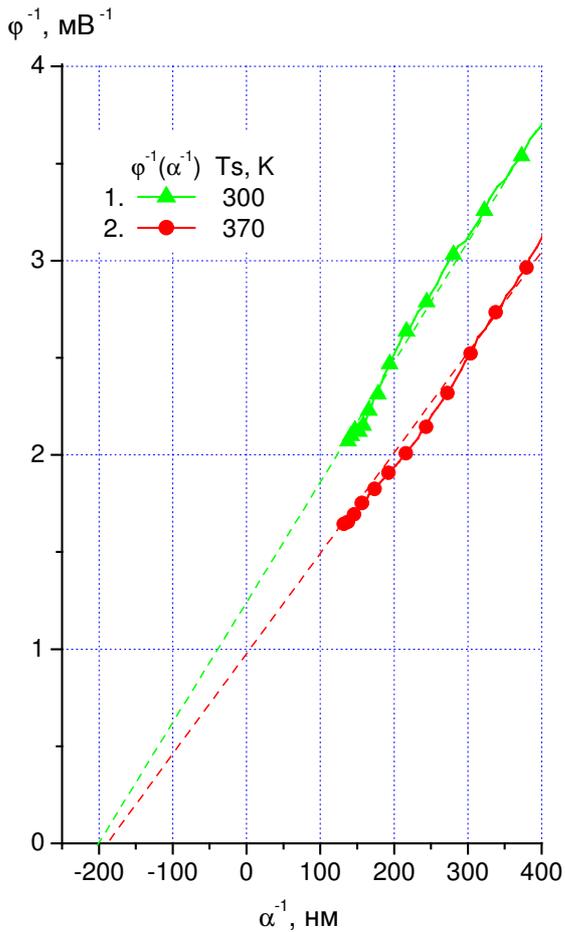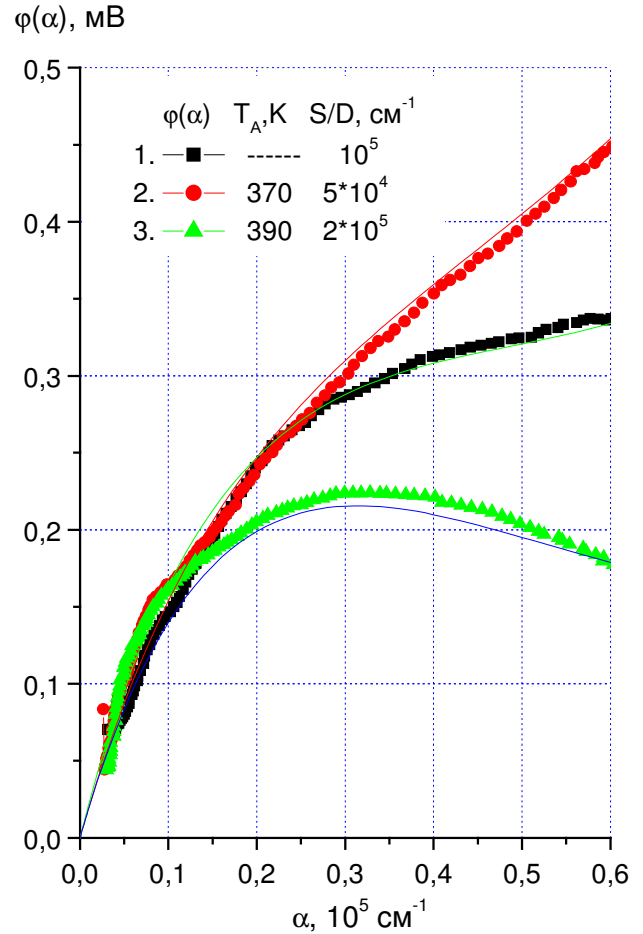

Рис. 3.19  Прямолінійна екстраполяція залежностей $\varphi_S^{-1}(\alpha^{-1})$ в області 1.45-1.75 еВ отриманих для плівок НТР нанесених при $T_S$ = 300 К (1) та 370 К (2).

Рис. 3.20  Залежності $\varphi_S(\alpha)$ в області 1.45-1.75 еВ для плівок НТР нанесених при $T_S$ = 300 К (1) та відпалених при $T_A$ = 370 К (2) та 390 К (3).

Отже, значення $L$ екситонів в плівках SnCl$_2$Pc в 3 рази більше, а в плівках НТР в 5 разів більше, ніж $L$ в плівках ZnPc [154] чи МРР, які широко використовуються для розробки органічних СЕ, що говорить про перспективність застосування досліджуваних нами структур для розробки ефективних фотоперетворювачів сонячного світла.

Розрахунок $\varphi(\alpha)$ за рівнянням (1.6), враховуючи відсутність верхнього шару, при отриманих значеннях $L$, показує хороше узгодження розрахованих залежностей $\varphi(\alpha)$ з експериментальними кривими. Як приклад, експериментальні залежності найкраще описуються розрахунковими даними для $\varphi_S(\alpha)$ для плівок НТР при $S/D \sim 10^5$ см$^{-1}$ до відпалу і $S/D \sim 5 \cdot 10^4$ і $2 \cdot 10^5$ см$^{-1}$ після відпалу при $T_A = 370$ і 390 К (рис. 3.20). Таким чином, проведений розрахунок узгоджується з експериментальними кривими $\varphi(\alpha)$.

Висновки до розділу 3

Отже, дослідження морфології та структури поверхні, оптичних та фотовольтаїчних властивостей термічно напилених полікристалічних плівок МРР, SnCl$_2$Pc та НТР показують, що параметри плівок залежать від $T_S$. Дослідження морфології та структури поверхні плівок показують, що форма та розміри кристалітів в цих плівках залежать від $T_S$. В плівках МРР утворюються два різних типи кристалітів: овальні та стрічкоподібні, при цьому величина кристалітів в плівках МРР збільшується з ростом $T_S$ [147,156]. В плівках SnCl$_2$Pc та НТР зі збільшенням $T_S$ відбувається рекристалізація круглих зерен в складні конгломерати, одночасно зі зменшенням шорсткості плівок.

В результаті дослідження оптичних властивостей визначено енергетичне положення СТ-станів в плівках SnCl$_2$Pc, обумовлених взаємодією центральних атомів Cl однієї молекули з периферійними С та Н атомами фталоціанінового кільця сусідніх молекул. Вклад СТ-станів в СП плівок SnCl$_2$Pc складає приблизно (15-20)%. Показано, що ефективність фотогенерації носіїв заряду при збуджені СТ-станів з енергіями 1.52±0.02 та 2.05±0.02 еВ більше, ніж при збуджені локалізованих екситонів [157-159].

Запропоновано схему енергетичної структури збуджених станів в плівках МРР, що враховує взаємодію між молекулами МРР в межах шару та між шарами [147].

Показано, що незважаючи на невелику зміну $\alpha$ плівок МРР та НТР зі збільшенням $T_S$ від 300 до 370 K, або в результаті відпалу при $T_A = 350$-370K відбувається суттєве зростання ФЕ практично у всій досліджуваній спектральній області. При цьому зростання ФЕ при відпалі плівок значно менше (40-60%), ніж зі збільшенням $T_S$ при тих же температурах. Таким чином, в результаті збільшення $T_S$ від 300 до 370 K ФЕ плівок МРР зростає в 4 рази, плівок НТР – вдвічі, тоді як для плівок SnCl$_2$Pc ФЕ практично не змінюється в досліджуваній області $T_S$. При цьому зростання ФЕ в досліджуваних плівках може бути обумовлено, як збільшенням $L$ екситонів, так і зменшенням $S$ внаслідок покращення структурної впорядкованості плівок при збільшенні $T_S$ до 370 K [147,156,160,161].

Визначено, що $L$ локалізованих екситонів Френкеля в термічно напилених плівках НТР складає $(200 \pm 50)$ нм [160], SnCl$_2$Pc – $(130 \pm 30)$ нм [159] і практично не змінюється з підвищенням $T_S$. В плівках МРР $L$ для локалізованих екситонів складає $(25 \pm 5)$ нм, і збільшується в два рази зі зростанням $T_S$ від 300 до 370 K; в спектральній області, де переважно поглинають CT-стани, значення $L = (10 \pm 5)$ нм для плівок МРР нанесених при 300 K, і $L = (25 \pm 5)$ нм для плівок нанесених при 370 K. Таким чином для плівок МРР $L$ при збудженні CT-станів приблизно в два рази менше, ніж при збудженні екситонів Френкеля [147].

Таким чином, представлені результати дослідження впливу $T_S$ та відпалу на морфологію, оптичні та фотовольтаїчні властивості плівок МРР, SnCl$_2$Pc та НТР можуть бути використані при оптимізації ефективності перетворення світла структурами на їх основі.

# РОЗДІЛ 4

## ДОСЛІДЖЕННЯ ВЛАСТИВОСТЕЙ АНІЗОТИПНИХ ГЕТЕРОСТРУКТУР

### 4.1. Спектральні особливості компонент анізотипних гетероструктур

Необхідною умовою для виготовлення ефективних та фоточутливих анізотипних (p-n) ГС є генерація нерівноважних носіїв заряду в результаті поглинання світла, а також ефективне розділення утворених носіїв заряду внутрішнім електричним полем біля ГР компонент ГС (п. 2.1). В якості компонент анізотипних ГС були вибрані не тільки широко досліджені фоточутливі плівки p-типу провідності Pn та PbPc, а й описані в попередньому розділі шари n-типу $SnCl_2Pc$, MPP та p-типу НТР.

В СП шарів MPP, Pn, $SnCl_2Pc$, НТР та CuI, які використовувались для виготовлення досліджуваних ГС, з підвищенням $T_S$ від 300 до 370 К відбувається невелика зміна інтенсивності та ширини смуг поглинання, але загальний вигляд спектрів зберігається (рис. 4.1, кр. 1-6). Проте підвищення $T_S$ призводить до суттєвої зміни СП плівок PbPc в області 1.2-1.6 еВ (рис. 4.1, криві 6,7), внаслідок зміни вмісту поліморфних модифікацій в плівках PbPc з ростом $T_S$ [85]. Адже як згадувалось в п. 1.4 в плівках PbPc, крім аморфної поліморфної модифікації, формуються моноклінна, триклінна та мало вивчена високотемпературна поліморфні модифікації [81].

З рис. 4.1 видно, що двошарові анізотипні ГС на основі плівок n-типу провідності MPP та p-типу провідності Pn, НТР, PbPc добре поглинають частину сонячного випромінювання, доля якої зростає в ряду: MPP/Pn → MPP/HTP → MPP/PbPc. А оскільки при поглинанні світла у вищезгаданих компонентах утворюються нерівноважні носії заряду та формується ФЕ (рис. 4.2), то зі збільшенням долі поглинутого світла повинен розширюватись і спектральний діапазон фоточутливості ГС. Також з рис. 4.1 видно, що анізотипні ГС $SnCl_2Pc$/Pn також поглинають значну частину

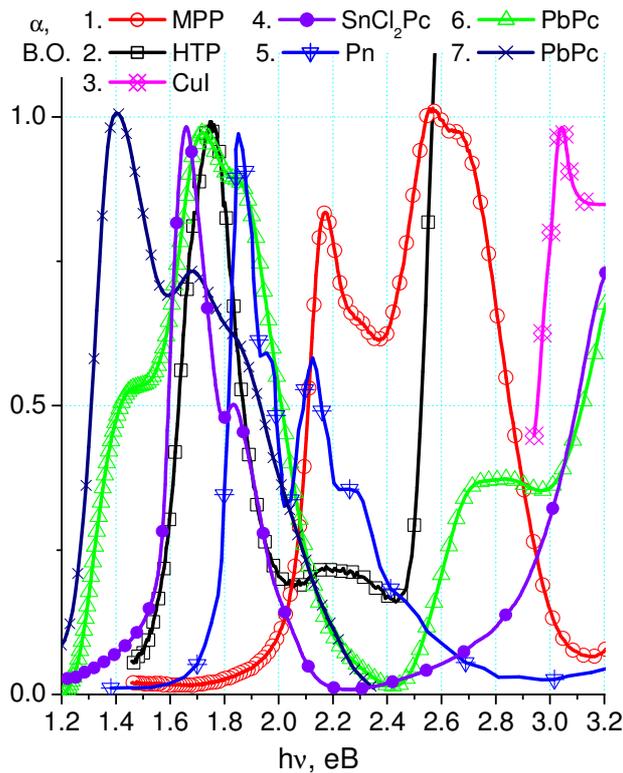 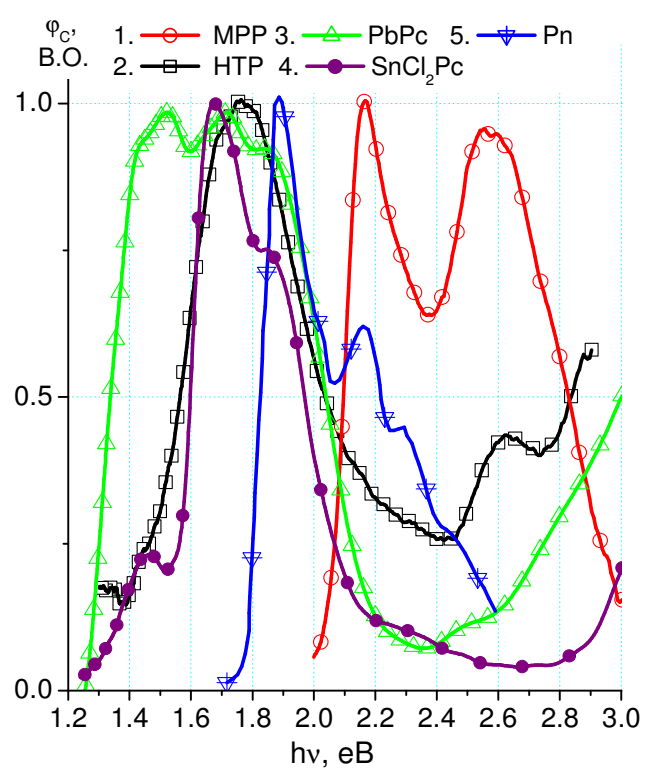

Рис. 4.1 СП тонких плівок
(компонент ГС): MPP (1), HTP (2),
CuI (3), SnCl₂Pc (4), Pn (5), PbPc (6)
напилених при $T_S$ = 300 К, і PbPc (7)
напилених при $T_S$ = 430 К.

Рис. 4.2 Спектри ФЕ при освітленні
через ITO-електрод ($\varphi_C$) тонких плівок
(компонент ГС): MPP (1), HTP (2),
PbPc (3), SnCl₂Pc (4), Pn (5)
виготовлених при $T_S$ = 300 К.

сонячного випромінювання. Тому ці структури можуть бути перспективними для розробки органічних фотоперетворювачів сонячного світла. Анізотипні ГС MPP/CuI поглинають меншу частину сонячного світла, і тому є менш перспективними для розробки СЕ. Проте ГС на основі CuI можуть бути модельною структурою для дослідження фотопроцесів на ГР НП, оскільки положення $E_f$ в плівках CuI практично не залежить від умов виготовлення і рівне 6.0±0.02 eB [109].

4.2. Спектральні властивості анізотипних гетероструктур

4.2.1. Вплив верхнього шару CuI на спектральні залежності фото-ерс шарів метил-заміщеного периленового барвника (МРР)

Дослідження оптичних та фотовольтаїчних властивостей плівок МРР (п. 3.3) показує, що на вільній поверхні шарів МРР формується велика концентрація центрів рекомбінації носіїв заряду, які можуть сильно впливати на фоточутливість ГС, виготовлених на основі шарів МРР. Для визначення ефективності шарів МРР, як компоненти ГС, були проведені дослідження впливу верхнього шару CuI нанесеного на вільну поверхню плівок МРР і утворення анізотипних ГС МРР/CuI. Оскільки плівки МРР поглинають в області прозорості шарів CuI (2-3 еВ), то при виготовленні ГС МРР/CuI повинен чітко проявлятися вплив верхнього шару (CuI) на спектральні залежності ФЕ плівок МРР.

Порівняння спектрів ФЕ структур ITO/МРР до і після нанесення шару CuI при освітленні вільної поверхні МРР чи шару CuI (рис. 4.3), показує, що нанесення на поверхню МРР шару CuI, тобто утворення ГС МРР/CuI, призводить до суттєвого збільшення ФЕ (від 6 до 10 разів при $T_S = 300$ К та від 10 до 15 разів при $T_S = 370$ К). Збільшення ФЕ може бути обумовлено як збільшенням згину зон біля ГР МРР/CuI, так і зменшенням $S$, при нанесенні шару CuI на вільну поверхню шарів МРР. Різниця в зростанні ФЕ для структур отриманих при $T_S = 300$ К та 370 К пов'язана з покращенням структурної впорядкованості не тільки шару МРР, а і шарів CuI зі збільшенням $T_S$ від 300 до 370 К.

Після нанесення шару CuI на вільну поверхню МРР і утворення анізотипної ГС МРР/CuI залежність $\varphi(\alpha)$ при освітленні сторони CuI в області 2.05-2.15 еВ стає практично лінійною. При цьому збільшується кут нахилу залежності $\varphi(\alpha)$ (рис. 4.4). Отже, нанесення на поверхню МРР тонкої плівки CuI призводить до суттєвого збільшення згину зон та зменшення $S$

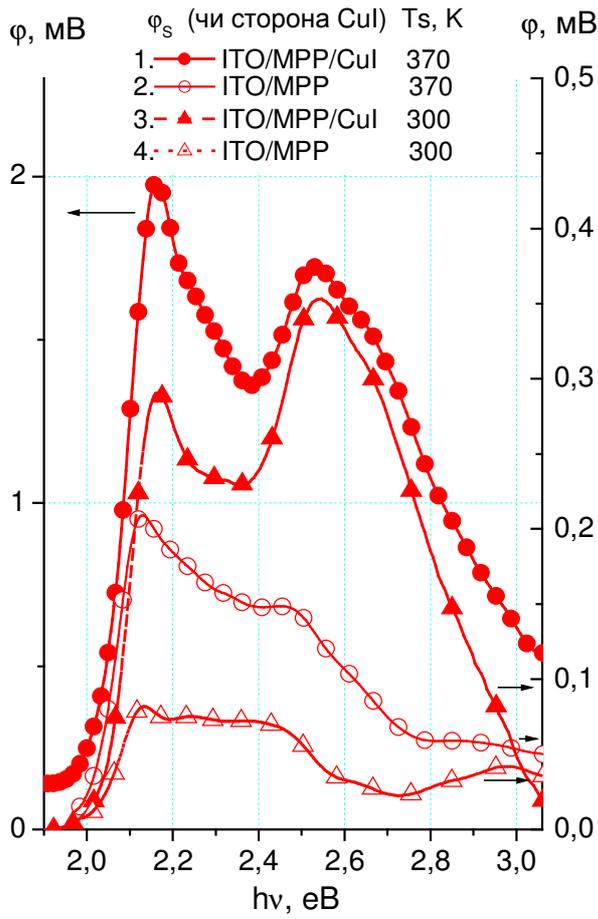 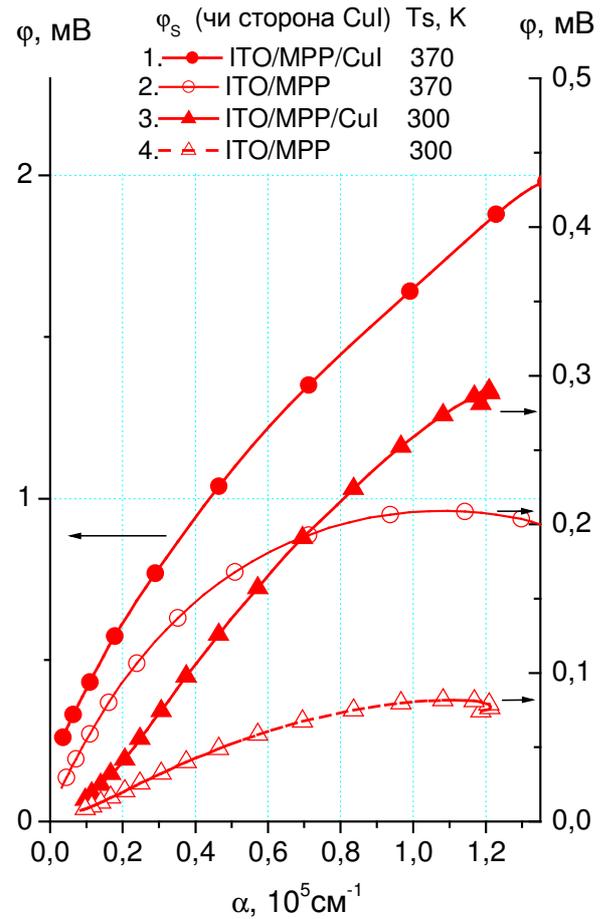

Рис. 4.3 Спектральні залежності ФЕ ($\varphi$) ГС ITO/MPP/CuI (1,3) та ITO/MPP (2,4) отриманих при $T_S = 370$ К (1,2) і $T_S = 300$ К (3,4) при освітленні сторони CuI (1,3) і вільної поверхні MPP (2,4).

Рис. 4.4 Залежності ФЕ ($\varphi$) від коефіцієнта поглинання ($\alpha$) для ГС ITO/MPP/CuI (1,3) і ITO/MPP (2,4) отриманих при $T_S = 370$ (1,2) та 300 К (3,4) при освітленні зі сторони CuI (1,3) і вільної поверхні MPP (2,4).

біля ГР. Це пояснюється сильним зменшенням концентрації поверхневих центрів захоплення та рекомбінації носіїв заряду на вільній поверхні шару MPP при відкачці (перед термічним напиленням шару CuI у вакуумі) адсорбованих молекул повітря на поверхні MPP. В результаті цього утворюється запірний згин зон на ГР шарів МРР (n-тип) і CuI (p-тип) з незначною *S*.

### 4.2.2. Фотовольтаїчні властивості гетероструктур метил-заміщений периленовий барвник / пентацен (MPP/Pn)

Спектральна залежність ФЕ анізотипних ГС MPP/Pn наведена на рис. 4.5 (кр.1). Нанесення шару Pn на поверхню МРР при $T_S$ = 370 К також призводить до значного збільшення ФЕ в порівнянні з $\varphi_S$ шарів МРР.

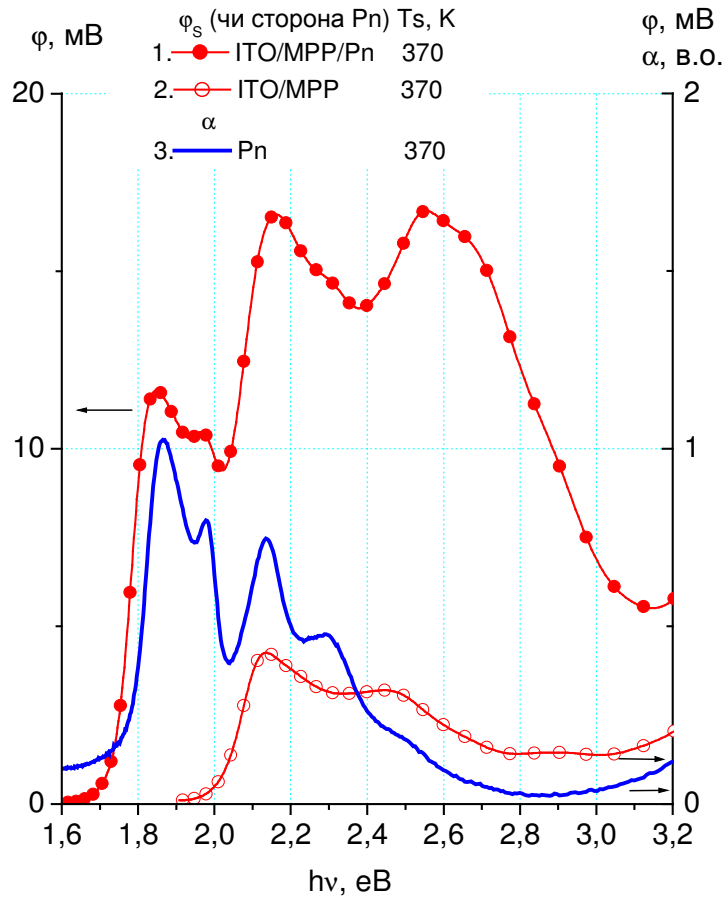

Рис. 4.5 Спектральні залежності ФЕ ($\varphi$) ГС ITO/Pn/MPP при освітленні зі сторони МРР (1), структури ITO/MPP при освітленні вільної поверхні (2) та СП плівки Pn (3). Всі структури отримані при $T_S$ = 370 К.

Зростання ФЕ становить від 20 до 90 разів в залежності від hv та напрямку освітлення (сторона вільної поверхні чи ITO контакт) ГС MPP/Pn. Суттєве збільшення ФЕ при формуванні ГС MPP/Pn вказує на утворення значного згину зон на ГР шарів MPP (n-тип) та Pn (p-тип), який суттєво збільшує ефективність розділення фотогенерованих екситонів або нерівноважних носіїв заряду в обох шарах ГС MPP/Pn.

На спектральних залежностях ФЕ ГС MPP/Pn чітко проявляються смуги поглинання шарів Pn при hv = 1.85 та 1.98 еВ та смуги поглинання шарів MPP при 2.15 та 2.55 еВ (рис. 4.5, кр.1). Тобто спектральні залежності ФЕ ГС MPP/Pn чітко корелюють зі СП компонент як в області слабкого, так і в області сильного поглинання. При тому, що на спектральній залежності ФЕ для шарів MPP при освітленні вільної поверхні плівок спостерігається антикореляція ФЕ та $\alpha$. Таким чином, відкачка адсорбованих молекул повітря з поверхні MPP у вакуумі, термічне нанесення шару Pn та утворення ГС призводить до сильного зменшення концентрації поверхневих центрів захоплення та рекомбінації носіїв заряду на вільній поверхні шару MPP. Це свідчить про формування біля ГР ГС MPP/Pn запірного бар'єра з малою концентрацією центрів захоплення та поверхневої рекомбінації носіїв заряду, що говорить про перспективність ГС MPP/Pn для створення ефективних органічних СЕ.

### 4.2.3. Особливості фотовольтаїчних властивостей гетероструктур $SnCl_2$ фталоціанін / пентацен ($SnCl_2Pc/Pn$)

ФЕ ГС $SnCl_2Pc/Pn$ при освітленні вільної поверхні (шару Pn) $\varphi$(hv) на порядок більше, ніж $\varphi_S$(hv) структур ITO/$SnCl_2Pc$ (рис. 4.6), як в областях переважного поглинання шарів $SnCl_2Pc$ (1.2-1.7 еВ), так і шарів Pn (1.8-2.6 еВ). Збільшення ФЕ структур ITO/$SnCl_2Pc$ після нанесення шару Pn

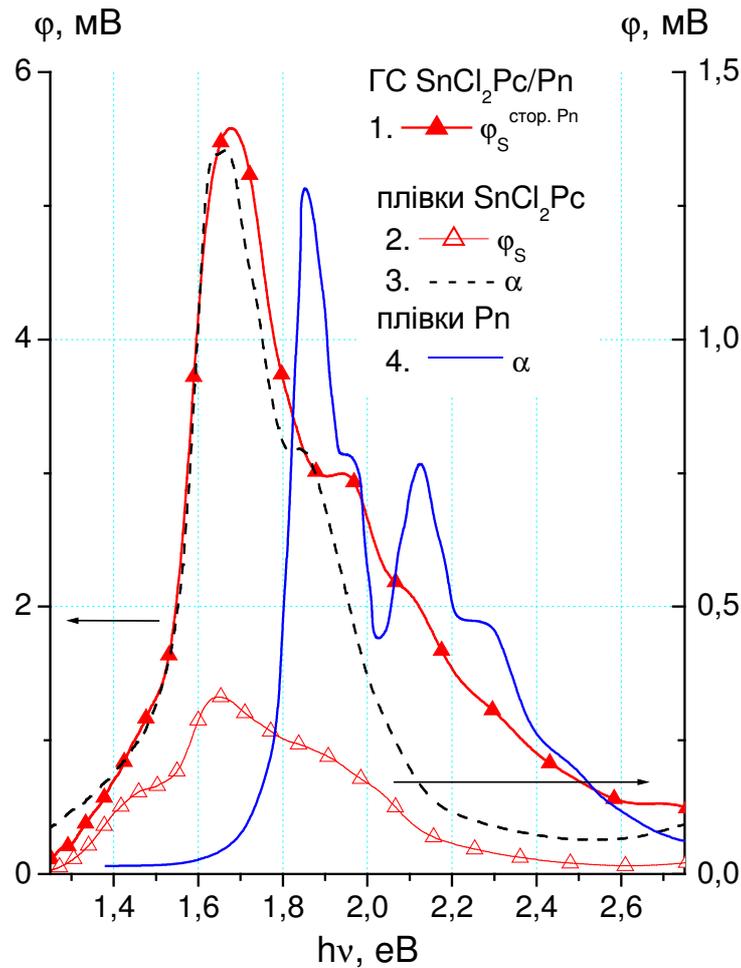

Рис. 4.6  Спектральні залежності ФЕ ($\varphi_S$) при освітленні
вільної поверхні для ГС SnCl$_2$Pc/Pn (1), структури
ITO/SnCl$_2$Pc (2) та коефіцієнта поглинання нормованого
на максимум ($\alpha/\alpha_{max}$) для плівок SnCl$_2$Pc (3) та Pn (4).

свідчить про утворення на ГР SnCl$_2$Pc/Pn суттєвого запірного бар'єру, який
збільшує ефективність розділення фотогенерованих носіїв заряду. Також
спектр ФЕ ГС SnCl$_2$Pc/Pn чітко корелює зі СП компонент: SnCl$_2$Pc та Pn,
оскільки на спектральних залежностях ФЕ ГС SnCl$_2$Pc/Pn проявляються піки
поглинання як шарів SnCl$_2$Pc (hv = 1.65 eB), так і шарів Pn (hv = 1.95 eB).

В області 1.20-1.65 eB залежність $\varphi(\alpha)$ структури ITO/SnCl$_2$Pc при
освітленні вільної поверхні прямує до насичення при $\alpha > 4 \cdot 10^4$ см$^{-1}$ (рис. 4.7),

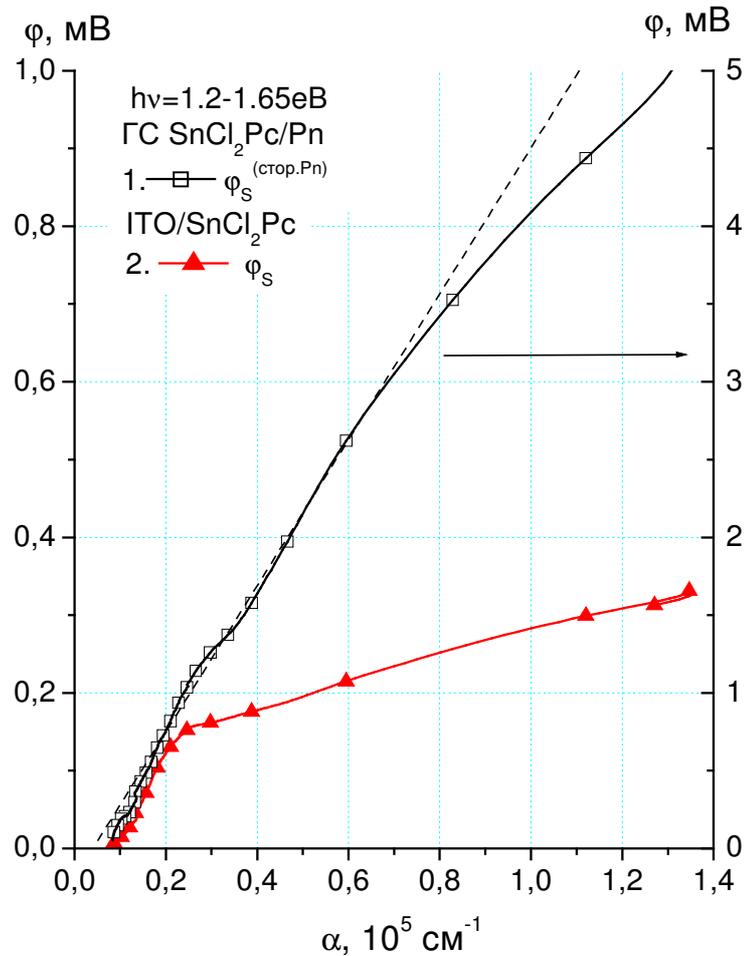

Рис. 4.7  Залежності ФЕ ($\varphi$) від коефіцієнта поглинання ($\alpha$)

для ГС SnCl$_2$Pc/Pn (1) та тонких плівок SnCl$_2$Pc (2) при

освітленні вільної поверхні структур.

а для ГС SnCl$_2$Pc/Pn спостерігається практично лінійна залежність $\varphi(\alpha)$ до

значень $\alpha \sim 10^5$ см$^{-1}$. Оскільки залежність $\varphi(\alpha)$ ГС SnCl$_2$Pc/Pn є практично

лінійною (рис. 4.7), то $S$ біля ГР SnCl$_2$Pc/Pn незначна. При цьому на вільній

поверхні плівок SnCl$_2$Pc утворюється невелика концентрація центрів

рекомбінації носіїв заряду (п. 3.3). Це свідчить про зменшення концентрації

центрів рекомбінації носіїв заряду та, відповідно, значення параметру $S$ біля

ГР SnCl$_2$Pc/Pn при відкачці повітря перед нанесенням шару Pn на вільну поверхню плівки SnCl$_2$Pc. Тому ГС SnCl$_2$Pc/Pn також є перспективними для розробки органічних СЕ.

4.2.4. Особливості фото-ерс в гетероструктурах гексатіопентацен /
метил-заміщений перилеиовий барвник (HTP/MPP)

Нанесення шару MPP на поверхню HTP і утворення ГС HTP/MPP також
призводить до значного збільшення ФЕ в порівнянні з $\varphi_S$ шарів HTP. Так, в
спектральній області 1.55-1.95 еВ, де основний вклад в ФЕ дають носії заряду
фотогенеровані в шарі HTP, ФЕ ГС HTP/MPP зростає приблизно в три рази в
порівнянні з ФЕ плівок HTP (рис. 4.8). Це свідчить про зростання величини
згину зон або потенціального бар'єру біля ГР при утворенні ГС HTP/MPP.

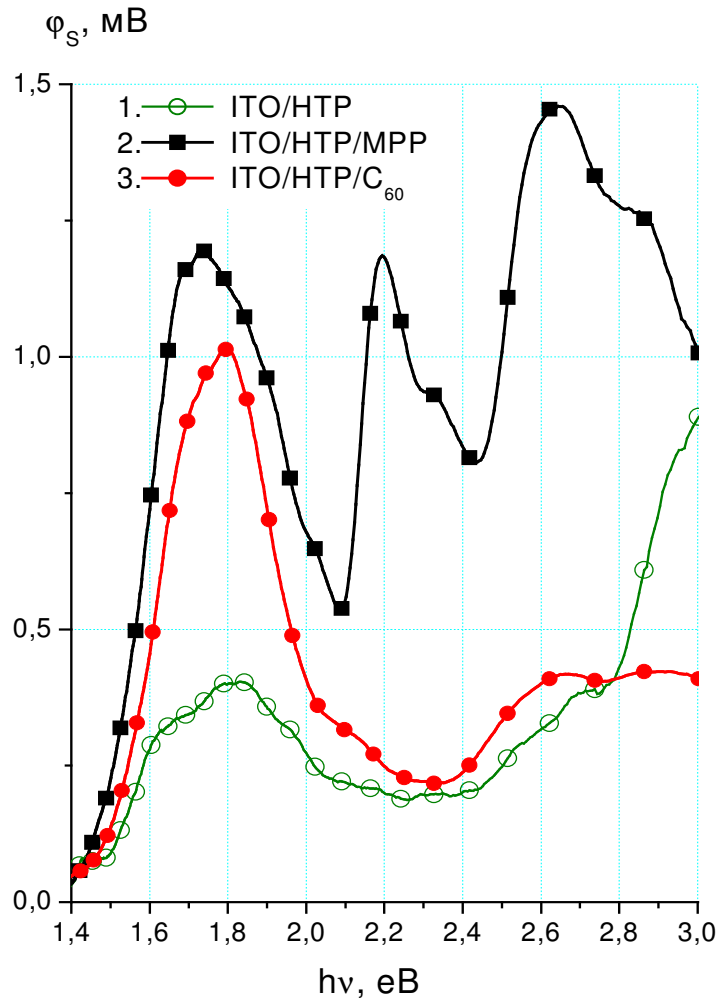

Рис. 4.8  Спектри ФЕ при освітленні зі сторони вільної поверхні
($\varphi_S$) структур ITO/HTP (1), ГС HTP/MPP (2), HTP/$C_{60}$ (3).

Також з рис. 4.8 також видно, що як фоточутливість в області прозорості НТР, так і інтегральна фоточутливість {площа під кривими $\varphi(h\nu)$} для досліджуваних анізотипних ГС НТР/МРР більше, ніж для структур ІТО/НТР та анізотипних ГС НТР/C$_{60}$.

ФЕ ГС НТР/МРР також чітко корелює зі СП компонент як в області сильного, так і в області слабкого поглинання: для шарів НТР в області 1.4-2.1 еВ, а для шарів МРР в області 2.1-3.0 еВ (рис. 4.9). Розрахункова залежність $\varphi(\alpha)$ для ГС НТР/МРР в області 1.45-1.75 еВ (сильного

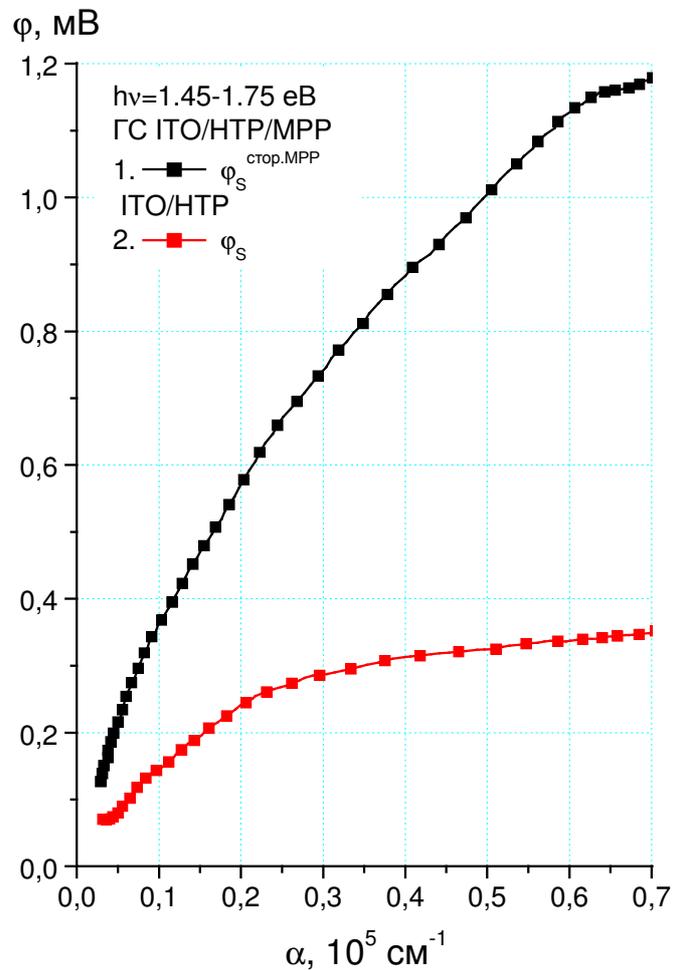

Рис. 4.9  Залежності ФЕ ($\varphi$) від коефіцієнта поглинання ($\alpha$)

для ГС НТР/МРР (1) та тонких плівок НТР (2).

поглинання плівок НТР) за рівнянням (1.6) узгоджується з експериментальними кривими при $S/D \leq 10^4$ см$^{-1}$. Порівняння значень $S/D$ для ГС НТР/МРР ($S/D \leq 10^4$ см$^{-1}$) та плівок НТР (до відпалу $S/D \sim 10^5$ см$^{-1}$) [160] свідчить про зменшення $S$ в ГС більше, ніж на порядок (при умові, що коефіцієнт дифузії ($D$) для шару НТР не змінюється при утворенні ГС). Тобто ефективність рекомбінації біля ГР компонент ГС НТР/МРР значно менше, ніж біля вільної поверхні плівок НТР. Отже, найбільш перспективною ГС на основі НТР є ГС НТР/МРР, яка є фоточутливою в області 1.55-3.1 еВ та має невелику концентрацію центрів рекомбінації носіїв заряду на ГР компонент.

### 4.2.5. Фотовольтаїчні властивості гетероструктур фталоціанін свинцю / метил-заміщений периленовий барвник (PbPc/MPP)

Спектральні залежності ФЕ ГС PbPc/MPP наведені на рис. 4.10 (кр. 1,2). При нанесенні шару МРР на поверхню PbPc і утворенні ГС PbPc/MPP також відбувається значне збільшення ФЕ (від 10 до 30 разів) в порівнянні з $\varphi_S$ як шарів МРР, так і PbPc. Це свідчить про формування значного запірного потенціального бар'єру на ГР PbPc/MPP, який сприяє збільшенню фото-генерації носіїв заряду, а, отже, і фоточутливості структур PbPc/MPP в порівнянні з одношаровими структурами ITO/MPP та ITO/PbPc.

На спектральній залежності ФЕ ГС PbPc/MPP (рис. 4.10, кр. 1) чітко проявляються максимуми при hv = 1.4 та 1.75 еВ, які відповідають смугам поглинання плівок PbPc, та максимум при 2.16 еВ, що узгоджується з піком поглинання шарів МРР. Тобто ФЕ ГС PbPc/MPP також корелює зі СП компонент в області сильного поглинання, як це спостерігалося в інших досліджуваних ГС на основі МРР. Це підтверджує, що $S$ біля ГР PbPc/MPP невелика і збільшення ФЕ відбувається в основному за рахунок збільшення висоти потенціального бар'єру.

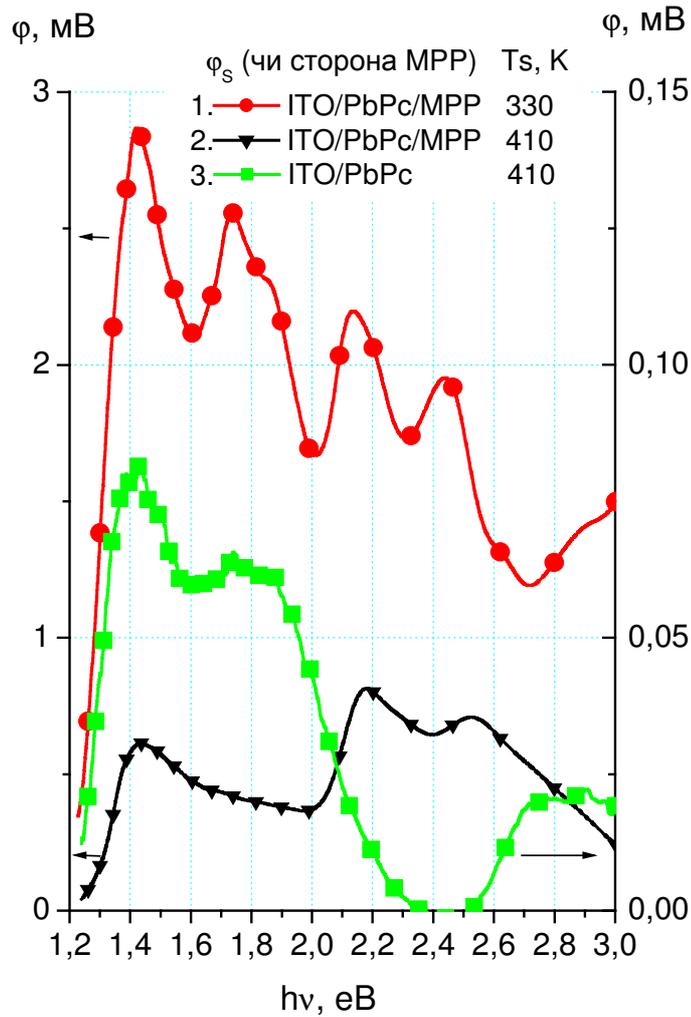

Рис. 4.10  Спектральні залежності ФЕ ($\varphi$) ГС ITO/PbPc/MPP,
отриманих при $T_S = 330$ К, при освітленні зі сторони MPP (1) і
отриманих при $T_S = 410$ К, при освітленні зі сторони PbPc (2),
і шару PbPc при освітленні вільної поверхні (3).

Спектральні залежності ФЕ ГС PbPc/MPP сильно залежать від $T_S$ (рис. 4.10, кр. 1,2). Крім того, структура поверхні нижнього шару PbPc впливає на структуру верхнього шару MPP при утворенні ГС PbPc/MPP. За літературними даними [91] при виготовленні фотоперетворювачів на основі PbPc найбільш оптимальні $T_S \approx 430$ та 330 К (п. 1.4.2.). Але для більш детального аналізу впливу поліморфних модифікацій плівок PbPc на ФЕ ГС створених на основі шарів PbPc, необхідні додаткові (більш широкі) дослідження ГС отриманих при різних $T_S$.

Слід відмітити, що більш детальний аналіз фотовольтаїчних властивостей ГС PbPc/MPP є актуальною задачею, оскільки двошарова структура PbPc/MPP ефективно поглинає сонячне світло в широкій спектральній області сильного сонячного випромінювання (рис. 4.11). При

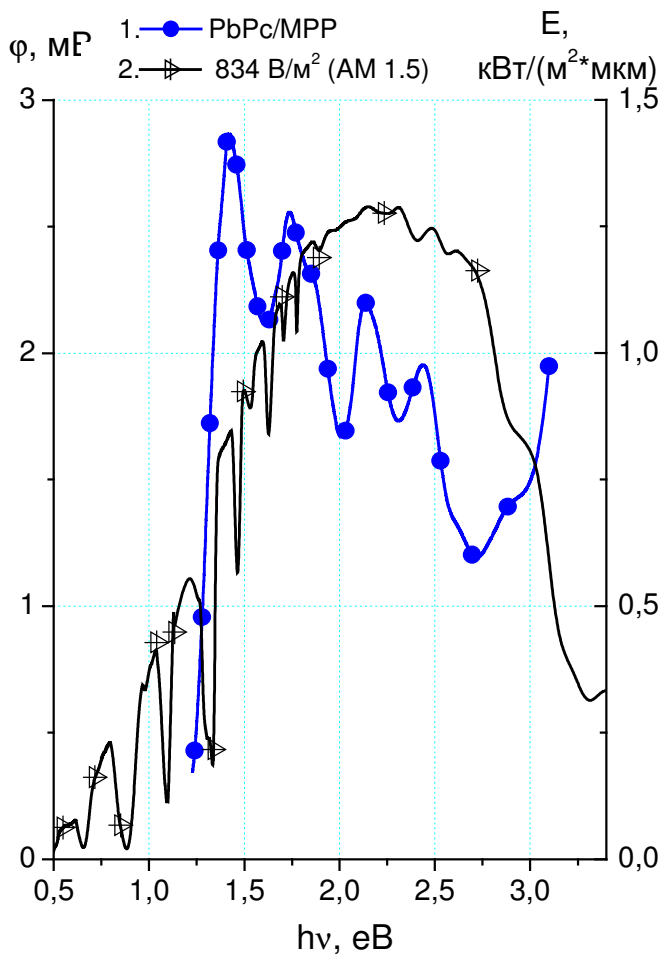

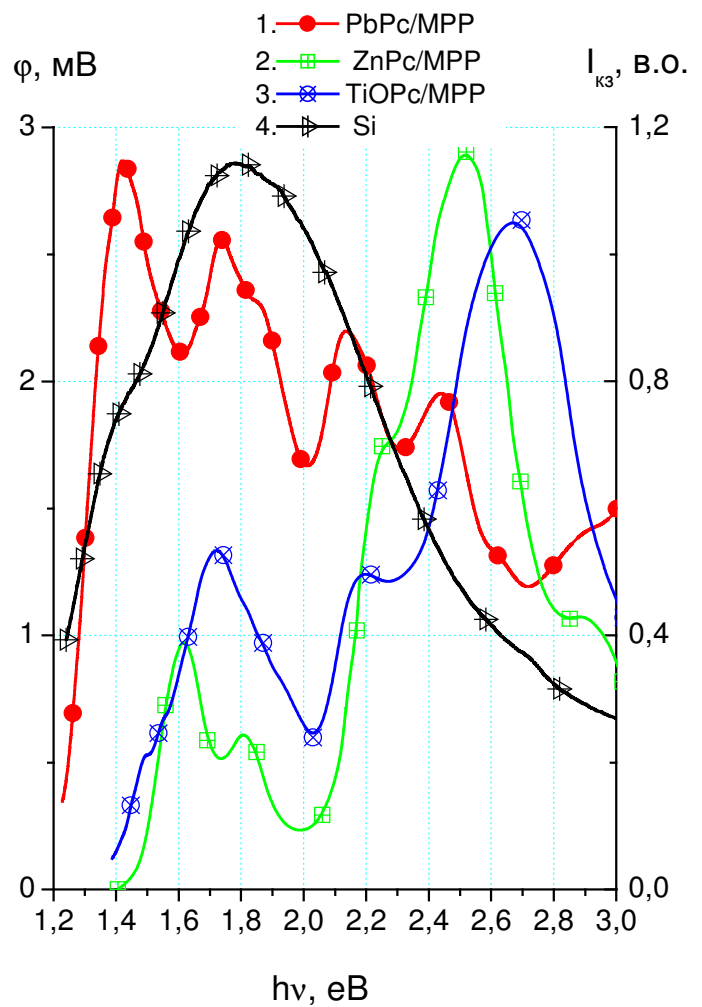

Рис 4.11 Спектр ФЕ ($\varphi$) ГС PbPc/MPP (1) та спектральні розподіли сонячного випромінювання ($E$) при інтенсивності освітлення 834 В/м² – АМ 1.5 (2) [162].

Рис 4.12 Спектри ФЕ ($\varphi$) (1) та струму короткого замикання ($I_{кз}$) (2-4) ГС PbPc/MPP (1), ZnPc/MPP (2) [43], TiOPc/MPP(3) [41] та кремнієвого фотоелемента серії ФД-288А (4).

цьому спектральний діапазон фоточутливості ГС PbPc/MPP (1.2-3.2 eB) є більшим, ніж інших анізотипних ГС досліджуваних в рамках цієї роботи: MPP/CuI (2.1-3.0 eB), Pn/MPP (1.8-3.0 eB), SnCl$_2$Pc/Pn (1.5-2.4 eB) та HTP/MPP (1.55-3.1 eB).

При порівнянні спектральних залежностей фоточутливості досліджуваних ГС PbPc/MPP з літературними даними фоточутливості ГС на основі MPP та інших Pc (рис. 4.12) бачимо, що спектральний діапазон фоточутливості ГС PbPc/MPP на 0.3 eB ширший (в сторону менших енергій), ніж спектр фоточутливості ГС TiOPc/MPP [41] та ZnPc/MPP [43]. При цьому в області 1.5-2.2 eB відносна ФЕ ГС PbPc/MPP в 2-3 рази більше, ніж відносні значення $I_{кз}$ ГС TiOPc/MPP та ZnPc/MPP (рис. 4.12).

Отже, спектральна область фоточутливості ГС PbPc/MPP більше, ніж раніше досліджених органічних ГС на основі MPP (TiOPc/MPP [41] та ZnPc/MPP [43]), і є близькою до спектральної фоточутливості кремнієвих фотоелементів (рис. 4.12). Таким чином порівнюючи отримані результати показано, що найбільш ефективні СЕ можуть бути виготовлені на основі ГС MPP/Pn, для яких спостерігалась найбільша ФЕ. Проте, при подальшій оптимізації технологічного режиму (вмісту різних поліморфних модифікацій в плівках PbPc), також перспективними можуть бути ГС MPP/PbPc, які ефективно поглинають сонячне світло в більш широкій спектральній області, ніж інші досліджувані ГС.

4.3. Порівняння отриманих результатів з енергетичною схемою гетеропереходу за моделлю Андерсона.

Важливою умовою отримання ефективних СЕ є необхідність створення на ГР двох компонент ГС значного ($\geq 0.5$ eB) запірного бар'єру (згину зон) з невеликою кількістю поверхневих станів (центрів захоплення та рекомбінації носіїв заряду). Спектральні залежності ФЕ досліджуваних анізотипних ГС PbPc/MPP, MPP/CuI, HTP/MPP, MPP/Pn і SnCl$_2$Pc/Pn (рис. 4.13) показують, що в результаті поглинання сонячного світла та утворення нерівноважних носіїв заряду, відбувається ефективне розділення носіїв заряду внутрішнім електричним полем поблизу ГР ГС, що також є необхідною умовою створення СЕ. Виникнення внутрішнього електричного поля біля ГР компонент гетеропереходу повязано з утворенням згину зон при контакті двох НП з різними значеннями $E_f$.

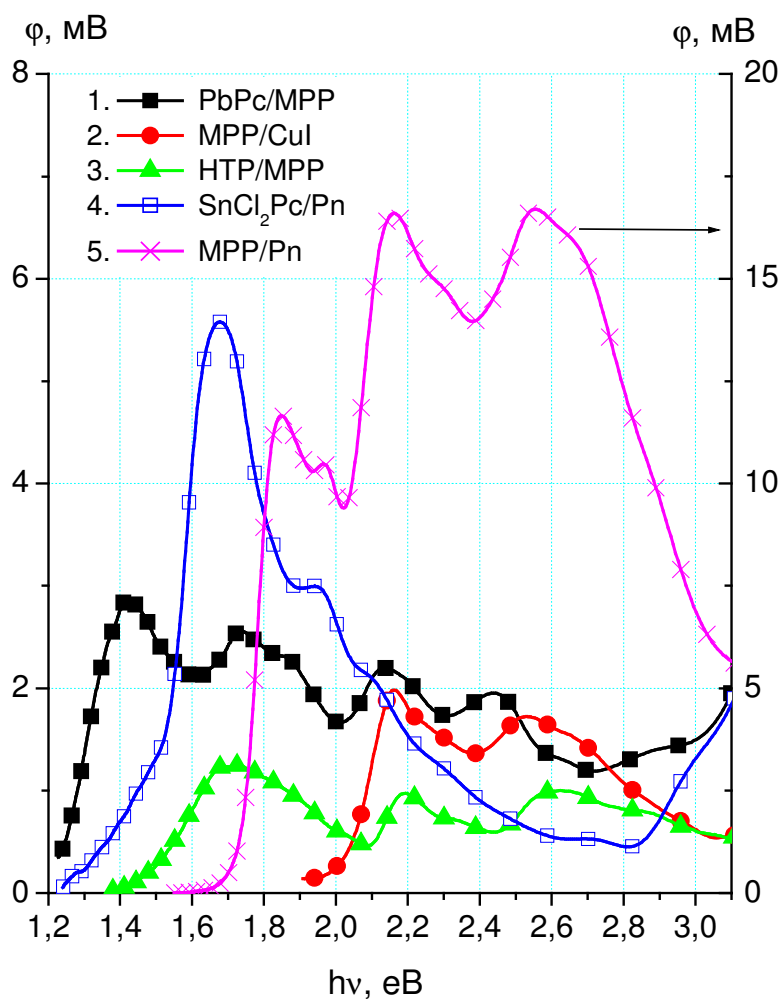

Рис. 4.13  Спектральні залежності ФЕ ($\varphi$) ГС PbPc/MPP (1), MPP/CuI (2), HTP/MPP (3), SnCl$_2$Pc/Pn (4) і MPP/Pn (5).

Для оцінки величини запірного згину зон біля ГР компонент анізотипних ГС, було використано енергетичну діаграму гетеропереходів запропоновану Андерсоном для двошарових ГС на основі неорганічних НП (п. 1.2) [3].

Модель Андерсона не враховує наявність поверхневих станів, тому застосування цієї моделі обмежено випадком невеликої концентрації поверхневих станів на ГР компонент ГС. Наближення Андерсона повинно виконуватись і для органічних НП, так як на їх поверхні відсутні обірвані ковалентні зв'язки, які в неорганічних НП є центрами захоплення та рекомбінації носіїв заряду. При цьому згідно з отриманими результатами (п. 4.2) біля ГР досліджуваних ГС формується невелика концентрація центрів захоплення та рекомбінації носіїв заряду. Тому модель Андерсона може бути використана для побудови енергетичних діаграм досліджуваних нами органічних ГС.

Для побудови енергетичних діаграм за моделлю Андерсона використовувались експериментальні значення енергетичних параметрів $I_C$, $E_g$, $\chi$ та $E_f$ досліджуваних нами матеріалів (див. табл. 2). Приклад енергетичних діаграм для ГС на основі шарів МРР наведено на рис. 4.14.

На основі параметрів енергетичних діаграм було визначено інтервал, в якому можуть знаходитись значення потенціального бар'єру ($V_D$) на ГР анізотипних ГС (див. табл. 4). $V_D$ рівний різниці $E_f$ шарів p- та n-типу провідності: $V_D = E_{f(\text{p})} - E_{f(\text{n})}$. Розкид значень $V_D$ в основному обумовлений як різними експериментальними даними для $E_f$ в різних роботах, так і можливими похибками експерименту. При цьому слід враховувати, що $E_f$

Таблиця 4.

Значення параметрів енергетичних діаграм для ГС на основі шарів МРР

| p/n | p-Pn/n-MPP | p-PbPc/ n-MPP | p-HTP/n-MPP | p-CuI/n-MPP |
|---|---|---|---|---|
| $\Delta Ec$ | 1.0 | 0.57 | 1.12±0.05 | 0.77 |
| $\Delta E_V$ | 1.4 | 1.57 | 1.72±0.05 | 0.37 |
| $V_D$ | (-0.10)-0.45 | (-0.2)-0.25 | (-0.1)-0.15 | 1.6-1.64 |
| $V_D$ | 0.175±0.275 | 0.025±0.225 | 0.025±0.125 | 1.62±0.02 |
| $\varphi_i$, a.u. | 17.8 | 3.6 | 1.3 | 1.5 |

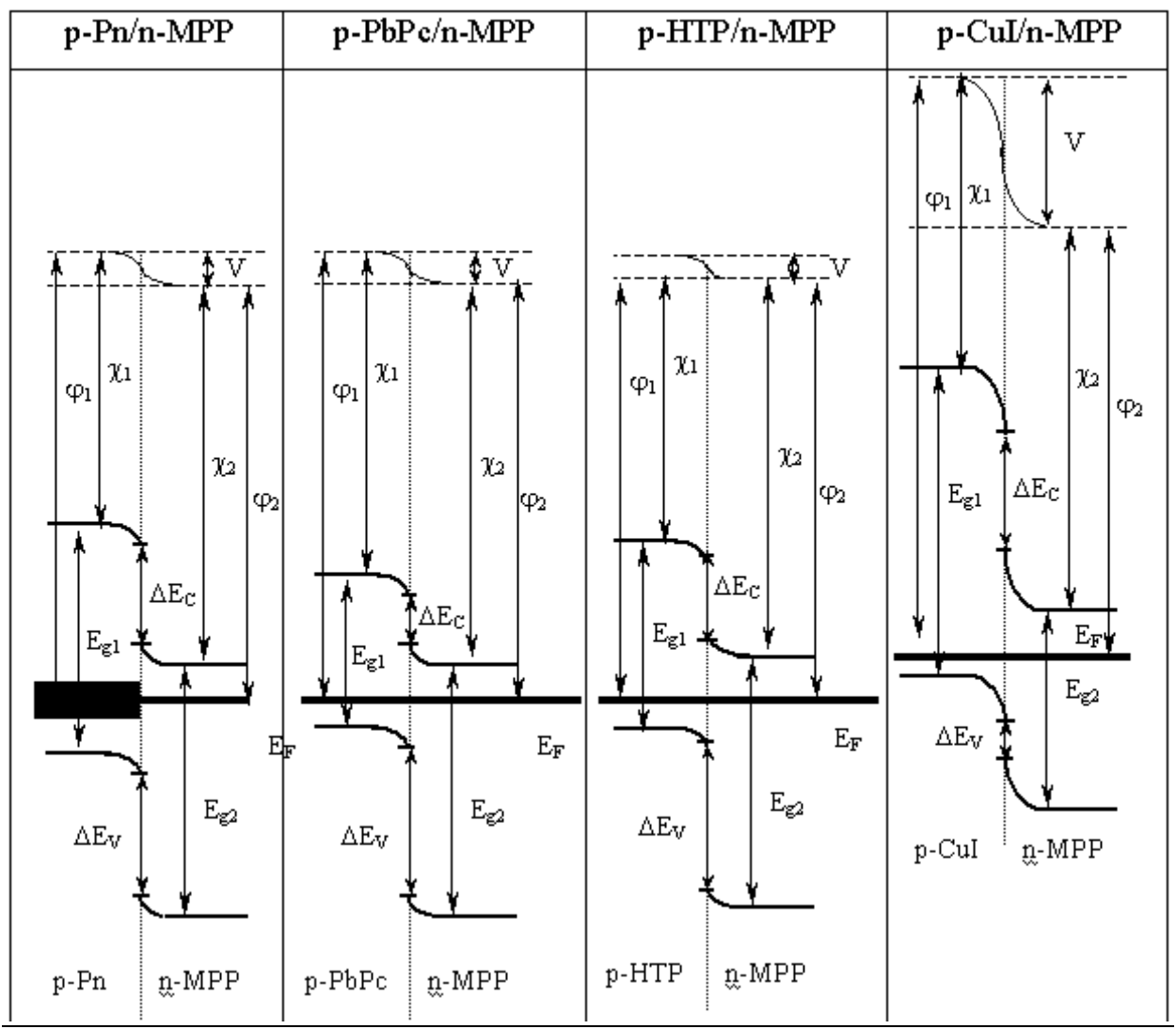

Рис 4.14  Діаграми енергетичних зон для p-n гетеропереходу в умовах рівноваги (індекси 1 і 2 відносяться до НП p - і n -типу, відповідно). Наведено випадок максимально досяжних бар'єрів.

може зсуватися в залежності від умов виготовлення та освітлення структур. Також визначені значення розривів зон провідності ($\Delta E_C = \chi_{(n)} - \chi_{(p)}$) та валентних зон ($\Delta E_V = I_{C(n)} - I_{C(p)}$) для ГС на основі МРР (див. табл. 4).

Для порівняння параметрів моделі Андерсона з експериментальними даними отриманими в цій роботі, визначено інтегральні значення ФЕ ($\varphi_i$) шляхом інтегрування спектральних залежностей ФЕ ГС на основі МРР в області 1.2-3.2 еВ (див. табл. 4).

Для перевірки кореляції між експериментальними значеннями ФЕ та $V_D$ на рис. 4.15 приведені залежності $V_D$ від $\varphi_i$ та максимального значення ФЕ в спектральній області 1.2-3.1 еВ ($\varphi_m$) – області сильного сонячного випромінювання. Як бачимо з табл. 4 та рис. 4.15 експериментально отримані значення $\varphi_i$ для ГС МРР/Pn, PbPc/МРР і НТР/МРР та значення $V_D$, визначенні за моделлю Андерсона, добре корелюють між собою.

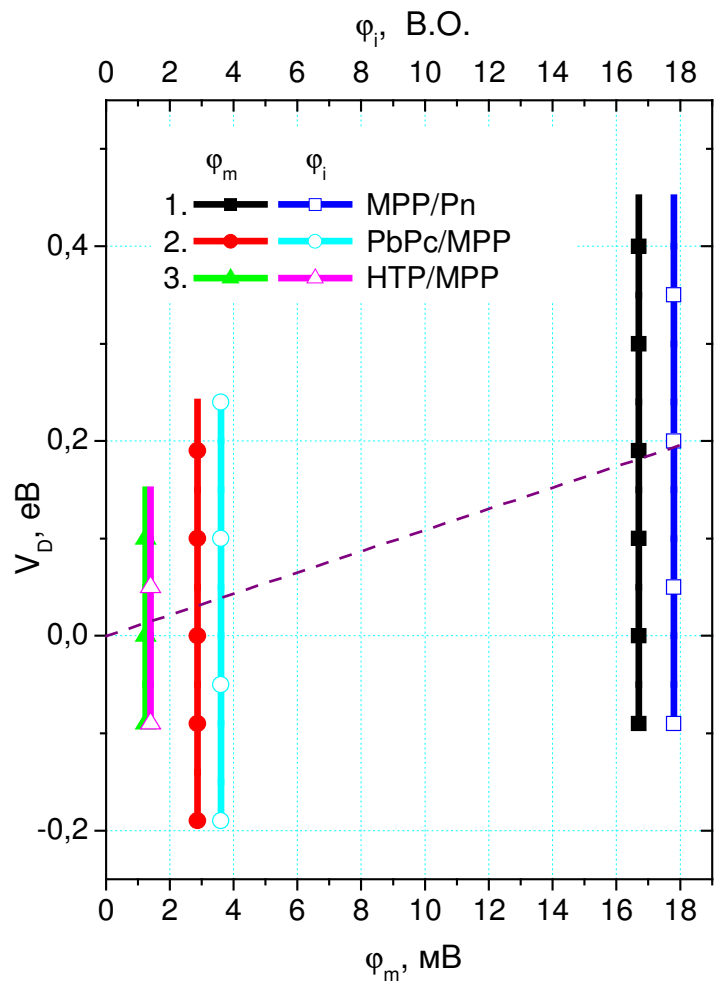

Рис. 4.15  Залежність висоти потенціального бар'єру ($V_D$) від максимального ($\varphi_m$) та інтегрального ($\varphi_i$) значення ФЕ.

Крім того, спостерігається лінійна залежність між середніми значеннями $V_D$ і $\varphi_m$ (або $\varphi_i$) в спектральній області сильного сонячного випромінювання для досліджуваних органічних ГС MPP/Pn, PbPc/MPP і HTP/MPP. (рис. 4.15). Проте отримані значення $\varphi_m$ та $\varphi_i$ для ГС MPP/CuI значно менші від передбачуваних моделлю Андерсона значень $V_D$ (рис. 4.14). Відхилення отриманих результатів від параметрів енергетичної схеми в ГС з CuI можна пояснити дифузією йоду в органічну плівку.

Отже, модель Андерсона для енергетичних діаграм анізотипних ГС, яка не враховує поверхневі стани, може бути використана для оцінки параметрів та ефективності двошарових органічних анізотипних ГС.

Висновки до розділу 4

Отже, спектральні залежності ФЕ досліджуваних анізотипних ГС показують, що поблизу ГР ГС відбувається ефективне розділення фотогенерованих носіїв заряду внутрішнім електричним полем. Так, наприклад, утворення ГС SnCl$_2$Pc/Pn [163] чи PbPc/MPP [164] призводить до збільшення ФЕ ГС більше, ніж на порядок, а для ГС MPP/Pn [164] майже на два порядки в порівнянні з ФЕ окремих компонент. Причому зростання ФЕ спостерігається в області сильного поглинання обох компонент ГС. Збільшення ФЕ при створенні ГС свідчить про формування на ГР компонент суттєвого запірного бар'єру (згину зон) та відповідного внутрішнього електричного поля. В результаті чого збільшується ефективність розділення носіїв заряду біля ГР компонент гетеропереходу [163,164].

Спектральні залежності ФЕ досліджуваних ГС чітко корелюють зі СП компонент навіть в області сильного поглинання, що свідчить про формування невеликої концентрації центрів рекомбінації екситонів та вільних носіїв заряду біля ГР цих ГС. Підтвердженням цього є порівняння залежностей ФЕ від $\alpha$ для ГС та компонент, що також говорить про суттєве зменшення $S$ на вільній поверхні плівок $SnCl_2Pc$ і MPP при відкачці повітря, термічного нанесення верхнього шару та утворення ГС [163,164].

Результати дослідження фотовольтаїчних властивостей двошарових анізотипних органічних ГС якісно узгоджуються з параметрами енергетичної схеми, побудованої за моделлю Андерсона, яка не враховує поверхневі стани [164,165].

Найбільша ФЕ серед досліджуваних структур спостерігалась для ГС MPP/Pn, нанесених при $T_S = 370$К. Тому найбільш ефективні СЕ можуть бути виготовлені на основі ГС MPP/Pn. Також перспективними є ГС PbPc/MPP, які мають меншу величину величину $\Delta E_C$ та ефективно поглинають сонячне світло з утворенням нерівноважних носіїв заряду в більш широкій спектральній області, ніж інші досліджувані ГС. Отже, двошарові анізотипні ГС MPP/Pn та PbPc/MPP отримані при $T_S$ до 370К є перспективними елементами для розробки органічних фотоперетворювачів, в тому числі СЕ [164,165].

# РОЗДІЛ 5

# ФОТОВОЛЬТАЇЧНІ ВЛАСТИВОСТІ ОРГАНІЧНИХ ІЗОТИПНИХ ГЕТЕРОСТРУКТУР

Виготовлення та дослідження ізотипних ГС пов'язані з необхідністю покращення ефективності органічних СЕ. Однією з причин низької ефективності органічних СЕ є великий послідовний опір структур, який для неорганічних СЕ суттєво зменшують при формуванні p-p+ та n-n+ переходів шляхом легування відповідними домішками [24]. Для органічних СЕ такий спосіб зменшення послідовного опору є неефективним і технологічно складним. Щоб зменшити послідовний опір органічних СЕ запропоновано [166] використовувати ізотипні ГС (p-p+ або n-n+ типу), які можна отримувати в однакових технологічних умовах і одночасно з виготовленням анізотипних ГС. Крім того за допомогою ізотипних ГС можна збільшити ефективність поглинання сонячного світла з подальшим утворенням носіїв заряду в СЕ, якщо на ГР ізотипних ГС не утворюється велика концентрація поверхневих станів. Таким чином використання ізотипних ГС при розробці органічних СЕ може призвести до збільшення ефективності фотоперетворення, а, отже, і до зростання ККД СЕ.

Проте, на даний час фотовольтаїчні властивості органічних ізотипних ГС мало досліджено. Тому метою даної роботи було дослідження фотовольтаїчних властивостей ізотипних ГС на основі фоточутливих органічних НП, встановлення можливості отримання ізотипних ГС з малою $S$ біля ГР компонент і розширення спектральної області збирання сонячного світла.

В якості основного модельного об'єкта було вибрано p-p+ ГС на основі Pn, одного з найбільш фоточутливих органічних НП, який уже використовується для розробки органічних СЕ [167-169]. Іншими компонентами p-типу провідності ізотипних ГС використовувались шари НТР та PbPc. Спектральна область фоточутливості плівок PbPc та НТР зсунута в довгохвильову область в порівнянні з областю фоточутливості плівок Pn (рис. 4.2). Це дозволяє розширити спектральну область фоточутливості при створенні ГС Pn/PbPc та Pn/HTP в порівнянні з областю фоточутливості Pn. Внаслідок такого вибору компонент двошарові структури Pn/PbPc та Pn/HTP поглинають сонячне світло у видимій та ближній IЧ області спектру, що є необхідною умовою для виготовлення ефективних СЕ.

## 5.1. Дослідження ізотипних гетероструктур пентацен – фталоціанін свинцю (Pn/PbPc)

Для визначення оптимальних умов виготовлення ізотипних ГС Pn/PbPc були проведені детальні дослідження властивостей цих ГС. Показано, що для ГС Pn/PbPc отриманих при $T_S$ = 300 К спектральна залежність ФЕ корелює зі СП та ФЕ плівок PbPc в області 1.2-1.8 еВ (фактично в області прозорості плівок Pn). В області 1.8-2.5 еВ (область сильного поглинання плівок Pn) ФЕ змінює знак на протилежний (рис. 5.1). Тобто спектр ФЕ ГС Pn/PbPc подібний до різниці спектрів ФЕ шарів PbPc та Pn, що і призводить до формування спектральної залежності ФЕ різних знаків в областях сильного поглинання шарів PbPc та Pn.

Згідно з моделлю Ван Опдорп, запропонованою для неорганічних ізотипних ГС [24], така переміна знаку в спектрах ФЕ свідчить про формування біля ГР ізотипних ГС Pn/PbPc великої концентрації поверхневих станів (центрів захоплення та рекомбінації носіїв заряду). Наявність заряджених станів на ГР ізотипних ГС призводить до утворення двох збіднених шарів. В результаті чого, ізотипний гетероперехід можна описати еквівалентною схемою двох діодів Шотткі, послідовно з'єднаних назустріч один одному.

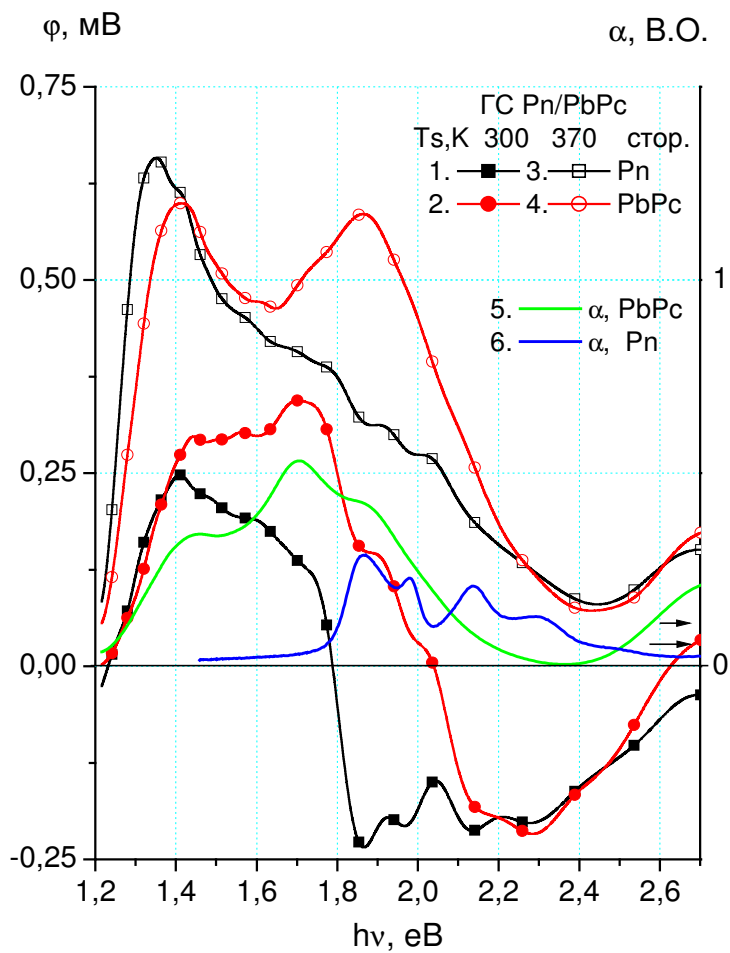

Рис. 5.1 Спектри ФЕ ($\varphi$) ізотипних ГС Pn/PbPc отриманих при $T_S$ = 300 (1,2) та 370 К (3,4) при освітленні зі сторони шару Pn (1,3) та шару PbPc (2,4) та СП шарів PbPc (5) та Pn (6) виготовлених при $T_S$ = 300 К.

Для ГС Pn/PbPc, отриманих при $T_S$ = 370 К, відсутня переміна знаку на спектрах ФЕ, а максимальне значення ФЕ на спектральній залежності в досліджуваній області 1.2-3.2 еВ майже в два рази більше, ніж ФЕ ГС Pn/PbPc, отриманих при $T_S$ = 300 К (рис. 5.1), і майже на порядок більше поверхневої ФЕ структур ITO/PbPc. Це свідчить про наявність невеликої концентрації центрів рекомбінації носіїв заряду на ГР ізотипних ГС Pn/PbPc отриманих при $T_S$ = 370 К. Спектри ФЕ ГС Pn/PbPc подібні до спектрів ФЕ плівок PbPc. Але в області смуги з максимумом при 1.85 еВ відносна інтенсивність ФЕ ГС Pn/PbPc на 30% більше, ніж плівок PbPc. Це обумовлено вкладом в формування ФЕ нерівноважних носіїв заряду, фотогенерованих в шарі Pn.

Отже, особливістю ГС Pn/PbPc є зменшення концентрації центрів рекомбінації носіїв заряду на ГР зі збільшенням $T_S$ виготовлення структур від 300 до 370 К. Однією з причин зменшення концентрації центрів рекомбінації носіїв заряду може бути більш ефективна дегазація молекул чи атомів повітря з вільної поверхні шарів Pn при $T_S = 370$ К, ніж при $T_S = 300$ К під час термічного нанесення у вакуумі верхнього шару PbPc. Схожа ситуація спостерігалась для анізотипних ГС на основі шарів n-типу провідності SnCl$_2$Pc та MPP.

## 5.2. Вплив підсвітки на параметри ізотипних гетероструктур пентацен / фталоціанін свинцю (Pn/PbPc)

Застосування еквівалентної схеми двох діодів Шотткі для опису фотовольтаїчних властивостей ізотипних ГС підтверджується дослідженнями впливу тильної немодульованої підсвітки на ФЕ ізотипних ГС. Оскільки внаслідок дії тильної підсвітки випромінюванням відповідних світлодіодів змінюються властивості бар'єрів локалізованих в шарах компонент ізотипних ГС, то це впливає на спектральні залежності ФЕ досліджуваних ГС.

Амплітуда від'ємної компоненти ФЕ ГС Pn/PbPc, отриманих при $T_S = 300$ К, зменшується зі збільшенням інтенсивності тильної підсвітки в області сильного поглинання шару Pn (hν = 2.2 еВ), яка переважно збуджує нерівноважні носії заряду в шарі Pn. При інтенсивності немодульованої підсвітки на порядок більше інтенсивності основного модульованого випромінювання від'ємна компонента ФЕ ГС Pn/PbPc зникає (рис. 5.2, кр. 3) і спектральна залежність ФЕ ГС стає схожою до СП плівки PbPc. Це вказує на зменшення потенціального бар'єру біля ГР ГС Pn/PbPc зі сторони Pn і, відповідно, зменшення вкладу шару Pn в ФЕ ГС.

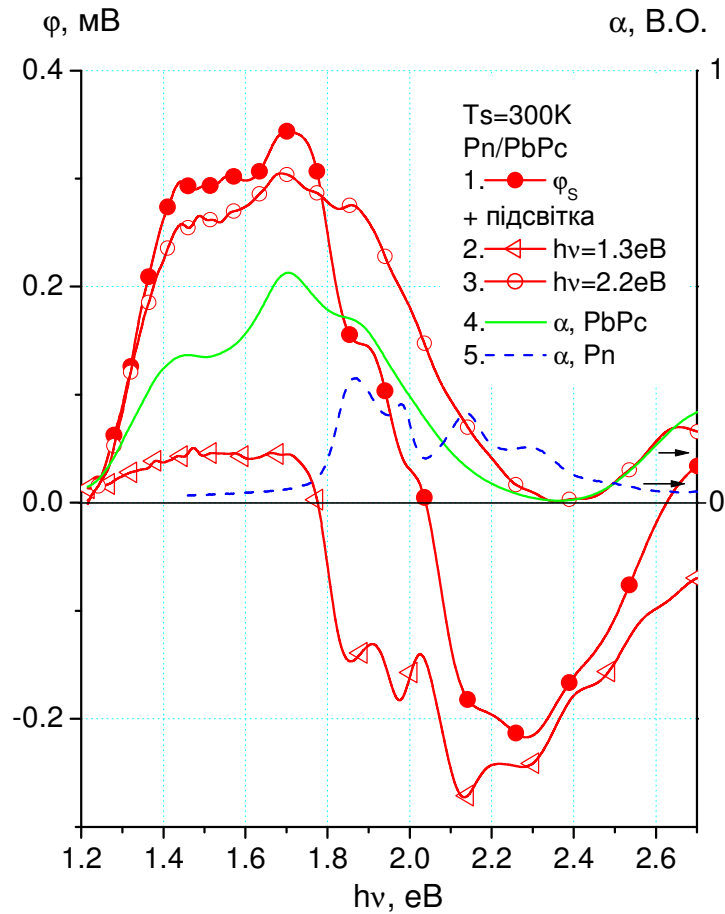

Рис. 5.2  Спектри ФЕ ($\varphi$) ГС Pn/PbPc при освітленні зі сторони шару PbPc без підсвітки (1) та з тильною підсвіткою світлодіодами з енергіями квантів h$\nu$ = 1.3 еВ (2), h$\nu$ = 2.2 еВ (3) та СП шарів PbPc (4) та Pn (5) при $T_S$ = 300 К.

А зі збільшенням інтенсивності тильної монохроматичної немодульованої підсвітки випромінюванням ІЧ світлодіода (h$\nu$ = 1.3 еВ) в області прозорості Pn і в області сильного поглинання шару PbPc зменшується амплітуда додатної компоненти ФЕ ізотипної ГС Pn/PbPc. При цьому амплітуда від'ємної компоненти ФЕ ГС Pn/PbPc зростає тільки на 20% зі збільшенням інтенсивності тильної підсвітки з h$\nu$ = 1.3 еВ (рис. 5.2, кр. 2), яка поглинається тільки плівкою PbPc, в той час як амплітуда додатної компоненти ФЕ ГС Pn/PbPc зменшується більше, ніж в 6 разів. Таким чином при дії підсвітки в області сильного поглинання шару PbPc спектр ФЕ ГС Pn/PbPc корелює зі СП або спектром ФЕ плівок Pn (рис. 5.2, кр. 2). Тобто під впливом підсвітки відбувається зменшення вкладу шару PbPc, внаслідок зменшення потенціального бар'єру зі сторони PbPc.

Для ГС Pn/PbPc виготовлених при $T_S$ = 370 K в результаті тильної підсвітки немодульованим світлом з hν = 1.3 eB (рис. 5.3), яка поглинається тільки шаром PbPc, інтенсивність смуги ФЕ ГС Pn/PbPc з максимумом при

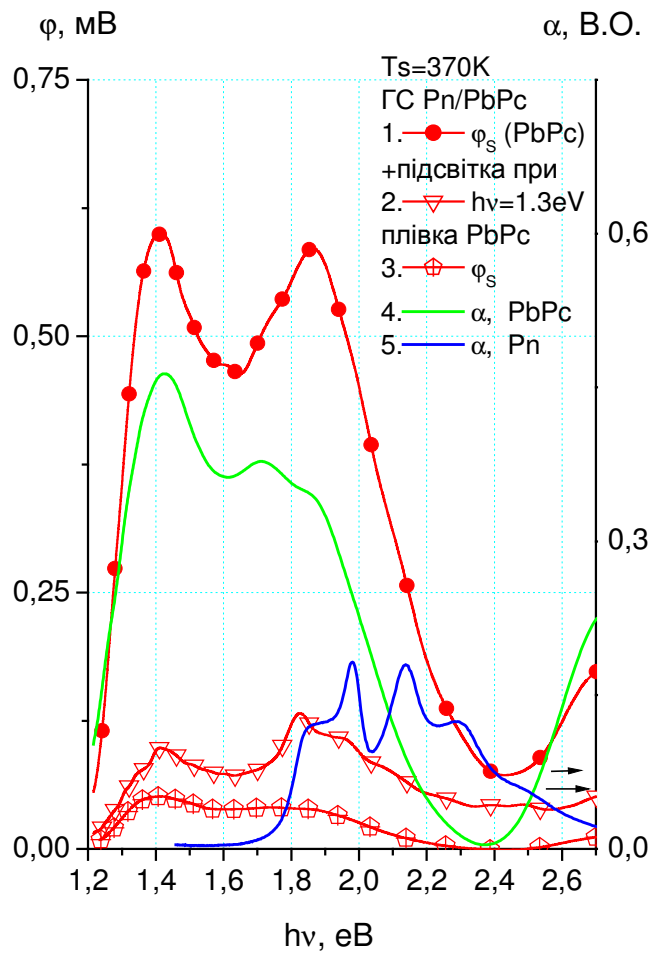

Рис. 5.3  Спектри ФЕ ($\varphi$) ГС Pn/PbPc при освітленні зі сторони шару PbPc без підсвітки (1) та при тильній підсвітці з hν = 1.3 eB (2), спектри ФЕ ($\varphi$) плівки PbPc при освітленні зі сторони вільної поверхні (3) та СП шарів PbPc (4) та Pn (5). Всі структури отримані при $T_S$ = 370 K.

1.85 еВ стає більшою інтенсивності смуги з максимумом при 1.4 еВ. Це обумовлено зменшенням вкладу шару PbPc в ФЕ ГС Pn/PbPc і наглядно показує вклад шару Pn в формування ФЕ ГС Pn/PbPc (рис. 5.3, кр. 2).

Згідно з моделлю Ван Опдорп для ізотипних ГС [24] при великій концентрації поверхневих станів (центрів рекомбінації носіїв заряду) на ГР ізотипних ГС, таку ГС можна описати еквівалентною схемою двох діодів Шотткі з'єднаних назустріч один одному, а $U_{xx}$ (або ФЕ) представити у вигляді:

$$\varphi = \varphi_1 \cdot \ln(1 + \frac{I_{ph1}}{I_{S1}}) - \varphi_2 \cdot \ln(1 + \frac{I_{ph2}}{I_{S2}}) =$$

$$= \varphi_1 \cdot \ln(1 + \gamma_1 \cdot P) - \varphi_2 \cdot \ln(1 + \gamma_2 \cdot P) , \qquad (5.1)$$

де $I_{ph1}$ та $I_{ph2}$ – густини фотострумів через відповідні діоди Шотткі (тобто діоди у відповідних компонентах ГС), а $I_{S1}$ та $I_{S2}$ – густини темнових струмів насичення. При цьому величина $I_{ph}$ прямопропорційна потужності освітлення ($P$) структур. З наведеного виразу видно, що у випадку збудження тільки шару PbPc в ГС Pn/PbPc (наприклад, ІЧ випромінювання) залежність $\varphi(P)$ повинна описуватися тільки одним членом (в нашому випадку першим – додатнім). А при збуджені світлом, яке поглинають обидві компоненти (червоний чи зелений світлодіод), залежність $\varphi(P)$ буде описуватися двома членами рівняння (5.1) і залежність $\varphi(P)$ може змінити знак. Якщо вклад другого (від'ємного) члена рівняння (5.1) нейтралізувати підсвіткою, зменшивши вклад відповідної компоненти (в нашому випадку вклад Pn зменшується за допомогою підсвітки з hν = 2.2 еВ), то основний вклад знову буде давати тільки перший (позитивний) член рівняння (5.1).

Для перевірки можливості використання моделі двох фотодіодів з'єднаних назустріч один одному для досліджуваних ГС Pn/PbPc, були проведені вимірювання залежностей $\varphi(P)$ в спектральній області, де вклад в формування ФЕ дають обидві компоненти ГС як при відсутності, так і при наявності тильної немодульованої підсвітки. При порівнянні залежностей $\varphi(P)$ без підсвітки та з підсвіткою видно, що тильна ІЧ підсвітка з hν = 1.3 еВ, яка поглинається і зменшує згин зон тільки в шарі PbPc, призводить до збільшення величини ФЕ в ГС Pn/PbPc, отриманих при $T_S$ = 300К (рис. 5.4, кр. 1,3). Тобто при відсутності підсвітки складові ФЕ, що виникають в ГС, компенсують одна одну. При зменшенні згину зон в шарі PbPc одна з складових ФЕ сильно зменшується, що призводить до зростання результуючої ФЕ. Це еквівалентно, схемі двох включених назустріч фотодіодів.

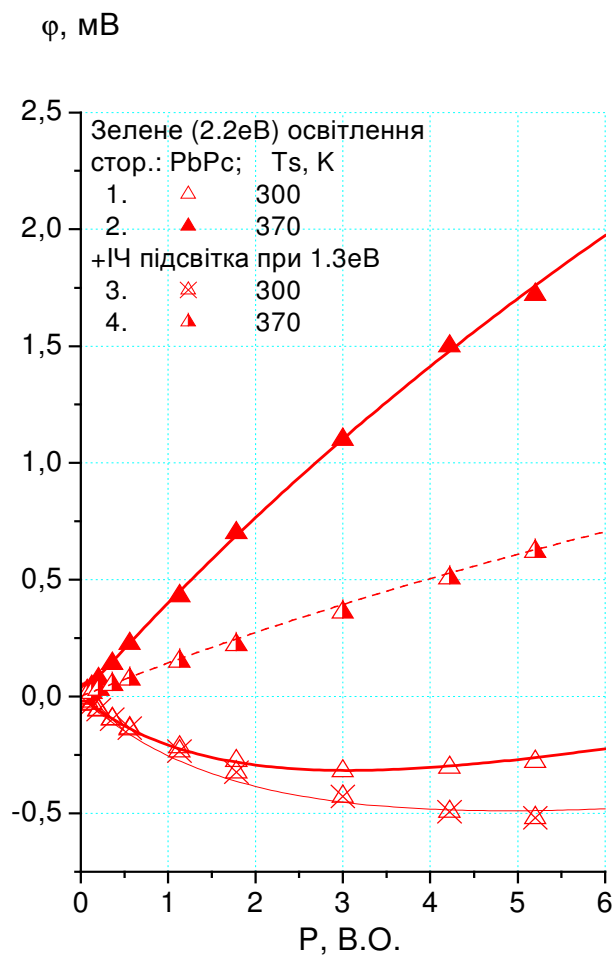

Рис. 5.4  Залежність ФЕ ($\varphi$) ГС Pn/PbPc від інтенсивності освітлення зеленим (hν = 2.2 еВ) світлодіодом зі сторони шару PbPc без підсвітки (1,2) та з (3,4) тильною підсвіткою (hν = 1.3 еВ); при $T_S$ = 300 К (1,3) та 370 К (2,4).

В той час як для ГС Pn/PbPc, отриманих при $T_S$ = 370 К, при дії аналогічної тильної підсвітки (hν = 1.3 eB), сигнали ФЕ зменшуються (рис. 5.4, кр. 2,4). Тобто при відсутності підсвітки складові ФЕ, які виникають в компонентах ГС, складаються, що еквівалентно схемі двох фотодіодів включених в одному напрямку [24]. Схема двох фотодіодів з'єднаних в одному напрямку використовується для опису ФЕ анізотипних ГС (див. п. 1.2) і в цьому випадку залежності $\varphi(P)$ описуються наступним виразом:

$$\varphi = \varphi_0 \cdot \ln\left[1 + \frac{I_{ph}}{I_S}\right] = \varphi_0 \cdot \ln[1 + \gamma \cdot P].$$ (5.2)

Оскільки основними процесами, що призводять до появи фотоструму в ізотипних ГС при відсутності станів на ГР та без врахування двоступінчатих механізмів фотогенерації носіїв заряду, є фотоемісія носіїв заряду через перехід в вузькозонному матеріалі та генерація електронно-діркових пар в широкозонному матеріалі. То загальний фотострум в ізотипних ГС є сумою фотострумів двох компонент ГС: $I_{ph} = I_{ph1} + I_{ph2}$ [24]. На відміну від ГС Pn/PbPc, отриманих при $T_S$ = 300 К, які мають велику концентрацію станів на ГР і згідно з рівнянням (5.1) можуть змінювати знак ФЕ в залежності від вкладу окремих компонент в результуючу ФЕ, то ГС Pn/PbPc, отримані при $T_S$ = 370 К, мають невелику концентрацію поверхневих станів і згідно з рівнянням (5.2) не змінюють знак ФЕ. Що і підтверджує наявність незначної $S$ на ГР шарів Pn та PbPc в ГС Pn/PbPc при $T_S$ = 370 К.

Отже, апроксимація експериментальних залежностей ФЕ від інтенсивності освітлення $\varphi(P)$ для ГС Pn/PbPc за допомогою рівнянь (5.1) та (5.2), підтверджує можливість застосування моделі Ван Опдорп для органічних ізотипних ГС (рис. 5.4). Так, залежності ФЕ від $P$ для ГС Pn/PbPc, отриманих при $T_S$ = 300 К, описуються за допомогою як додатної, так і від'ємної частини рівняння (5.1), що говорить про утворення двох різних бар'єрів на ГР компонент ізотипних ГС при великій концентрації поверхневих станів. При цьому згідно з виразом (1.7) в двошарових ГС при наявності великої *S* на ГР складова ФЕ широкозонної компоненти також може змінювати знак [24].

Як відомо, підсвітка зменшує величину потенціального бар'єру (згину зон) переважно в тій компоненті, яка поглинає випромінювання цієї підсвітки [125,126]. Очевидно, що ефективність зменшення ФЕ буде більшою в компоненті ГС, де величина згину зон більша. Тому за допомогою впливу немодульованої підсвітки на величину ФЕ (рис. 5.5) можна оцінити розподіл величини просторового заряду біля ГР в різних компонентах ГС.

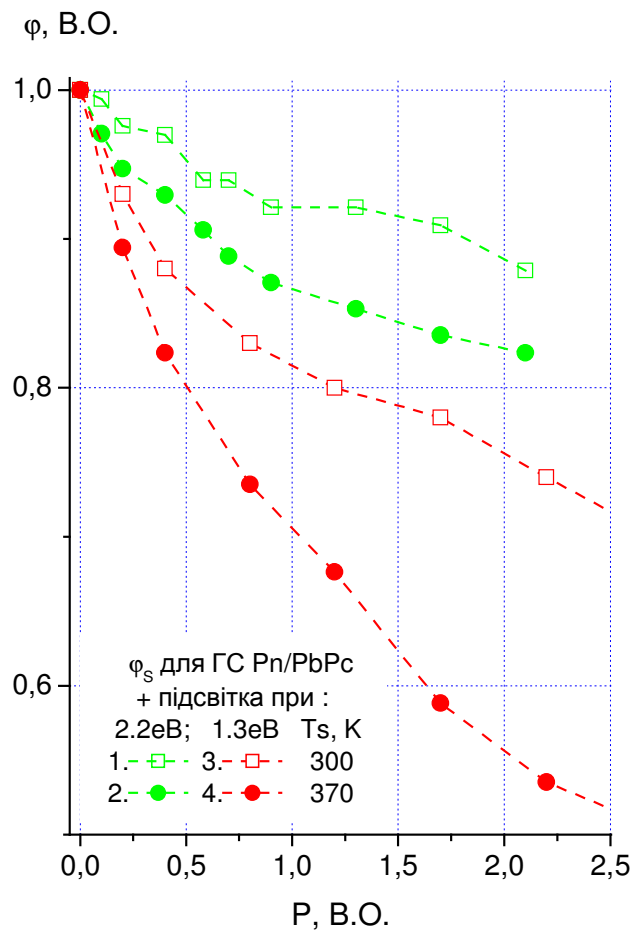

Рис. 5.5  Залежність відносної ФЕ ($\varphi$) для ГС Pn/PbPc при (модульованому) освітленні зі сторони шару PbPc (з hv = 1.45 еВ) від потужності немодульованої тильної підсвітки зеленим (hv = 2.2еВ) (1,2) та ІЧ (hv = 1.3 еВ) (3,4) світлодіодом. Для ГС Pn/PbPc отриманих при $T_S$ = 300 К (1,3) та 370 К (2,4).

З рис. 5.5 видно, що величина ФЕ ГС Pn/PbPc при підсвітці з hv = 1.3 eB сильніше зменшується, ніж при підсвітці з hv = 2.2 eB. Тобто немодульована тильна підсвітка при hv = 1.3 eB сильніше впливає на величину ФЕ ГС Pn/PbPc, ніж підсвітка при hv = 2.2 eB. При цьому слід відмітити, що обидві підсвітки (зеленим з hv = 2.2 eB (1,2) та ІЧ з hv = 1.3 eB (3,4) світлодіодом) більш ефективно впливають на ФЕ в структурах, отриманих при 370К. Оскільки підсвітка при hv = 1.3 eB поглинається тільки в шарі PbPc, а підсвітка в області hv = 2.2 eB поглинається переважно в шарі Pn. То звідси слідує, що в ізотипних ГС Pn/PbPc отриманих при $T_S$ = 300 та 370 К просторовий заряд в основному локалізований в шарі PbPc, а величина згину зон біля ГР більша в ГС Pn/PbPc, отриманих при $T_S$ = 370К.

## 5.3. Властивості ізотипних гетероструктур пентацен / гексатіопентацен (Pn/HTP)

Як було показано вище, в ГС Pn/PbPc при збільшенні $T_S$ від 300 до 370 К суттєво зменшується концентрація центрів рекомбінації носіїв заряду на ГР під час термічного нанесення у вакуумі шару PbPc на поверхню шарів Pn. Для перевірки існування цього ефекту в інших структурах ізотипні ГС Pn/HTP також були виготовлені при $T_S$ = 370 К. Видно, що в спектрах ФЕ (рис. 5.6) ГС Pn/HTP отриманих при $T_S$ = 370 К не спостерігається переміни знаку, як і для ГС Pn/PbPc, виготовлених при $T_S$ = 370 К. Це підтверджує припущення про зменшення концентрації центрів рекомбінації носіїв на вільній поверхні плівок Pn при $T_S$ = 370 К в процесі термічного напилення верхнього шару при утворенні ГС.

При порівнянні спектру ФЕ ГС Pn/HTP зі спектром ФЕ структур ITO/HTP (рис. 5.6 кр. 1,2), отриманих при $T_S$ = 370 К, видно, що нанесення верхнього шару HTP на поверхню Pn призводить до зростання ФЕ ГС

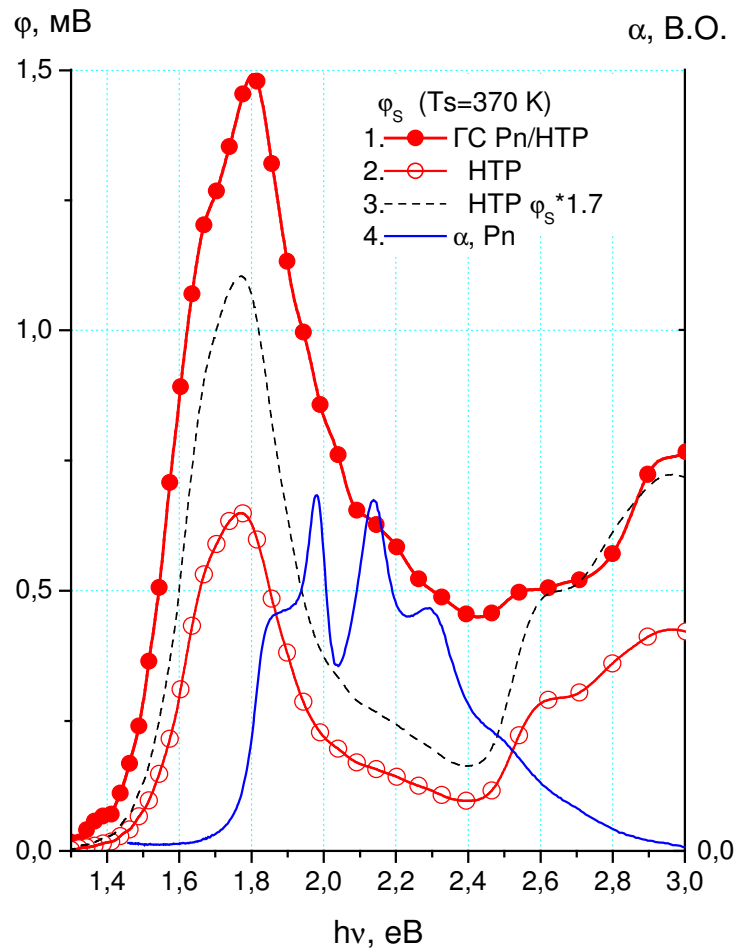

Рис. 5.6  Спектри ФЕ ($\varphi$) ізотипних ГС Pn/HTP при освітленні зі сторони шару HTP (1), структур ITO/HTP при освітленні вільної поверхні (2), спектральна залежність ФЕ структур ITO/HTP помножена на два (3) та СП шару Pn (4). Всі структури отримані при $T_S = 370$ К.

в області сильного поглинання шару HTP (1.4-1.8 еВ) майже в 2 рази в порівнянні з ФЕ шару HTP (рис. 5.6, кр. 3). А в області сильного поглинання шару Pn (1.8-2.4 еВ) при утворенні ГС Pn/HTP ФЕ збільшується приблизно в 3 рази, при цьому найбільше зростання ФЕ відбувається в області максимуму інтенсивної смуги поглинання Pn при hv = 2.10-2.15 еВ (рис. 5.6, кр. 4). Це свідчить про наявність фотогенерації носіїв заряду в обох шарах ГС Pn/HTP.

Для підтвердження формування просторового заряду в ГС Pn/HTP і його локалізації в різних компонентах ГС були проведені дослідження впливу немодульваної підсвітки з енергією випромінювання 1.9 і 2.2 еВ на спектральні залежності ФЕ ГС Pn/HTP. Спектри ФЕ ГС Pn/HTP під впливом підсвітки приведені на рис. 5.7. В результаті дії підсвітки з hν = 1.9 еВ, яка поглинається обома компонентами ГС Pn/HTP, ФЕ значно зменшується в спектральній області 1.4-2.3 еВ (рис. 5.7). Це свідчить про наявність просторового заряду в обох компонентах ГС Pn/HTP біля ГР. В той же час

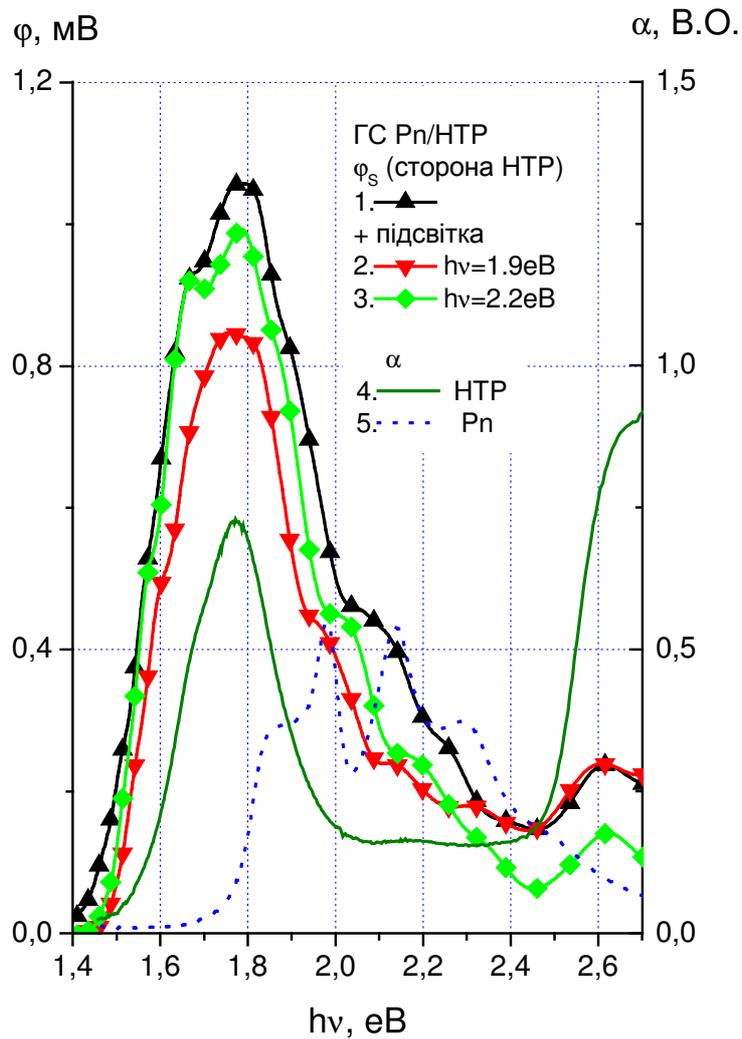

Рис. 5.7  Спектри ФЕ ($\varphi$) ГС Pn/HTP при освітленні зі сторони шару HTP без підсвітки (1) та при тильній підсвітці з hν = 1.9 еВ (2) та 2.2 еВ (3) та СП шарів HTP (4) та Pn (5). Всі структури отримані при $T_S$ = 370 К.

ефективність дії підсвітки з hv = 2.2 еВ, яка поглинається переважно шаром Pn, значно менша, ніж для підсвітки з hv = 1.9 еВ, що поглинається обома компонентами. Це говорить про локалізацію просторового заряду ГС Pn/НТР переважно в шарі НТР.

5.4. Порівняння результатів для ізотипних гетероструктур на основі пентацену

Порівняння спектрів ФЕ ізотипних ГС Pn/PbPc та Pn/НТР отриманих при $T_S$ = 370 К показує (рис. 5.8), що спектральна область фоточутливості ГС Pn/PbPc та Pn/НТР розширюється в довгохвильову область в порівнянні з

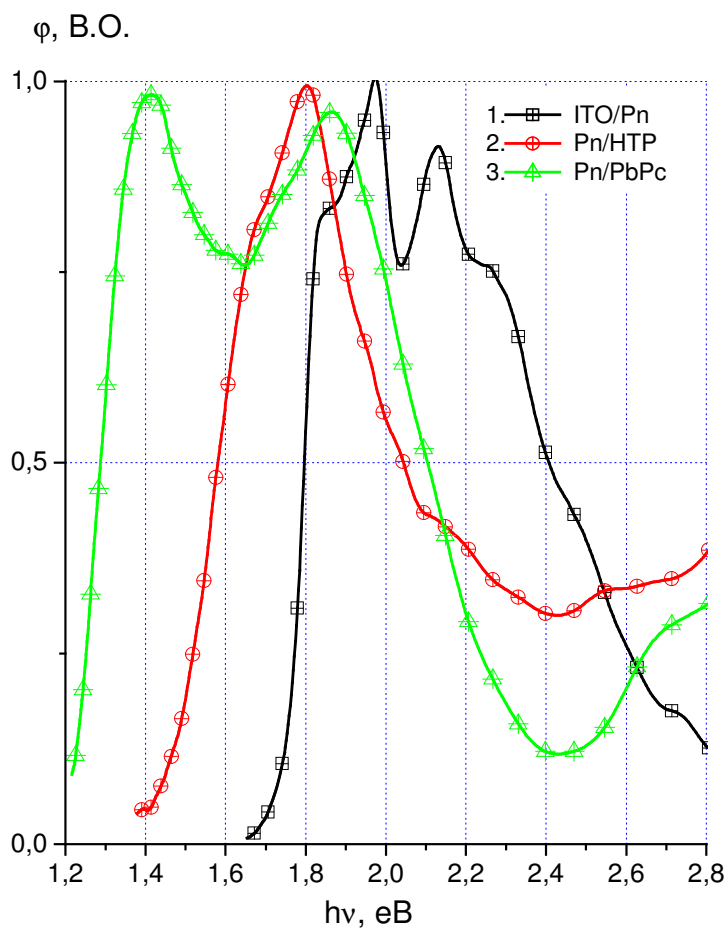

Рис. 5.8  Спектральні залежності ФЕ ($\varphi$) плівок Pn (1) та ізотипних ГС Pn/НТР (2), Pn/PbPc (3) виготовлених при $T_S$ = 370 К.

областю фоточутливості Pn відповідно на 0.3 і 0.5 еВ. В результаті чого значно збільшується ефективність поглинання сонячного світла та, відповідно, утворення неосновних носіїв заряду.

Слід відмітити, що інтегральні значення ФЕ ($\varphi_i$), отримані шляхом інтегрування спектрів ФЕ в області 1.2-3.1 еВ, для ГС Pn/HTP більше, ніж для ГС Pn/PbPc. Це узгоджується зі значеннями потенціального бар'єру ($V_D = E_{f(1)} - E_{f(2)}$.) енергетичних діаграм ізотипних гетеропереходів (без врахування поверхневих станів) у випадку максимально досяжних $V_D$, при використанні літературних даних (див. табл. 2.1). Проте, згідно з отриманими результатами, область фоточутливості ізотипних ГС Pn/PbPc розширюється найбільше, що робить їх перспективними структурами для покращення ефективності органічних СЕ.

Висновки до розділу 5

Біля ГР ізотипних ГС Pn/PbPc, отриманих термічним напиленням при $T_S = 300$ К, формується велика концентрація центрів рекомбінації носіїв заряду. Внаслідок цього на спектральній залежності ФЕ ГС Pn/PbPc спостерігається переміна знаку і властивості цих структур можна описати моделлю (запропонованою Ван Опдорпом для неорганічних ГС) з'єднаних назустріч двох діодів Шотткі, що свідчить про формування на ГР компонент цих ГС великої концентрації центрів рекомбінації носіїв заряду [170,171].

Для ГС Pn/PbPc отриманих при 370К не спостерігається переміна знаку ФЕ, а величина ФЕ більше, ніж для ГС Pn/PbPc, отриманих при $T_S = 300$К. Що вказує на зменшення концентрації поверхневих станів біля ГР для ГС Pn/PbPc зі збільшенням $T_S$ від 300 до 370 К. Отже, ізотипні ГС Pn/PbPc, напиленні на підкладки з $T_S = 370$ К, дозволяють розширити спектральну область поглинання сонячного світла, збільшити коефіцієнт збирання носіїв заряду, що може бути використано для підвищення ефективності багатошарових органічних СЕ [170,171].

В спектрах ФЕ ГС Pn/HTP отриманих при $T_S$ = 370 К не спостерігається переміна знаку, як і для ГС Pn/PbPc, виготовлених при $T_S$ = 370 К. Що говорить про зменшення концентрації центрів рекомбінації носіїв на вільній поверхні плівок Pn при $T_S$ = 370 К в процесі термічного напилення верхнього шару ГС. Для ГС Pn/HTP розширення спектру в довгохвильову область менше, а інтегральне значення ФЕ більше, ніж для ГС Pn/PbPc.

Отже, ізотипні ГС Pn/PbPc отримані при $T_S$ = 370К є перспективними елементами для створення p-p$^+$ переходів в багатошарових органічних СЕ з метою збільшення поглинання сонячного світла з подальшим утворенням носіїв заряду, оскільки область фоточутливості ізотипних ГС Pn/PbPc розширюється найбільше [170,171].

ЗАГАЛЬНІ ВИСНОВКИ

1. Визначено, що довжина дифузії локалізованих екситонів Френкеля в термічно напилених плівках НТР складає (200 ± 50) нм, $SnCl_2Pc$ – (130 ± 30) нм і практично не змінюється з підвищенням температури підкладки. В плівках МРР довжина дифузії локалізованих екситонів складає (25 ± 5) нм, і збільшується в два рази зі зростанням температури підкладки від 300 до 370 К.

2. Встановлено енергетичне положення СТ-станів в плівках $SnCl_2Pc$, обумовлених взаємодією центральних атомів Cl однієї молекули з периферійними С та Н атомами сусідніх молекул. Вклад СТ-станів в спектри поглинання плівок $SnCl_2Pc$ складає приблизно (15-20)%. Показано, що ефективність фотогенерації носіїв заряду при збуджені СТ-станів з енергіями 1.52±0.02 та 2.05±0.02 еВ більше, ніж при збуджені локалізованих екситонів.

3. Запропоновано схему енергетичної структури збуджених станів в плівках МРР, яка враховує взаємодію між молекулами МРР в межах шару та між шарами.

4. Показано, що швидкість поверхневої рекомбінації носіїв заряду в плівках МРР та НТР можна суттєво зменшити відпалом плівок при температурі 370 К або термічним напиленням при температурах підкладки 370 К.

5. Встановлено, що результати дослідження фотовольтаїчних властивостей анізотипних органічних ГС якісно узгоджуються з параметрами енергетичної схеми, побудованої за моделлю Андерсона, яка не враховує поверхневі стани.

6. Визначено, що властивості органічних ізотипних ГС Pn/PbPc, отриманих при температурі підкладки 300 К, добре описуються еквівалентною схемою двох з'єднаних назустріч діодів Шотткі, що говорить про формування на ГР компонент цих ГС великої концентрації центрів рекомбінації носіїв заряду.

7. Показано, що при підвищені температури підкладки до 370 К швидкість поверхневої рекомбінації носіїв заряду в ізотипних ГС Рn/PbPc сильно зменшується, а фоточутливість зростає. Тому ці ізотипні ГС можуть бути використані для створення p-p$^+$ переходів в багатошарових органічних СЕ для збільшення поглинання сонячного світла з подальшим утворенням носіїв заряду.

8. Встановлено, що досліджувані ізотипні та анізотипні ГС на основі органічних НП отримані при температурі підкладки до 370 К є перспективними елементами для розробки органічних фотоперетворювачів, в тому числі органічних СЕ.

СПИСОК ВИКОРИСТАНИХ ДЖЕРЕЛ